%% file: main_new.tex
\documentclass[sigconf,screen,nonacm]{acmart} 
\usepackage{algorithmic}
\usepackage{multicol}
\usepackage{graphicx}
\usepackage{textcomp}
\def\BibTeX{{\rm B\kern-.05em{\sc i\kern-.025em b}\kern-.08em
    T\kern-.1667em\lower.7ex\hbox{E}\kern-.125emX}}
\usepackage{dblfloatfix}   
\usepackage{balance}
\usepackage{xspace}
\usepackage[normalem]{ulem}
\usepackage{fancyhdr}
\usepackage[labelfont=bf, textfont=bf, skip=2pt]{caption}
\usepackage[english]{babel}
\usepackage{booktabs}
\usepackage{multirow}
\usepackage{pifont}
\usepackage{footnote}
\usepackage{enumitem}
\usepackage{ltablex}
\usepackage{makecell}
\usepackage{xargs} 
\usepackage{color, colortbl}
\usepackage[multiple]{footmisc}
\usepackage{marginnote}
\usepackage{clipboard}
\usepackage{subcaption}
\usepackage{mathtools}
\usepackage[compact]{titlesec}
\usepackage{wrapfig}
\usepackage{imakeidx}
\usepackage{soul}
\usepackage[keeplastbox]{flushend}
\usepackage[acronym,nonumberlist,nowarn]{glossaries}
    \glsdisablehyper
     \loadglsentries{acronyms}
\usepackage{imakeidx}
\usepackage{graphbox}
\usepackage{cuted,capt-of}
    
\setcopyright{none}

\begin{document}
\input{macros}

\title[DAMOV: A New Methodology and Benchmark Suite for Evaluating Data Movement Bottlenecks]{DAMOV: A New Methodology and Benchmark Suite for\\ Evaluating Data Movement Bottlenecks}

\newcommand{\tsc}[1]{\textsuperscript{#1}} 
\newcommand{\affilETH}{\tsc{1}}
\newcommand{\affilUIUC}{\tsc{2}}
\newcommand{\affilUT}{\tsc{3}}
\newcommand{\affilUM}{\tsc{4}}
\settopmatter{authorsperrow=1} 

\author{
 {
  Geraldo F. Oliveira\affilETH \qquad
  Juan Gómez-Luna\affilETH \qquad
  Lois Orosa\affilETH \qquad
  Saugata Ghose\affilUIUC
 }
}
\author{
 {
  Nandita Vijaykumar\affilUT \qquad
  Ivan Fernandez\affilETH$^,$\affilUM \qquad
  Mohammad Sadrosadati\affilETH \qquad
  Onur Mutlu\affilUIUC
 }
}
\affiliation{
\institution{
      \vspace{8pt}
      \affilETH ETH Zürich \qquad
      \affilUIUC University of Illinois Urbana-Champaign \qquad
      \affilUT University of Toronto \qquad 
      \affilUM University of Malaga
  }
  \country{}
}
 
\renewcommand{\authors}{Geraldo F. Oliveira, Juan Gómez-Luna, Lois Orosa, Saugata Ghose, Nandita Vijaykumar, Ivan Fernandez, Mohammad Sadrosadati, and Onur Mutlu}

\begin{abstract}
    \input{00_sec_abstract}

\end{abstract}


\renewcommand{\shortauthors}{G. F. Oliveira et al.}
\renewcommand{\shorttitle}{DAMOV: A New Methodology and Benchmark Suite for Evaluating Data Movement Bottlenecks}

\maketitle

\input{01_sec_introduction}

\input{02_sec_methodology}
\input{04_sec_scalability_analysis}
\input{05_sec_case_studies}

\input{06_sec_related_work}

\input{07_sec_conclusion}

\section*{Acknowledgments}
    \geraldorevi{We thank the SAFARI Research Group members for \gfii{valuable} feedback and the stimulating intellectual environment they provide. \juan{We acknowledge support from the SAFARI \gfiv{Research Group}'s industrial partners, especially ASML, Facebook, \gfii{Google,} Huawei, Intel, Microsoft, VMware, and \gfiv{the} Semiconductor Research Corporation. This research was partially supported by the ETH Future Computing Laboratory.}} \gfviii{An earlier version of this work was posted on arxiv.org (\url{https://arxiv.org/pdf/2105.03725.pdf}) on May 8, 2021. Talk videos for this work are available on YouTube, including a short talk video (\url{https://youtu.be/HkMYuYMuZOg}), a long talk video (\url{https://youtu.be/GWideVyo0nM}), and a tutorial on the DAMOV framework and benchmarks (\url{https://youtu.be/GWideVyo0nM?t=8028}).}

\balance 
\bibliographystyle{IEEEtran}
\bibliography{references}

\clearpage
\balance
\appendix
\noindent \textbf{\Large APPENDIX}
\section{Application Functions in the \bench Benchmark Suite}
\label{sec:benchlist}
\input{08_appendixA}

\clearpage
\section{Representative Application Functions}
\input{08_appendixB}

\label{sec:selectedworkloads}

\clearpage
\section{Complete List of Evaluated Applications}
\label{sec:evaluatedapp}
\input{08_appendixC}


\nobalance 
 \begin{wrapfigure}{l}{20mm} 
    \includegraphics[width=1in,height=1.25in,clip,keepaspectratio]{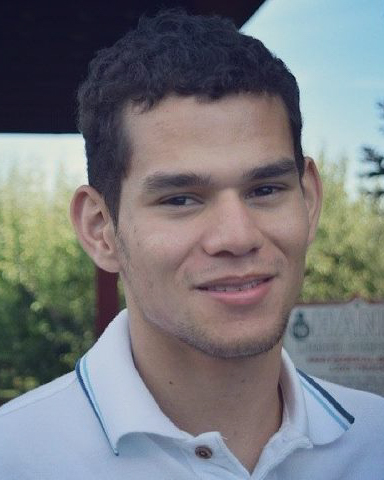}
  \end{wrapfigure}\par \noindent
  \textbf{\uppercase{Geraldo F. Oliveira}}  received a B.S. degree in computer science from the Federal University of Viçosa, Viçosa, Brazil, in 2015, and an M.S. degree in computer science from the Federal University of Rio Grande do Sul, Porto Alegre, Brazil, in 2017. Since 2018, he has been working toward a Ph.D. degree with Onur Mutlu at ETH Zürich, Zürich, Switzerland. His current research interests include system support for processing-in-memory and processing-using-memory architectures, data-centric accelerators for emerging applications, approximate computing, and emerging memory systems for consumer devices. He has several publications on these topics.\par

\vspace{10pt}
 \begin{wrapfigure}[8]{l}{20mm} 
    \includegraphics[width=1in,height=1.25in,clip,keepaspectratio]{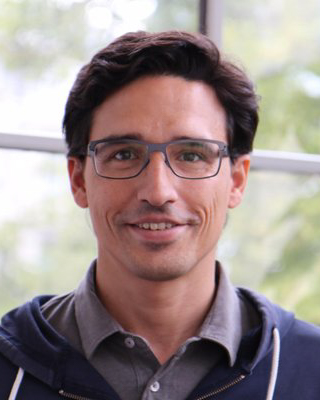}
  \end{wrapfigure}\par \noindent
  \textbf{\uppercase{Juan Gómez-Luna}} is a senior researcher and lecturer at SAFARI Research Group @ ETH Zürich. He received the BS and MS degrees in Telecommunication Engineering from the University of Sevilla, Spain, in 2001, and the PhD degree in Computer Science from the University of Córdoba, Spain, in 2012. Between 2005 and 2017, he was a faculty member of the University of Córdoba. His research interests focus on processing-in-memory, memory systems, heterogeneous computing, and hardware and software acceleration of medical imaging and bioinformatics. He is the lead author of PrIM (\url{https://github.com/CMU-SAFARI/prim-benchmarks}), the first publicly-available benchmark suite for a real-world processing-in-memory architecture, and Chai (\url{https://github.com/chai-benchmarks/chai}), a benchmark suite for heterogeneous systems with CPU/GPU/FPGA.\par

\vspace{10pt}
 \begin{wrapfigure}{l}{20mm} 
    \includegraphics[width=1in,height=1.25in,clip,keepaspectratio]{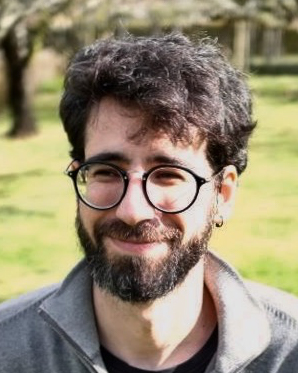}
  \end{wrapfigure}\par \noindent
  \textbf{\uppercase{Lois Orosa}} is a senior researcher in the SAFARI research group at ETH Zurich. His current research interest are in computer architecture, hardware security, memory systems, and machine learning  accelerators. He obtained his PhD from the University of Santiago de Compostela, and he was a PostDoc in the Institute of Computing  at University of Campinas. He was a visiting scholar at University of Illinois at Urbana-Champaign and Universidade NOVA de Lisboa, and he acquired industrial experience at several companies.\par
  
 \begin{wrapfigure}{l}{20mm} 
    \includegraphics[width=1in,height=1.25in,clip,keepaspectratio]{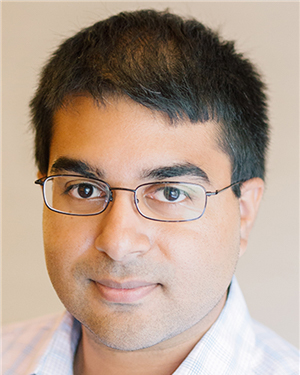}
  \end{wrapfigure}\par \noindent
  \textbf{\uppercase{Saugata Ghose}} is an assistant professor in the Department of Computer Science at the University of Illinois Urbana-Champaign.  He holds M.S. and Ph.D. degrees in electrical and computer engineering from Cornell University, and dual B.S. degrees in computer science and in computer engineering from Binghamton University, State University of New York.  Prior to joining Illinois, he was a postdoc and later a systems scientist at Carnegie Mellon University.  He received the best paper award from DFRWS-EU in 2017 for work on solid-state drive forensics, and was a 2019 Wimmer Faculty Fellow at CMU.   His current research interests include data-oriented computer architectures and systems, new interfaces between systems software and architectures, low-power memory and storage systems, and architectures for emerging platforms and domains.  For more information, please visit his website at \url{https://ghose.web.illinois.edu/}.\par

\vspace{15pt}
 \begin{wrapfigure}{l}{20mm} 
    \includegraphics[width=1in,height=1.25in,clip,keepaspectratio]{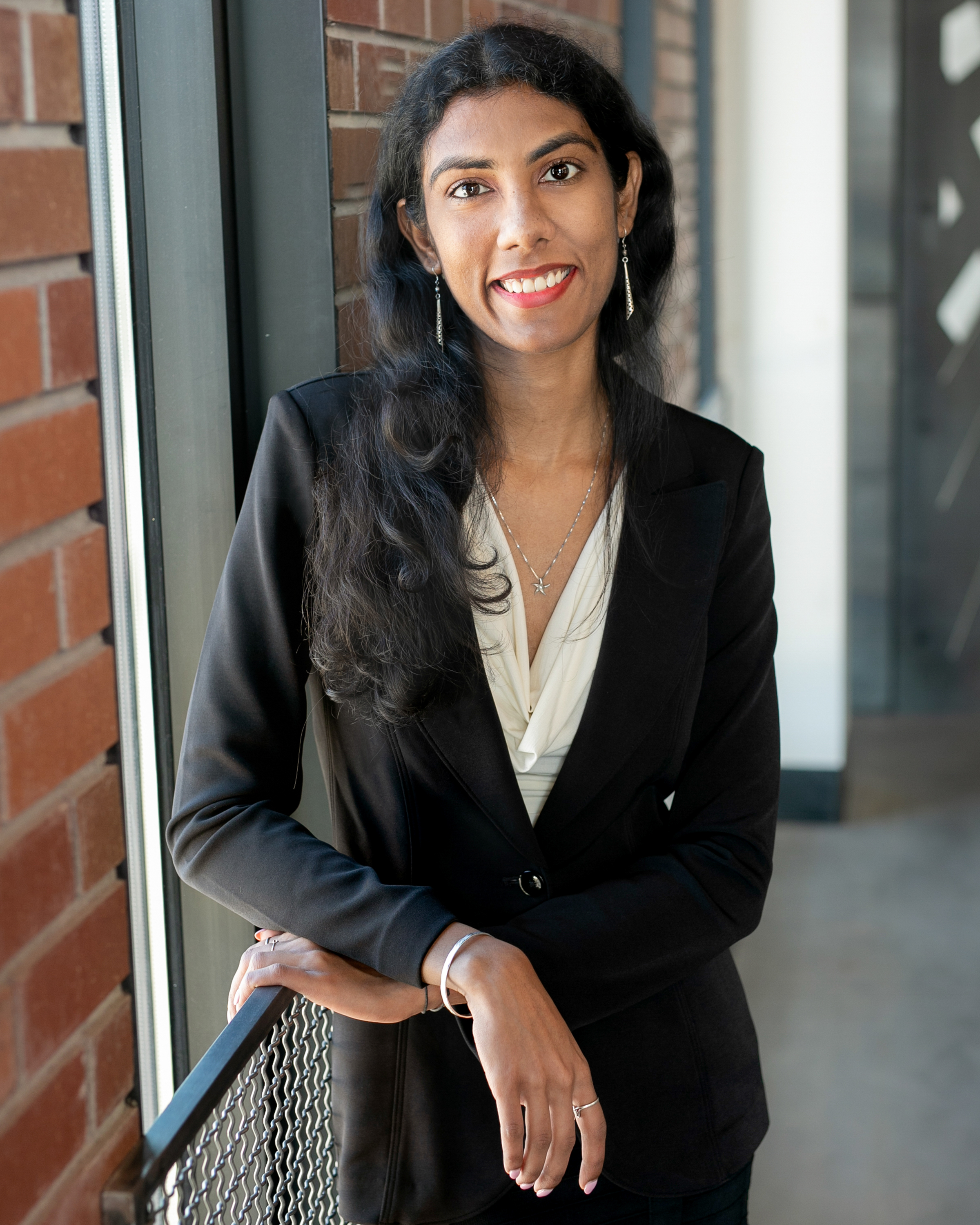}
  \end{wrapfigure}\par \noindent
  \textbf{\uppercase{Nandita Vijaykumar}} is an assistant professor in the Computer Science Department at the University of Toronto and the Department of Computer and Mathematical Sciences at the University of Toronto Scarborough. She is also affiliated with the Vector Institute for Artificial Intelligence. Before joining the University of Toronto, she was a research scientist in the Memory Architecture and Accelerator Lab at Intel Labs. She received her Ph.D. and M.S. in 2019 from Carnegie Mellon University where she was advised by Prof. Onur Mutlu and Prof. Phil Gibbons. She also worked with the Systems Group in the Computer Science Department at ETH Zurich as a visiting student. In the past, She have also worked for AMD, Intel, Microsoft, and Nvidia. Her research interests lie in the general area of computer architecture, compilers, and systems with a focus on the interaction between programming models, systems, and architectures. Her current interests are in the system-level and programming challenges of robotics and large-scale machine learning. For more information, please visit his website at \url{http://www.cs.toronto.edu/~nandita/}.\par
  
  \vspace{15pt}
 \begin{wrapfigure}{l}{20mm} 
    \includegraphics[width=1in,height=1.25in,clip,keepaspectratio]{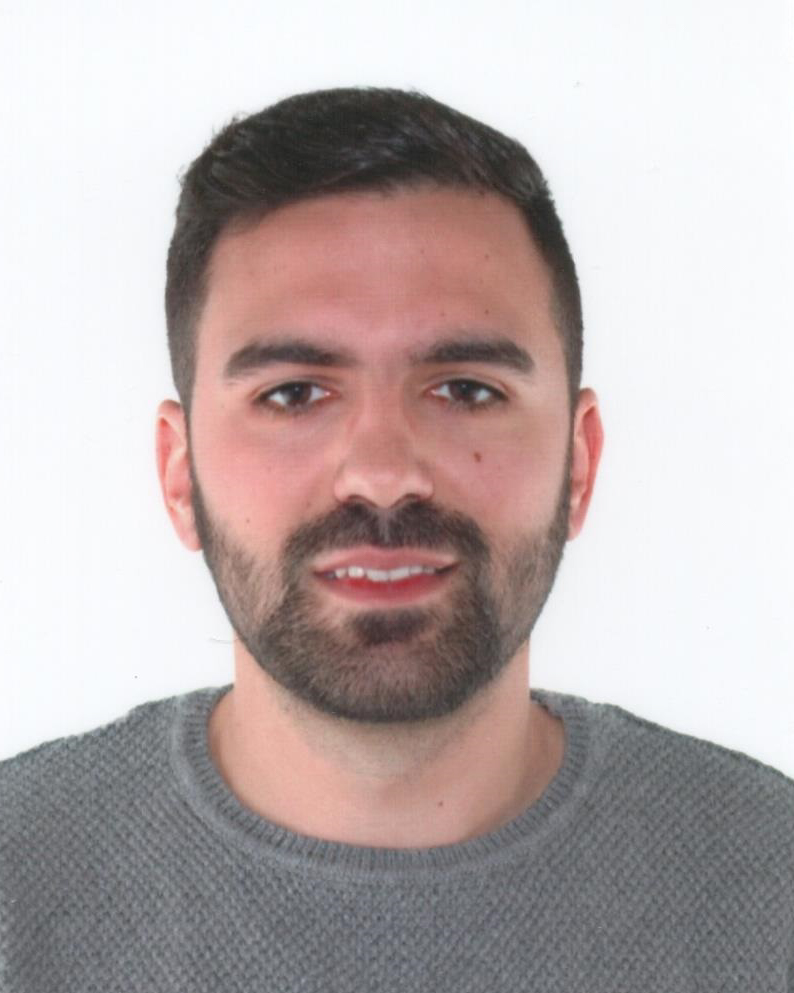}
  \end{wrapfigure}\par \noindent
  \textbf{\uppercase{Ivan Fernandez}} received his B.S. degree in computer engineering and his M.S. degree in mechatronics engineering from University of Malaga in 2017 and 2018, respectively. He is currently working toward the Ph.D. degree at the University of Malaga. His current research interests include processing in memory, near-data processing, stacked memory architectures, high-performance computing, transprecision computing, and time series analysis.\par
  
  \vspace{18pt}
 \begin{wrapfigure}{l}{20mm} 
    \includegraphics[width=1in,height=1.25in,clip,keepaspectratio]{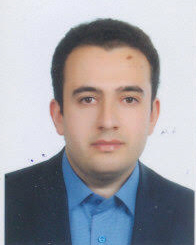}
  \end{wrapfigure}\par \noindent
  \textbf{\uppercase{Mohammad Sadrosadati}} received the B.Sc., M.Sc., and Ph.D. degrees in Computer Engineering from Sharif University of Technology, Tehran, Iran, in 2012, 2014, and 2019, respectively. He spent one year from April 2017 to April 2018 as an academic guest at ETH Zurich hosted by Prof. Onur Mutlu during his Ph.D. program. He is currently a postdoctoral researcher at ETH Zurich working under the supervision of Prof. Onur Mutlu. His research interests are in the areas of heterogeneous computing, processing-in-memory, memory systems, and interconnection networks. Due to his achievements and impact on improving the energy efficiency of GPUs, he won Khwarizmi Youth Award, one of the most prestigious awards, as the first laureate in 2020, to honor and embolden him to keep taking even bigger steps in his research career.\par
  
\vspace{15pt}
 \begin{wrapfigure}{l}{20mm} 
    \includegraphics[width=1in,height=1.25in,clip,keepaspectratio]{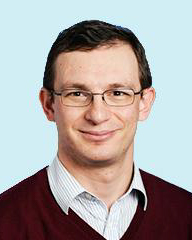}
  \end{wrapfigure}\par \noindent
  \textbf{\uppercase{Onur Mutlu}} is a Professor of Computer Science at ETH Zurich. He is also a faculty member at Carnegie Mellon University, where he previously held the Strecker Early Career Professorship. His current broader research interests are in computer architecture, systems, hardware security, and bioinformatics. A variety of techniques he, along with his group and collaborators, has invented over the years have influenced industry and have been employed in commercial microprocessors and memory/storage systems. He obtained his PhD and MS in ECE from the University of Texas at Austin and BS degrees in Computer Engineering and Psychology from the University of Michigan, Ann Arbor. He started the Computer Architecture Group at Microsoft Research (2006-2009), and held various product and research positions at Intel Corporation, Advanced Micro Devices, VMware, and Google. He received the IEEE High Performance Computer Architecture Test of Time Award, the IEEE Computer Society Edward J. McCluskey Technical Achievement Award, ACM SIGARCH Maurice Wilkes Award, the inaugural IEEE Computer Society Young Computer Architect Award, the inaugural Intel Early Career Faculty Award, US National Science Foundation CAREER Award, Carnegie Mellon University Ladd Research Award, faculty partnership awards from various companies, and a healthy number of best paper or ``Top Pick'' paper recognitions at various computer systems, architecture, and security venues. He is an ACM Fellow, IEEE Fellow for, and an elected member of the Academy of Europe (Academia Europaea). His computer architecture and digital logic design course lectures and materials are freely available on YouTube (\url{https://www.youtube.com/OnurMutluLectures}), and his research group makes a wide variety of software and hardware artifacts freely available online (\url{https://safari.ethz.ch/}). For more information, please see his webpage at \url{https://people.inf.ethz.ch/omutlu/}.\par

\end{document}

%% file: acronyms.tex
\newacronym{ADI}{ADI}{Alternating Direction Implicit}
\newacronym{AGU}{AGU}{Address Generation Unit}
\newacronym[longplural={Access History Tables}]{AHT}{AHT}{Access History Table}
\newacronym{AHT-R}{AHT-R}{Access History Table - Read}
\newacronym{AHT-W}{AHT-W}{Access History Table - Write-Back}
\newacronym{AIP}{AIP}{Access Interval Predictor}
\newacronym{AL}{AL}{Added Latency for column accesses}
\newacronym{ASIC}{ASIC}{Application Specific Integrated Circuit}
\newacronym{AVX}{AVX}{Advanced Vector Extensions}
\newacronym[longplural={Basic Block Vectors}]{BBV}{BBV}{Basic Block Vector}
\newacronym{BHT}{BHT}{Branch History Table}
\newacronym{HBM}{HBM}{Hybrid Bandwidth Memory}
\newacronym{DDR4}{DDR4}{Double-Data Rate 4}

\newacronym{NN}{NN}{Neural Network}
\newacronym{BTB}{BTB}{Branch Target Buffer}
\newacronym{CAS}{CAS}{Column Address Strobe}
\newacronym{AMAT}{AMAT}{Average Memory Access Time}
\newacronym{CCD}{CCD}{Column to Column Delay}
\newacronym{AI}{AI}{Arithmetic Intensity}
\newacronym[longplural={Columnar Database Systems}]{CDBS}{CDBS}{Columnar Database System}
\newacronym[longplural={Chip Multiprocessors}]{CMP}{CMP}{Chip Multiprocessor}
\newacronym{CMT}{CMT}{Chip Multithreading}
\newacronym[longplural={Coarse-Grain Reconfigurable Arrays}]{CGRA}{CGRA}{Coarse-Grain Reconfigurable Array}
\newacronym{C-RAM}{C-RAM}{Computational-RAM}
\newacronym{CWD}{CWD}{Column Write Delay}
\newacronym{DBI}{DBI}{Dynamic Binary Instrumentation}
\newacronym[longplural={Database Management Systems}]{DBMS}{DBMS}{Database Management System}
\newacronym{DDR}{DDR}{Double Data Rate}
\newacronym{DEWP}{DEWP}{Dead Line and Early Write-Back Predictor}
\newacronym{DRAM}{DRAM}{Dynamic Random Access Memory}
\newacronym{DSBP}{DSBP}{Dead Sub-Block Predictor}
\newacronym{DSM}{DSM}{Decomposed Storage Manage}
\newacronym{FAW}{FAW}{Four row Activation Window}
\newacronym{FP}{FP}{Floating-Point}
\newacronym{EDP}{EDP}{Energy-Delay Product}
\newacronym{FCFS}{FCFS}{First-Come First-Serve}
\newacronym{FIFO}{FIFO}{Fist In, First Out}
\newacronym{FSM}{FSM}{Finite State Machine}
\newacronym{FPGA}{FPGA}{Field-Programmable Gate Array}
\newacronym[longplural={Functional Units}]{FU}{FU}{Functional Unit}
\newacronym{GAg}{GAg}{Global Adaptive branch prediction using one Global PHT}
\newacronym{GAs}{GAs}{Global Adaptive branch prediction using per-Set PHT}
\newacronym{GCC}{GCC}{GNU Compiler Collection}
\newacronym{GEMS}{GEMS}{General Execution-driven Multiprocessor Simulator}
\newacronym[longplural={General Purpose Processors}]{GPP}{GPP}{General Purpose Processor}
\newacronym{HMC}{HMC}{Hybrid Memory Cube}
\newacronym{HIVE}{HIVE}{HMC Instruction Vector Extensions}
\newacronym{IATAC}{IATAC}{Inter-Access Time per Access Count}
\newacronym{ILP}{ILP}{Instruction Level Parallelism}
\newacronym{IPC}{IPC}{Instructions per Cycle}
\newacronym{ISA}{ISA}{Instruction Set Architecture}
\newacronym{IRAM}{IRAM}{Intelligent RAM}
\newacronym{KIPS}{KIPS}{Kilo Instructions per Second}
\newacronym{LDS}{LDS}{Linked Data Structure}
\newacronym{LFSR}{LFSR}{Linear Feedback Shift Register}
\newacronym{LFMR}{LFMR}{Last-to-First Miss-Ratio}
\newacronym{MLP}{MLP}{Memory-Level Parallelism}

\newacronym{LLC}{LLC}{last-level cache}
\newacronym{LRU}{LRU}{Least Recently Used}
\newacronym{LSU}{LSU}{Load Store Unit}
\newacronym{LTP}{LTP}{Last-Touch Predictor}
\newacronym{LvP}{LvP}{Live-time Predictor}
\newacronym{LWP}{LWP}{Last Write Predictor}
\newacronym{MARSS}{MARSS}{Micro Architectural and System Simulator}
\newacronym{McPAT}{McPAT}{Multi-core Power, Area, and Timing}
\newacronym{MIPS}{MIPS}{Microprocessor without Interlocked Pipeline Stages}
\newacronym{MOB}{MOB}{Memory Order Buffer}
\newacronym{MMU}{MMU}{Memory Management Unit}
\newacronym{MPKI}{MPKI}{Misses per Kilo-Instruction}
\newacronym{MSHR}{MSHR}{Miss-Status Handling Registers}
\newacronym{NAS}{NAS}{Numerical Aerodynamic Simulation}
\newacronym{NDP}{NDP}{Near-Data Processing}
\newacronym{NMP}{NMP}{Near Memory Processor}
\newacronym{NoC}{NoC}{Network-on-Chip}
\newacronym{NPB}{NPB}{NAS Parallel Benchmark}
\newacronym{NUCA}{NUCA}{Non-Uniform Cache Architecture}
\newacronym{NUMA}{NUMA}{Non-Uniform Memory Access}
\newacronym{OoO}{OoO}{Out-of-Order}
\newacronym{OpenMP}{OpenMP}{Open Multi-Processing}
\newacronym{OS}{OS}{Operating System}
\newacronym{PAg}{PAg}{Per-address Adaptive branch prediction using one Global PHT}
\newacronym{PAs}{PAs}{Per-address Adaptive branch prediction using per-Set PHT}
\newacronym{PC}{PC}{Program Counter}
\newacronym[longplural={Pointer-Chasing Engines}]{PCE}{PCE}{Pointer-Chasing Engine}
\newacronym{PCM}{PCM}{Performance Counter Monitor}
\newacronym{PIM}{PIM}{Processing-in-Memory}
\newacronym[longplural={Pattern History Tables}]{PHT}{PHT}{Pattern History Table}
\newacronym{RAPL}{RAPL}{Running Average Power Limit}
\newacronym{RAS}{RAS}{Row Address Strobe}
\newacronym{RAT}{RAT}{Registers Alias Table}
\newacronym{RC}{RC}{Row Cycle}
\newacronym{RCD}{RCD}{RAS to CAS Delay}
\newacronym[longplural={Row-based Database Systems}]{RDBS}{RDBS}{Row-based Database System}
\newacronym{ROB}{ROB}{Reorder Buffer}
\newacronym{RRD}{RRD}{Row to Row activation Delay}
\newacronym{RP}{RP}{Row Precharge}
\newacronym{RTL}{RTL}{Register Transfer Level}
\newacronym{RTP}{RTP}{Read To Precharge}
\newacronym{RVU}{RVU}{Reconfigurable Vector Unit}
\newacronym{SDP}{SDP}{Skewed Dead-Block Predictor}
\newacronym{SESC}{SESC}{Superescalar Simulator}
\newacronym{SSE}{SSE}{Streaming SIMD Extensions}
\newacronym{SFP}{SFP}{Spatial Footprint Predictor}
\newacronym{SiNUCA}{SiNUCA}{Simulator of Non-Uniform Cache Architectures}
\newacronym{SMT}{SMT}{Simultaneous Multi-Threading}
\newacronym{SIMD}{SIMD}{Single Instruction Multiple Data}
\newacronym{SPEC}{SPEC}{Standard Performance Evaluation Corporation}
\newacronym{SoC}{SoC}{System-on-Chip}
\newacronym{SPP}{SPP}{Spatial Pattern Predictor}
\newacronym{SRAM}{SRAM}{Static Random Access Memory}
\newacronym{SSOR}{SSOR}{Symmetric Successive Over-Relaxation}
\newacronym{SSV}{SSV}{Search Set Vector}
\newacronym{TLB}{TLB}{Translation Look-aside Buffer}
\newacronym{TLP}{TLP}{Thread Level Parallelism}
\newacronym[longplural={Through-Silicon Vias}]{TSV}{TSV}{Through-Silicon Via}
\newacronym{VWQ}{VWQ}{Virtual Write Queue}
\newacronym{WTR}{WTR}{Write Recovery time}
\newacronym{WR}{WR}{Write To Read delay time}

%% file: macros.tex
\definecolor{Gray}{gray}{0.9}
\definecolor{darkolivegreen}{rgb}{0.33, 0.42, 0.18}
\definecolor{plum}{rgb}{0.56, 0.27, 0.52}
\definecolor{islamicgreen}{rgb}{0.0, 0.56, 0.0}

\newcommand{\tempcommand}[1]{\renewcommand{\arraystretch}{#1}}
\newcommand{\circled}[1]{\tikz[baseline=(char.base)]{\node[shape=circle,draw,inner sep=0pt,fill=black, text=white] (char) {#1};}}
\definecolor{lightyellow}{rgb}{0.980, 0.956, 0.623}
\sethlcolor{lightyellow}
\definecolor{dgreen}{rgb}{0.00, 0.80, 0.00}
\definecolor{ddgreen}{rgb}{0.00, 0.50, 0.00}

\newif\ifsubmission
\submissiontrue

\ifsubmission
\newcommand\geraldo[1][0]{}
\newcommand\sg[1][0]{}
\newcommand\revision[1][0]{}
\else
\newcommand{\geraldo}[1]{\textcolor{blue}{#1}}
\newcommand{\sg}[1]{\textcolor{orange}{#1}}
\fi

\newif\ifrevisionri
\revisionrifalse

\ifrevisionri
    \newcommand\geraldorevi[1][0]{}
\else
    \newcommand{\geraldorevi}[1]{\textcolor{black}{#1}} 
    \newcommand{\geraldorevii}[1]{\textcolor{black}{#1}} 
    \newcommand{\gfii}[1]{\textcolor{black}{#1}} 
    \newcommand{\gfiii}[1]{\textcolor{black}{#1}} 
    \newcommand{\gfiv}[1]{\textcolor{black}{#1}} 
    \newcommand{\gfv}[1]{\textcolor{black}{#1}} 
    \newcommand{\gfvi}[1]{\textcolor{black}{#1}}
    \newcommand{\gfvii}[1]{\textcolor{black}{#1}} 

    \newcommand{\gfviii}[1]{\textcolor{black}{#1}} 

    \newcommand{\sgv}[1]{\textcolor{black}{#1}} 

    \newcommand{\sgrvi}[1]{\textcolor{black}{#1}} 
    \newcommand{\juan}[1]{\textcolor{black}{#1}} 

    \newcommand{\jgl}[1]{}
    \newcommand{\jgll}[1]{[{\color{ddgreen}JGL: #1}]}
    \newcommand{\juang}[1]{\textcolor{black}{#1}}  

    \newcommand{\juangg}[1]{\textcolor{black}{#1}}
    \newcommand{\juanggg}[1]{\textcolor{black}{#1}} 
    \newcommand{\gf}[1]{[{\color{blue}GF: #1}]}
    \newcommand{\om}[1]{[{\color{red}OM: #1}]}
\fi

\DeclareRobustCommand{\hlcyan}[1]{{\sethlcolor{cyan}\hl{#1}}}

\newcommand{\bench}{{DAMOV}\xspace}

\DeclarePairedDelimiter\ceil{\lceil}{\rceil}
\DeclarePairedDelimiter\floor{\lfloor}{\rfloor}

\setlength{\textfloatsep}{4pt plus 1.0pt minus 2.0pt}
\setlength{\dbltextfloatsep}{4pt plus 1.0pt minus 2.0pt}
\setlength{\dblfloatsep}{4pt plus 1.0pt minus 2.0pt}
\setlength{\intextsep}{5pt plus 1.0pt minus 2.0pt}

 \titlespacing\section{0pt}{5pt plus 2pt minus 2pt}{0pt plus 2pt minus 2pt}
 \titlespacing\subsection{0pt}{5pt plus 2pt minus 2pt}{0pt plus 2pt minus 2pt}
 \titlespacing\subsubsection{0pt}{5pt plus 2pt minus 2pt}{0pt plus 2pt minus 2pt}

%% file: 00_sec_abstract.tex
%
Data movement between the CPU and main memory is a first-order obstacle \geraldorevi{against} improv\geraldorevi{ing} performance, scalability, and energy efficiency in modern systems. Computer systems employ a range of techniques to reduce overheads tied to data movement, spanning from traditional mechanisms (e.g., deep \geraldorevi{multi-level} cache hierarch\geraldorevi{ies}, aggressive hardware prefetcher\geraldorevi{s}) to emerging techniques such as Near-Data Processing (NDP), where some computation is moved close to memory. Prior NDP works investigate the root causes of data movement bottlenecks using different profiling methodologies and tools. However, there is still a lack of understanding about the key metrics that can identify different data movement bottlenecks and their relation to traditional and emerging data movement mitigation mechanisms. Our goal is to methodically identify potential sources of data movement over a broad set of applications \geraldorevi{and to comprehensively compare traditional compute-centric data movement mitigation techniques (e.g., cach\gfiii{ing} and prefetch\gfiii{ing}) to more memory-centric techniques (e.g., NDP), \gfiii{thereby developing a rigorous understanding of} the best techniques \geraldorevii{to mitigate} each source of data movement.}

With this goal in mind, we perform the first large-scale characterization of \geraldorevii{a wide variety} \gfiii{of} \sg{applications, across a wide range of application domains,} to identify fundamental program properties that lead to data movement to/from main memory. We develop the first systematic methodology to classify \gfiii{applications} based on the sources contributing to data movement bottlenecks. From our large-scale characterization \gfiii{of 77K functions across 345 applications}, we select 144 \geraldorevii{functions} to form the first open-source benchmark suite \geraldorevii{(\bench)} for main memory data movement studies. We select a diverse range of \geraldorevii{functions} that (1)~represent different types of data movement bottlenecks, and (2)~come from a wide range of application domains. Using NDP as a case study, we identify new insights about the different data movement bottlenecks and use these insights to determine the most suitable data movement mitigation mechanism for a particular application. \gfii{We open-source DAMOV and the complete source code for our new characterization methodology at~\url{https://github.com/CMU-SAFARI/DAMOV}.}

%% file: 01_sec_introduction.tex
\section{Introduction}

Today's \gfiii{computing} systems \gfiii{require} mov\gfiii{ing} data from main memory (consisting of DRAM) to the CPU cores \gfiii{so that computation can take place on the data}. Unfortunately, this \emph{data movement} is a major bottleneck for \geraldorevi{system} performance and energy \geraldorevi{consumption}\gfiii{~\cite{boroumand2018google,mutlu2013memory}}. DRAM technology scaling is failing to keep up with the increasing memory demand from applications~\gfvii{\cite{kang2014co, hong2010memory, kanev_isca2015, mutlu2015research, mutlu2013memory,mutlu2015main,kim2020revisiting, kim2014flipping,mutlu2017rowhammer,ghose2018your,mutlu2019rowhammer,frigo2020trrespass,liu2013experimental,liu2012raidr,patel2017reach,qureshi2015avatar,mandelman2002challenges,khan2014efficacy,khan2016parbor,khan2017detecting, lee2015adaptive,lee2017design,chang2017understanding,chang.sigmetrics2016,chang2014improving,meza2015revisiting,david2011memory,deng2011memscale}}, resulting in significant latency and energy costs due to data movement~\cite{
mutlu2013memory,
mutlu2015research,
dean2013tail,
kanev_isca2015,
ferdman2012clearing,
wang2014bigdatabench,
mutlu2019enabling,
mutlu2019processing,
mutlu2020intelligent,
ghose.ibmjrd19,
mutlu2020modern,
boroumand2018google, 
wang2016reducing, 
pandiyan2014quantifying,
koppula2019eden,
kang2014co,
mckee2004reflections,
wilkes2001memory,
kim2012case,
wulf1995hitting,
ghose.sigmetrics20,
ahn2015scalable,
PEI,
hsieh2016transparent,
wang2020figaro}. High-performance systems have evolved to include mechanisms that \gfiii{aim to} alleviate data movement's impact on \geraldorevi{system performance} and energy consumption, such as deep cache hierarchies and aggressive prefetchers. However, such mechanisms \gfiii{not only come with significant hardware cost and complexity, but they also} often fail to hide the \gfiii{latency and energy} costs of accessing DRAM in many modern and emerging applications~\cite{boroumand2018google, kanev_isca2015,
jia2016understanding, 
tsai:micro:2018:ams, sites1996}. These applications' memory behavior can \geraldorevi{differ} significantly from more traditional applications since \geraldorevi{modern applications} often \geraldorevi{have} lower memory locality, more irregular access patterns, and larger working sets~\gfiii{\cite{ghose.sigmetrics20, ghose.ibmjrd19,ahn2015scalable,seshadri2015gather,nai2017graphpim,hsieh2016accelerating,ebrahimi2009techniques,mutlu2003runahead, hashemi2016continuous,kim2018grim,cali2020genasm,amiraliphd}}. One promising technique that aims to alleviate the data movement bottleneck in modern and emerging applications is \gls{NDP}~\gfvii{\cite{lee2016simultaneous, ahn2015scalable, nai2017graphpim, boroumand2018google, lazypim, top-pim, gao2016hrl, kim2018grim, drumond2017mondrian, RVU, NIM, PEI, gao2017tetris, Kim2016, gu2016leveraging, HBM, HMC2, boroumand2019conda, hsieh2016transparent, cali2020genasm,Sparse_MM_LiM, NDC_ISPASS_2014, farmahini2015nda,loh2013processing,pattnaik2016scheduling,akin2015data, hsieh2016accelerating,babarinsa2015jafar,lee2015bssync, devaux2019true,
Chi2016, Shafiee2016, seshadri2017ambit, seshadri2019dram, li2017drisa, seshadri2013rowclone, seshadri2016processing, deng2018dracc, xin2020elp2im, song2018graphr, song2017pipelayer,gao2019computedram, eckert2018neural, aga2017compute,dualitycache, fernandez2020natsa, hajinazarsimdram,syncron,boroumand2021polynesia,boroumand2021mitigating,amiraliphd,oliveira2017generic,kim2019d,kim2018dram,besta2021sisa,ferreira2021pluto,seshadri2016buddy,boroumand2017lazypim,kim2017grim,ghose2018enabling,seshadri2018rowclone,mutlu2019processing,mutlu2019enabling,ghose2019workload,olgun2021quactrng,gokhale1995processing,Near-Data,nair2015evolution}},\footnote{We use the term \gls{NDP} to refer to \emph{any} type of Processing-\geraldorevi{in}-Memory~\cite{mutlu2020modern}.} where the cost of data movement to/from main memory is reduced by placing computation \gfiii{capability} close to memory. In \gls{NDP}, the computational logic close to memory has access to data \gfii{that resides in main memory} with significantly higher memory bandwidth, lower latency, and lower energy consumption than the CPU has in existing systems. There is very high bandwidth available to the cores in the logic layer of 3D-stacked memories, as demonstrated by many past works (e.g., \cite{ahn2015scalable, smc_sim, lazypim, lee2016simultaneous, NIM, NDC_ISPASS_2014,RVU,top-pim, boroumand2019conda, boroumand2018google,gao2017tetris, kim2018grim,cali2020genasm, fernandez2020natsa}). \geraldorevi{To illustrate this,} we use the STREAM Copy~\cite{mccalpin_stream1995} workload to measure the peak memory bandwidth the host CPU and an NDP architecture with processing elements in the logic layer of a single 3D-stacked memory (e.g., Hybrid Memory Cube~\cite{HMC2}) can leverage.\footnote{\geraldorevi{See Section~\ref{sec:methodology} for our experimental evaluation methodology.}} We observe that the peak memory bandwidth that the NDP \gfiii{logic} can leverage (431 GB/s) is 3.7$\times$ the peak memory bandwidth that the host CPU can exploit (115 GB/s). This happens since the external memory bandwidth is bounded by the limited number of I/O pins available in the DRAM device\gfiii{~\cite{lee2015decoupled}}.

Many recent works explore how \gls{NDP} can benefit various application domains, such as graph processing~\cite{PEI, ahn2015scalable, nai2017graphpim, song2018graphr, lazypim, boroumand2019conda, zhang2018graphp, angizi2019graphs, matam2019graphssd, angizi2019graphide, zhuo2019graphq}, \gfiii{machine learning}~\gfiii{\cite{gao2017tetris, Kim2016,Shafiee2016,Chi2016, boroumand2018google,boroumand2021mitigating,amiraliphd}}, bioinformatics~\cite{kim2018grim,NIM, cali2020genasm}, databases~\gfiii{\cite{drumond2017mondrian, RVU, seshadri2017ambit, lazypim, boroumand2019conda,hsieh2016accelerating,boroumand2021polynesia,amiraliphd}}, security~\gfvii{\cite{gu2016leveraging,kim2019d,kim2018dram}}, data manipulation~\cite{seshadri2017ambit,li2017drisa,seshadri2013rowclone, wang2020figaro, chang2016low, li2016pinatubo, rezaei2020nom, seshadri2015fast}, \geraldorevi{and mobile workloads~\gfiii{\cite{boroumand2018google,amiraliphd}}}. These works demonstrate that simple metrics such as \gfii{last-level CPU cache} \gls{MPKI} and \gls{AI} are useful \gfiii{metrics that serve} as a proxy for the amount of data movement an application experiences. \gfii{These metrics can be used} as a potential guide for choosing when to apply data movement mitigation mechanisms such as \gls{NDP}. However, such metrics \gfiii{(and the corresponding insights)} are often extracted from a small set of applications, with similar \gfii{or not-rigorously-analyzed} data movement characteristics. Therefore, it is difficult to generalize the metrics and insights these works provide to a broader set of applications, making it unclear what different metrics can reveal about a new \gfii{(i.e., previously uncharacterized)} application's data movement behavior (and how to mitigate its associated data movement costs).

We illustrate this issue by highlighting the limitations of two \gfii{different} methodologies commonly used to identify memory bottlenecks and often used as a guide to justify the use of \gls{NDP} architectures \gfii{for an application}: (a)~analyzing a roofline model~\cite{williams2009roofline} of the \gfii{application}, and (b)~using \geraldorevi{last-level CPU cache} \gls{MPKI} as an indicator of \gls{NDP} suitability \gfii{of the application}. The roofline model correlates the computation requirements of an application with its memory requirements under a given system. The model contains two \emph{roofs}: (1)~a diagonal line (\emph{y = Peak Memory Bandwidth $\times$ Arithmetic Intensity}) called the \emph{memory roof}, and (2)~a horizontal line (\emph{y = Peak System Throughput}) called the \emph{compute roof}~\cite{williams2009roofline}. If an application lies under the memory roof, the application is classified as \emph{memory-bound}; if an application lies under the compute roof, it is classified as \emph{compute-bound}. \gfiii{Many prior} works~\gfiii{\cite{azarkhish2018neurostream,ke2019recnmp,asgari2020mahasim,Kim2018HowMC,liang2019ins,glova2019near, fernandez2020natsa, gu2020dlux, singh2020nero,juansigmetrics21,radulovic2015another,boroumand2021mitigating,yavits2021giraf,herruzo2021enabling,asgarifafnir}} employ this \gfii{roofline} model to identify memory-bound applications that can benefit from \gls{NDP} architectures. Likewise, many prior works~\gfiii{\cite{hsieh2016accelerating,nai2017graphpim,kim2015understanding,boroumand2018google, nai2015instruction, ghose.ibmjrd19, lim2017triple, tsai:micro:2018:ams,kim2018coda,hong2016accelerating,gries2019performance}} observe that applications with high \geraldorevi{last-level cache} \gls{MPKI}\footnote{Typically, an \gls{MPKI} value greater than 10 is considered \emph{high} by prior works~\cite{hashemi2016accelerating,chou2015reducing,kim2010atlas,kim2010thread,muralidhara2011reducing, subramanian2016bliss,usui2016dash}.} are good candidates for \gls{NDP}.

\begin{figure*}[ht]
    \centering
    \includegraphics[width=0.92\linewidth]{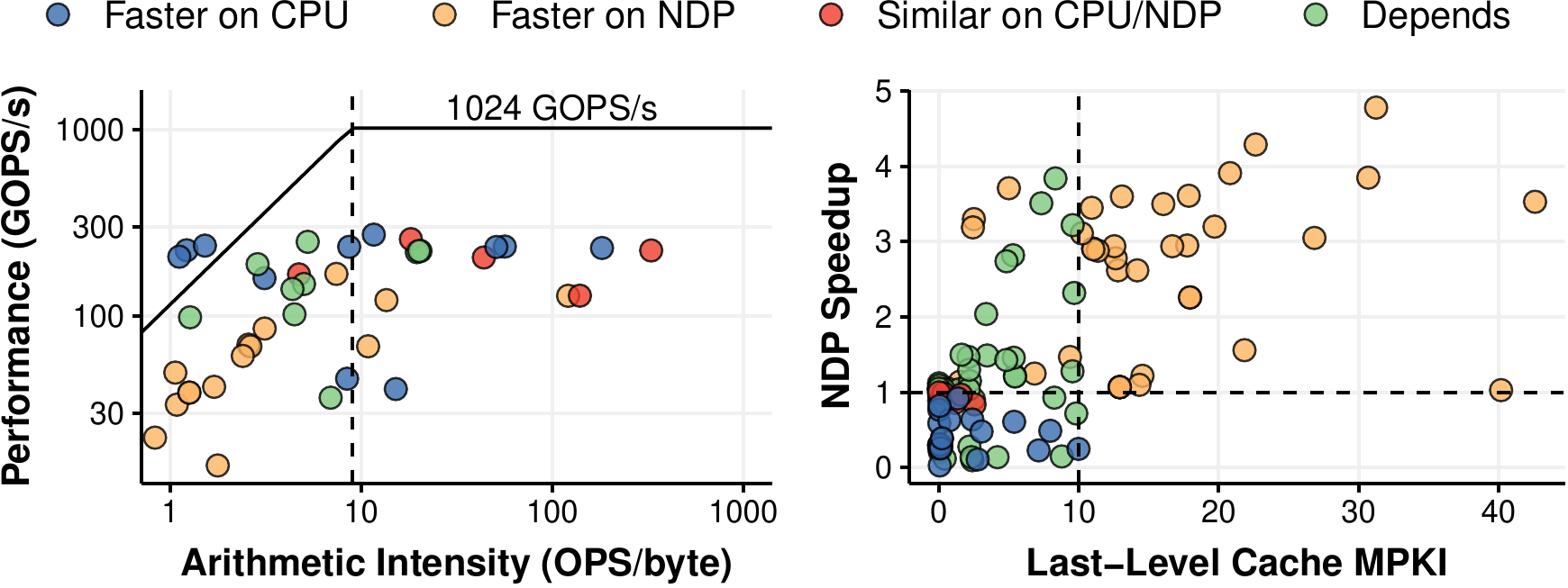}%
    \caption{Roofline (left) and \gfii{last-level cache} MPKI vs. NDP speedup (right) for \gfii{44} memory-bound applications. Applications are classified into four categories:
    (1)~those that experience performance degradation due to NDP (\gfii{blue; }\geraldorevi{Faster on CPU}), 
    (2)~those that experience performance improvement due to NDP (\gfii{yellow;} \geraldorevi{Faster on NDP}), 
    (\gfii{3})~\gfii{th\gfiii{o}se where the host CPU} and NDP performance are similar (\geraldorevi{\gfii{red;} Similar on CPU/NDP}),
    (\gfii{4})~those that experience either performance degradation or performance improvement due to NDP depending on the microarchitectural configuration (\geraldorevi{\gfii{green;} Depends}).}

    \label{fig_roofline_and_mpki}
\end{figure*}

Figure~\ref{fig_roofline_and_mpki} shows the roofline model (left) and a plot of \gls{MPKI} vs.\ speedup (right) of a system with general-purpose \gls{NDP} support over a baseline system without \gls{NDP} for a diverse set of 44~applications (see Table~\ref{tab:benchmarks}). In the \gls{MPKI} vs. speedup plot, the \gls{MPKI} corresponds to a baseline host \gfii{CPU} system. The speedup represents the performance improvement of a general-purpose \gls{NDP} system over the baseline (see Section~\ref{sec:step3} for our methodology). We make the following observations. First, analyzing the roofline model (Figure~\ref{fig_roofline_and_mpki}, left), we observe that most of the memory-bound applications (yellow dots) benefit from \gls{NDP}, as foreseen by prior works. We later observe (\geraldorevi{in} Section~\ref{sec_scalability_class1a}) that such applications are \gfii{DRAM} bandwidth-bound and are a natural fit for \gls{NDP}.  However, the \gfiii{roofline} model does \gfiii{\emph{not} accurately} account for \gfiii{the \gls{NDP} suitability of} memory-bound applications that 
(i)~benefit from \gls{NDP} \gfii{only} \geraldorevi{under} particular microarchitectural configurations, \gfii{e.g.}, either at low or high core counts (green dots, which are applications that are either bottlenecked by \gfii{DRAM} latency or suffer from \gfii{L3} cache contention; see Sections \mbox{\ref{sec_scalability_class1c}} and \mbox{\ref{sec_scalability_class2a}}); or 
(ii)~experience performance degradation when executed using NDP (blue dots, which are applications that suffer from the lack of a deep cache hierarchy in NDP architectures; see Section~\mbox{\ref{sec_scalability_class2c}}). Second, analyzing the \gls{MPKI} vs.\ speedup plot (Figure~\ref{fig_roofline_and_mpki}, right), we observe that while \geraldorevi{all} applications with high \gls{MPKI} benefit from \gls{NDP} (yellow dots \gfiii{with \gls{MPKI} higher than 10}), some applications with \emph{low} \gls{MPKI} can \geraldorevi{\emph{also}} benefit from \gls{NDP}  \gfiii{in \gfiv{all} of the \gls{NDP} microarchitecture configurations we evaluate (yellow dots with \gls{MPKI} lower than 10) or under specific \gls{NDP} microarchitecture configurations (green dots with \gls{MPKI} lower than 10)}. Thus, even though both the roofline model and \gls{MPKI} can identify \gfiii{some} specific sources of memory bottlenecks and can \gfiii{sometimes} be used as a proxy for \gls{NDP} suitability, \geraldorevi{they \gfiv{alone} cannot definitively determine \gls{NDP} suitability because they cannot} comprehensively identify different \gfiii{possible} sources of memory bottlenecks in a system.

Our \emph{goal} in this work is \geraldorevi{(1)} to understand the major sources of inefficiency that lead to data movement bottlenecks by observing \gfiii{and identifying} relevant metrics and \geraldorevi{(2)} to develop a benchmark suite for data movement that captures each of these sources. To this end, we develop a new three-step methodology to correlate application characteristics with the \emph{primary} sources of data movement bottlenecks and to determine the potential benefits of three example data movement mitigation mechanisms: (1) a deep cache hierarchy, (2) a hardware prefetcher, and (3) a general-purpose \gls{NDP} architecture.\footnote{We focus on these three data movement mitigation mechanisms for two different reasons: (1) deep cache hierarchies and hardware prefetchers are standard mechanisms in \geraldorevi{almost all modern} systems, and (2) \gls{NDP} represents a promising paradigm shift for many modern data-intensive applications.}  \geraldorevi{We use two main profiling strategies to gather key metrics from applications: (i) an architecture-independent profiling tool and (ii) an architecture-dependent profiling tool. The architecture-independent profiling tool provides metrics that characterize the application memory behavior independently of the underlying hardware. In contrast, the architecture-dependent profiling tool evaluates the impact of the system configuration \gfiii{(e.g., cache hierarchy)} on the memory behavior.} Our methodology has three steps. \geraldorevi{In \textit{Step~1}}, we use a hardware profiling tool to identify memory-bound functions \geraldorevi{across} many applications. This step allows for a quick first-level identification of many applications that suffer from memory bottlenecks \gfii{and functions that cause these bottlenecks}. \geraldorevi{In \textit{Step~2}, we use the architecture-independent profiling tool to collect metrics that provide insights about the memory access behavior of the \gfii{memory-bottlenecked} functions. In \textit{Step~3}, we collect architecture-\gfii{dependent} metrics and analyze the performance and energy of each function in an application when each of our three candidate data movement mitigation mechanisms is applied to the system.} By combining the data obtained from all three steps, we can systematically classify the leading causes of data movement bottlenecks \gfii{in an application or function} into different bottleneck classes. 

Using this \geraldorevi{new} methodology, we characterize a large\geraldorevi{,} heterogeneous set of applications (\geraldo{345~applications from \gfiii{37} different workload suites) across a wide range of domains.} \geraldo{\geraldorevi{Within} these applications,} we \gfiv{analyze 77K functions and} find \gfiii{a subset of} \geraldo{144}~functions \gfiii{from 74 different applications} that are memory\gfii{-}bound (and \geraldorevi{that} consume a significant fraction of the overall execution time). \geraldo{We} fully characterize \geraldo{this} set of 144 representative functions to serve as a core set of application kernel benchmarks\gfii{, which we release \gfiii{as the} open\gfiii{-}source \bench \gfiii{(\underline{DA}ta \underline{MOV}ement)} Benchmark Suite~\cite{damov}}. Our analyses reveal \gfii{six} new insights about the sources of memory bottlenecks and their relation \gfiii{to} \gls{NDP}:

\begin{enumerate}[noitemsep, leftmargin=*, topsep=0pt]
    \item \gfii{Applications with high last-level cache \gls{MPKI} and low temporal locality are \emph{DRAM bandwidth-bound}. These applications benefit from the large memory bandwidth available to the NDP system (Section~\ref{sec_scalability_class1a}).} 
    
    \item \gfii{Applications with low last-level cache \gls{MPKI} and low temporal locality are \emph{DRAM latency-bound}. These applications do \emph{not} benefit from L2/L3 caches. The NDP system improves performance and energy efficiency by sending L1 misses directly to DRAM (Section~\ref{sec_scalability_class1b}).} 
    
     \item \gfii{A second group of applications with low LLC MPKI and low temporal locality are \emph{bottlenecked by L1/L2 cache capacity}. These applications benefit from the NDP system at low core counts. However, at high core counts (and thus larger L1/L2 cache space), the caches capture most of the data locality in these applications, decreasing the benefits the NDP system provides  (Section~\ref{sec_scalability_class1c}). We make this observation using a \emph{new} metric that we develop, called \emph{last-to-first miss-ratio (LFMR)}, which we define as the ratio between the number of LLC misses and the total number of L1 cache misses. We find that this metric accurately identifies how efficient the cache hierarchy is in reducing data movement.}
        
    \item \gfii{Applications with high temporal locality and low LLC MPKI are \emph{bottlenecked by L3 cache contention} at high core counts. In such cases, the NDP system provides a cost-effective way to alleviate cache contention over increasing the L3 cache capacity (Section~\ref{sec_scalability_class2a}).}
    
    \item \gfii{Applications with high temporal locality, low LLC MPKI, and low \gls{AI} are bottlenecked by the \emph{L1 cache capacity}. The three candidate data movement mitigation mechanisms achieve similar performance and energy consumption for these applications (Section~\ref{sec_scalability_class2b}).}
    
    \item  \gfii{Applications with high temporal locality, low LLC MPKI, and high AI are \emph{compute-bound}. These applications benefit from a deep cache hierarchy and hardware prefetchers, but the NDP system degrades their performance (Section~\ref{sec_scalability_class2c}).}
    \\
\end{enumerate}

We publicly release our 144~\geraldorevi{\gfii{representative data movement bottlenecked} functions} \gfii{from 74 applications} as \gfii{the first} open-source benchmark suite for data movement\geraldorevi{, called \bench \gfii{Benchmark Suite}}, along with the \gfii{complete} source code for our new characterization methodology~\cite{damov}.

This work makes the following key contributions: 
\begin{itemize}[itemsep=0pt, topsep=0pt, leftmargin=*]
    \item We propose the first methodology to characterize data-intensive workloads based on the source of their data movement bottlenecks. This methodology is driven by insights obtained from a large-scale \geraldorevi{experimental} characterization of \geraldo{\sg{345}~applications from \sg{\gfii{37}} different benchmark suites} and an evaluation of \sg{the performance of memory-bound \gfii{functions} from these applications} with three data-movement mitigation mechanisms.
    \item We release \geraldorevi{DAMOV,} the first open-source \gfii{benchmark} suite for main memory data movement-related studies\gfiii{,} based on our systematic characterization methodology. This suite consists of 144~\gfii{functions} representing different sources of data movement bottlenecks and can be used as a baseline benchmark set for future data-movement mitigation research.
    \item \geraldorevi{We show how our \gfiii{\bench} benchmark suite can aid the study of open research problems for \gls{NDP} architectures\gfiii{, via} four case studies. \gfiii{In particular, we} evaluate (i) the impact of load balance and inter-vault communication \gfii{in \gls{NDP} systems}, (ii) the \gfiii{impact of} \gls{NDP} accelerators \gfiii{on our memory bottleneck analysis}, (iii) the \gfiii{impact of different core models on \gls{NDP} architectures}, and (iv) the potential benefits of identifying  simple \gls{NDP} \gfiii{instructions}. \gfii{We conclude that our benchmark suite and methodology can be employed to address many different open research \gfiii{and development} questions on data movement mitigation mechanisms, particularly topics related to \gls{NDP} systems and architectures.}}
\end{itemize}

%% file: 02_sec_methodology.tex
\glsresetall

\section{Methodology Overview}
\label{sec:methodology}

\begin{figure*}[h]
    \centering
    \includegraphics[width=\linewidth]{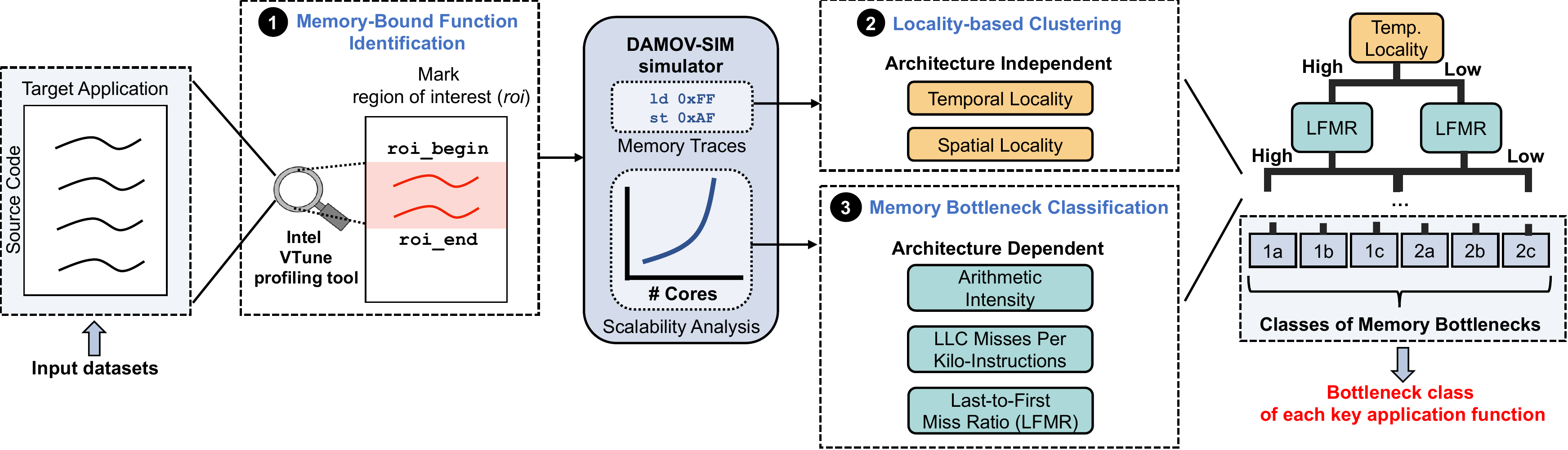}
    \caption{\geraldorevi{Overview of our three-step \gfii{workload characterization} methodology.}}
    \label{figure_methodology}
\end{figure*}

\sg{We develop a new} workload characterization methodology \sg{to analyze} data movement bottlenecks and the suitability \sg{of} different data movement mitigation mechanisms \sg{for these bottlenecks}, with a focus on \gls{NDP}. \sg{Our} methodology consists of three main steps, \geraldo{as Figure~\mbox{\ref{figure_methodology}} depicts:}
(1)~\sg{\emph{memory-bound function identification} using application profiling};
(2)~\emph{locality-based clustering} to analyze spatial and temporal locality in an architecture-independent manner; and
(3)~\emph{\juan{memory} bottleneck classification} using a scalability analysis to nail down the sources of memory boundedness\geraldorevi{, including architecture-dependent characterization}. \geraldo{Our methodology takes as input an application's source code and its \geraldorevi{input} datasets, \gfiii{and produces as output a classification of} the primary source of memory bottleneck of \gfiii{important functions in an application (i.e., bottleneck class of each key application function).}}
We illustrate the applicability of this methodology with a detailed characterization of 144 \sg{\gfii{functions}} that we select \sg{from} among \gfiv{77K analyzed functions of} \geraldo{345} \sg{characterized} applications. In this section, we give an overview of our workload characterization methodology. 
\sg{We use this methodology to drive the analyses \gfiii{we perform}} \juan{in Section~\ref{sec:characterization}}.

\subsection{Experimental Evaluation Framework}
\label{sec_damovsim}

\sg{As our scalability analysis depends on the \geraldorevi{hardware} architecture, we need \geraldorevi{a hardware} platform that can allow us to replicate and control all of our configuration parameters.  Unfortunately, such an analysis cannot be performed practically using real hardware, as
(1)~there are \geraldorevi{very} few available NDP hardware platforms, and the ones that currently exist do not allow us to comprehensively analyze our general-purpose NDP configuration \gfiii{in a controllable way} (as existing platforms \gfii{are} specialized \gfii{and non-configurable}); and
(2)~the configurations of real CPUs can vary significantly across the range of core counts that we want to analyze\gfiii{, eliminating the possibility of a carefully controlled study}.} \sg{As a result, we must rely on accurate simulation platforms to perform an accurate comparison across different configurations.
To this end, we build a framework that} integrates \sg{the ZSim CPU simulator~\gfiii{\cite{sanchez2013zsim}} with the Ramulator memory simulator~\gfiii{\cite{kim2016ramulator}}} to produce a fast, scalable, and cycle-accurate \gfii{open-source} simulator \geraldorevi{called \bench-SIM~\cite{damov}}. We use ZSim to simulate the core microarchitecture, cache hierarchy, coherence protocol, and prefetchers. \gfiii{W}e use Ramulator to simulate the DRAM \sg{architecture\gfii{, memory controllers,} and memory} accesses. To compute spatial and temporal locality, we modify ZSim to generate a single-thread memory trace for each \gfii{application}, which we use as input for the locality \gfiii{analysis} algorithm \gfiii{described} \sg{in Section~\ref{sec:step2} (which statically computes the temporal and spatial locality at word-level granularity)}.

\subsection{Step 1: Memory-Bound Function Identification}
\label{sec:step1}

The first step (\geraldorevi{labeled \ding{182} in \geraldo{Figure~\ref{figure_methodology})}} aims to identify \sg{\geraldorevi{the} functions of an application \geraldorevi{that} are \emph{memory-bound} (i.e., functions \geraldorevi{that} suffer from data movement bottlenecks)}.
These bottlenecks might be caused at any level of the memory hierarchy. There are various potential sources of memory boundedness, such as cache misses, cache coherence traffic, and long queu\geraldorevi{ing} latencies. Therefore, we need to take \geraldorevi{all such} potential causes into account. This step is optional if the \sg{\geraldorevi{application's \gfii{memory-bound functions (i.e.,} region\gfiii{s} of interest\gfii{, \emph{roi}, in Figure~\ref{figure_methodology})} \gfiii{are}} already known \emph{a priori}}.

Hardware profiling tools, both open-source and proprietary, are available to obtain hardware counters and metrics that characterize the application behavior on a computing system. 
In this work, we use \gfii{the} Intel VTune \gfiii{Profiler}~\cite{vtune}, which 
\juan{implements the well-known \gfiii{\emph{top-down analysis}}~\cite{yasin_ispass2014}\gfii{. \gfiii{T}op-down analysis \gfiii{uses} the available CPU hardware counters to hierarchically identify different sources of CPU system bottlenecks for an application}. \gfii{Among the \gfiii{various} metrics \gfiii{measured} by top-down analysis,} there is a relevant one} called \emph{Memory Bound}~\cite{MemoryBo20} that measures the percentage of CPU pipeline slots that are \gfiii{\emph{not}} utilized \gfii{due to} any issue related to data access. We employ this metric to identify functions that suffer from data movement bottlenecks
\sg{(\gfii{which we define as} functions where \emph{Memory Bound} is greater than 30\%)}.

\subsection{Step 2: Locality-Based Clustering}
\label{sec:meth-locality}
\label{sec:step2}

Two key properties of an \sg{application's memory access pattern are its} inherent spatial locality (i.e., \sg{the likelihood of accessing} nearby memory locations in the near future) and temporal locality (i.e., \sg{the likelihood of} accessing a memory location \gfiii{again} in the near future). \geraldorevi{These properties} are closely related to how well the application can exploit the memory hierarchy in computing systems and how accurate hardware prefetchers can be. Therefore, to understand the sources of memory bottlenecks for an application, we should analyze how much spatial and temporal locality its memory accesses \sg{inherently} exhibit. However, we should isolate these properties from particular configurations of the memory subsystem. Otherwise, it would be unclear if memory bottlenecks are due to the nature of the memory accesses or \sg{due to the characteristics and limitations} of the memory subsystem \gfiii{(e.g., limited cache size, too simple or inaccurate prefetching policies)}. \sg{As a result, in this step (\geraldorevi{labeled} \ding{183} in \geraldo{Figure~\ref{figure_methodology})}, we use \emph{architecture-independent} static} analysis to obtain spatial and temporal locality metrics for the functions selected in the previous step (Section~\ref{sec:step1}). Past works~\gfvii{\cite{Conte91,John98,weinberg2005quantifying, shao2013isa, zhong2009program,gu2009component, beard2017eliminating,lloyd2015memory,conte1991brief,conte1995advances}} propose different ways of analyzing \geraldorevi{spatial} and temporal locality in an architecture-independent manner. In this work, we use the definition of \gfiii{spatial and temporal} metrics presented in \cite{weinberg2005quantifying, shao2013isa}.

The spatial locality metric is calculated for a window of memory references\footnote{\gfiii{We compute both the spatial and temporal locality metrics at the word granularity. In this way, we keep our \gfiv{locality analysis} architecture-independent\gfiv{, using} \emph{only} properties of the application under study.}} of length \emph{W} \sg{using} Equation~\ref{eq:spatial}. First, \sg{for} every \emph{W} memory references, we calculate the minimum distance between any two addresses (\emph{stride}). Second, we create a histogram called \sg{the} \emph{stride profile}, where each bin \emph{i} stores how many times each \emph{stride} appears. Third, to calculate the spatial locality, we divide the \emph{percentage} of times stride $i$ is referenced ($stride\ profile(i)$) by the stride length $i$ \geraldorevi{and sum \gfii{the resulting} value across all \gfii{instances} of $i$}.

\begin{equation}
Spatial\ Locality = \sum_{i=1}^{\#bins } \frac{stride\ profile(i)}{i}
\label{eq:spatial}
\end{equation}

\noindent
A spatial locality \sg{value} close to 0 is caused by \geraldorevi{large} \emph{stride} values (e.g., regular accesses with \geraldorevi{large} strides\gfii{)} or random accesses, while a \sg{value} equal to 1 is caused by a \gfii{completely} sequential access pattern.

The temporal locality metric is calculated by using a histogram of reused addresses. First, we count the number of times each memory address is repeated in a window of \emph{L} memory references. Second, \gfii{we create a histogram called \textit{reuse profile}, where each bin $i$ represents the number of times a memory address is reused, expressed \juan{as} a power of 2.} For each memory address, we increment the bin that represents the corresponding number of repetitions. For example, \textit{reuse profile(0)} represents memory addresses that are reused only once. \textit{reuse profile(1)} represents memory addresses that are reused twice. Thus, if a memory address is reused \gfii{$N$} times, we increment \gfii{\textit{reuse profile($\floor{log_{2} N}$)}} by one. Third, we obtain the temporal locality metric with Equation~\ref{eq:temporal}.

\begin{equation}
Temporal\ Locality = \sum_{i=0}^{\#bins } \frac{ 2^{i} \times reuse\ profile(i)}{total\ memory\ accesses}
\label{eq:temporal}
\end{equation}

A temporal locality \sg{value} \gfii{of} 0 \sg{\gfiii{indicates}} no data reuse, \sg{while a value} close to 1 \sg{indicates \gfii{very} high data} reuse \geraldorevi{(i.e., a value equal to 1 means that the application accesses a \gfii{single} memory address continuously)}.

\sg{To calculate these metrics, we empirically select window lengths \emph{W} and \emph{L} \gfii{to} 32.  \gfii{We find that \gfiii{different values chosen for} \emph{W} and \emph{L} do not significantly change \gfiii{the conclusions of} our analysis. We observe that our \gfiii{conclusions} remain the same when we set both values to 8, 16, 32, 64, and 128.}}

\subsection{Step 3: Bottleneck Classification}
\label{sec:step3}

\sg{While Step~2 allows us to understand inherent application sources for memory boundedness, it is important to understand how \geraldorevi{hardware} architectural features can also result in memory bottlenecks.  As a result, in our third step (\geraldorevi{\ding{184} in} \geraldo{Figure~\ref{figure_methodology}}),} we perform a scalability analysis \gfii{of the functions selected in \emph{Step 1}, where we evaluate performance and energy scaling for three different system configurations. The scalability analysis} \sg{ makes use of} three \sg{\emph{architecture-dependent}} metrics: (1)~\textit{\gls{AI}}, (2)~\textit{\gls{MPKI}}, and (3)~a new metric called \textit{\gls{LFMR}}. We select these metrics for the following reasons. First, \sg{\gls{AI}} can measure the compute intensity of an application. Intuitively, we expect an application with high comput\geraldorevi{e} intensity to not suffer from severe data movement bottlenecks, as demonstrated by prior work~\cite{doerfler2016applying}. Second, \sg{\gls{MPKI} serves} as a proxy for the memory intensity of an application. It can also indicate the memory pressure \gfii{experienced} by the main memory \gfiii{system}~\gfiii{\cite{hashemi2016accelerating, PEI,mutlu2007stall,mutlu2008parallelism,kim2010atlas,hsieh2016transparent,subramanian2016bliss,ebrahimi2010fairness,hashemi2016continuous,ghose.sigmetrics20}}. Third, \sg{\gls{LFMR}, \gfii{a new metric we introduce} \geraldorevi{\gfii{and} is described in detail \juan{later} in this subsection}, indicates how efficient the cache hierarchy is in reducing} data movement. 

\sg{As part of our methodology development, we evaluate other metrics related to data movement, including} raw cache misses, coherenc\geraldorevi{e} traffic, and DRAM row misses/hits/conflicts. We observe that even though such metrics are useful for further characterizing an application (as we do in some of our \sg{later} analyses \geraldorevi{in Section~\ref{sec:scalability}}), they do not necessarily characterize a specific type of data movement \gfii{bottleneck}. We show in Section~\ref{sec_scalability_benchmark_diversity} that the \geraldorevi{three architecture-dependent and two architecture-independent} metrics we select for our classification are enough to \gfii{accurately characterize and} cluster the different types of data movement \gfii{bottlenecks} \gfiii{in a wide variety of applications}.

\subsubsection{Definition \gfiii{of Metrics.}} We define \gfiii{Arithmetic Intensity (\gls{AI})} as the number of \gfiii{arithmetic and logic} operations \gfiii{performed} per \geraldorevi{L1} cache line accessed.\footnote{\geraldorevi{We consider \gls{AI} \gfiii{to be} architecture-dependent since we consider the number of cache lines accessed by the application \gfii{(and hence the hardware cache block size)} to compute the metric. \sgv{This is the same definition of AI used by the hardware profiling tool \gfvi{we employ} in \emph{Step 1} (i.e., the Intel VTune Profiler~\cite{vtune}).}}} This metric indicates how much computation there \sg{is} per memory request. Intuitively, applications with high \gls{AI} are likely to be computationally intensive, while applications with low \gls{AI} tend to be memory intensive. We use \gls{MPKI} \sg{at the \gls{LLC}}, i.e., the number of \sg{\gls{LLC}} misses per one thousand instructions. This metric is considered \sg{to be} a good indicat\gfiii{or} of \gls{NDP} suitability by several \sg{prior} works~\cite{hsieh2016accelerating,nai2017graphpim,kim2015understanding,boroumand2018google, nai2015instruction, ghose.ibmjrd19, lim2017triple, tsai:micro:2018:ams,kim2018coda,hong2016accelerating}. We define the \gls{LFMR} of an application as the ratio between the number of \sg{\gls{LLC}} misses and the total number of L1 cache misses. We find that this metric \geraldorevi{accurately} identifies \geraldorevi{how much} an application benefits from the deep cache hierarchy of a \sg{contemporary} CPU. \sg{An \gls{LFMR} value} close to 0 means that the number of \sg{\gls{LLC}} misses is very small compared to the number of L1 misses, i.e., the L1 misses are likely to hit in \sg{the L2 or L3 caches.} However, \sg{an \gls{LFMR} value} close to 1 means that very few L1 misses \sg{\gfii{hit in} L2 or L3 cache\gfii{s}}, i.e., the application does not benefit \gfiii{much} from the deep cache hierarchy, \sg{and most L1 misses need to be serviced by main memory}.

\subsubsection{Scalability Analysis and System Configuration\gfiii{.}} \gfii{The goal of the scalability analysis \gfiii{we perform} is to nail down the specific sources
of data movement bottlenecks in the application. In \gfiii{this} analysis, we (i) evaluate the performance and energy scaling of an application in three different system configurations; and (ii) collect the key metrics for our bottleneck classification (i.e., \gls{AI}, \gls{MPKI}, and \gls{LFMR}).} \sg{During} scalability analysis, we simulate \gfii{three} \gfii{system} configurations of a general-purpose multicore processor:
\begin{itemize}[itemsep=0pt, topsep=0pt, leftmargin=*]
    \item \gfii{A host CPU with a deep cache hierarchy (i.e., private L1 (32~kB) and L2 (256~kB) caches, and a shared L3 (8~MB) cache with 16 banks). We call this configuration \textit{Host CPU}.}
    \item  \gfii{A host CPU with a deep cache hierarchy (same cache configurations as in \textit{Host CPU}), augmented  with a stream prefetcher~\cite{palacharla1994evaluating}. We call this configuration \textit{Host CPU with prefetcher}.}
    \item \gfii{An \gls{NDP} CPU with a single level of cache (only a private \gfiii{read-only}\footnote{\gfiii{We use read-only L1 caches to simplify the cache coherence model of the NDP system. Enabling efficient synchronization and cache coherence in NDP architectures is an open-research problem, as we discuss in Section~\ref{sec_scalability_limitations}.}} L1 cache (32~kB), as \gfiii{assumed} in many prior \gls{NDP} works~\gfiii{\cite{boroumand2019conda, lazypim, boroumand2018google,smc_sim, tsai:micro:2018:ams,syncron,fernandez2020natsa,singh2019napel,drumond2017mondrian,ahn2015scalable}}) and no hardware prefetcher. We call this configuration \textit{NDP}.}
\end{itemize}

\noindent The remaining components of the processor configuration are \sg{kept} the same (e.g., number of cores, instruction window size, branch predictor) \sg{to isolate the impact of only the caches\gfii{,} prefetcher\geraldorevi{s}\gfii{, and \gls{NDP}}}. This way, we expect that the performance and energy differences between \geraldorevi{the \gfii{three}} configurations \gfii{to} come \emph{exclusively} from the different data movement requirements. \sg{For \gfii{the three} configurations, we sweep the number of CPU cores in our analysis from 1 to 256, as previous works~\cite{drumond2017mondrian,santos2016exploring, ahn2015scalable} show that large core counts are necessary to saturate the bandwidth \geraldorevi{provided by} modern high-bandwidth memories, and because modern CPUs and NDP proposals can have varying core counts.  The core count sweep allows us to} observe \sg{(1)~}how \gfiii{an} \gfii{application's performance} changes when increasing the pressure on the memory subsystem, \sg{(2)~}how much \sg{\gls{MLP}~\gfiii{\cite{glew1998mlp,qureshi2006case,mutlu2008parallelism,mutlu2006efficient,mutlu2005techniques}}} the \gfii{application} has, and \sg{(3)~how much the cores} leverage the cache hierarchy and the available memory bandwidth. \gfii{We proportionally increase the size of the CPU's private L1 and L2 caches  when increasing the number of CPU cores in our analysis (\gfiii{e.g., when} scaling the CPU core \gfiii{count} from 1 to 4, we also scale the aggregated L1/L2 cache size by a factor of 4).} \gfii{We use out-of-order and in-order CPU  cores \gfiii{in} our analysis for all three configurations. In this way, we \geraldorevi{build confidence} that our trends and findings are independent of a specific underlying general-purpose core microarchitecture.} \sg{We simulate a memory architecture similar to the  \gls{HMC}~\cite{HMC2}, where \gfiii{(1)} the host CPU accesses memory through a high-speed off-chip link, and \gfiii{(2)} the \gls{NDP} logic \gfiii{resides} in the logic layer of the memory chip and has direct access to the DRAM banks (thus taking advantage of higher \gfiii{memory} bandwidth and lower \gfiii{memory} latency).} \sg{Table~\ref{table_parameters} lists \geraldorevi{the} \gfiii{parameters} \geraldorevi{of} our \geraldorevi{host CPU, \gfiii{host CPU with prefetcher,} and \gls{NDP}} \geraldorevi{baseline} configuration\geraldorevi{s}.}

\begin{savenotes}
\begin{table*}[!t]
    \tempcommand{1.3}
    \centering
    \caption{Evaluated Host CPU and NDP system configurations.}
    \label{table_parameters}
    \footnotesize
    \resizebox{0.85\textwidth}{!}{
        \begin{tabular}{|c  l|}
        \hline
        \rowcolor{Gray}
        \multicolumn{2}{|c|}{ \textbf{Host CPU Configuration}}\\
        \hline
        \hline 
        \multirow{4}{*}{\shortstack{\textbf{Host CPU}\\ \textbf{Processor}}} &  1, 4, 16, 64, and 256~cores @2.4~GHz, 32~nm; 4-wide out-of-order   \\
                                            & 1, 4, 16, 64, and 256~cores @2.4~GHz, 32~nm; 4-wide in-order   \\
                                            & Buffers: 128-entry ROB; 32-entry LSQ (each)\\
                                            & Branch predictor: Two-level GAs~\cite{yeh1991two}. 2,048~entry BTB; 1~branch per fetch    \\
                                            
        \hline
        \multirow{2}{*}{\shortstack{\textbf{Private}\\ \textbf{L1 Cache}}} & 32~KB, 8-way, 4-cycle; 64~B line; LRU policy \\
                                                   & Energy: 15/33~pJ per hit/miss~\cite{muralimanohar2007optimizing, tsai:micro:2018:ams}                                          \\
            
        \hline
       \multirow{3}{*}{\shortstack{\textbf{Private}\\ \textbf{L2 Cache}}} & 256~KB, 8-way, 7-cycle; 64~B line; LRU policy \\
                                                  & MSHR size: 20-request, 20-write, 10-eviction \\
                                                  & Energy: 46/93~pJ per hit/miss~\cite{muralimanohar2007optimizing, tsai:micro:2018:ams}                                            \\
       \hline
       \multirow{4}{*}{\shortstack{\textbf{Shared}\\ \textbf{L3 Cache}}}  & 8~MB (16-banks), 0.5~MB per bank, 16-way, 27-cycle \\
                                                  & 64~B line; LRU policy; Bi-directional ring~\cite{ausavarungnirun2014design}; Inclusive; MESI protocol~\cite{papamarcos1984low} \\
                                                  & MSHR size: 64-request, 64-write, 64-eviction \\
                                                  & Energy: 945/1904~pJ per hit/miss~\cite{muralimanohar2007optimizing, tsai:micro:2018:ams} \\
       \hline
       \hline 
       \rowcolor{Gray}
       \multicolumn{2}{|c|}{\textbf{Host CPU with Prefetcher Configuration}}\\
       \hline
       \hline 
       \multirow{3}{*}{\shortstack{\textbf{Processor,}\\ \textbf{Private L1 Cache, Private L2 Cache,} \\ \textbf{and Share L3 Cache}}} &  \\
                                                                                                   & Same as in Host CPU Configuration   \\
                                                                                                   & \\ 
       \hline 
       \textbf{L2 Cache Prefetcher}               & Stream prefetcher~\cite{palacharla1994evaluating,srinath2007feedback}: 2-degree; 16 stream buffers; 64 entries \\
       \hline
       \hline 
       \rowcolor{Gray}
       \multicolumn{2}{|c|}{\textbf{NDP Configuration}}\\ 
       \hline
       \hline
       \multirow{4}{*}{\shortstack{\textbf{NDP CPU}\\ \textbf{Processor}}} & 1, 4, 16, 64, and 256~cores @2.4~GHz, 32~nm; 4-wide out-of-order   \\
                                                     & 1, 4, 16, 64, and 256~cores @ 2.4~GHz, 32~nm; 4-wide in-order   \\ 
                                                     & Buffers: 128-entry ROB; 32-entry LSQ (each)\\
                                                    & Branch predictor: Two-level GAs~\cite{yeh1991two}. 2,048~entry BTB; 1~branch per fetch    \\ \hline
        \multirow{2}{*}{\shortstack{\textbf{Private}\\ \textbf{L1 Cache}}}                               &  32~KB, 8-way, 4-cycle; 64~B line; LRU policy; Read-only Data Cache\\ 
                                                     & Energy: 15/33~pJ per hit/miss~\cite{muralimanohar2007optimizing, tsai:micro:2018:ams}  \\
                                                     
       \hline
       \hline 
       \rowcolor{Gray}
       \multicolumn{2}{|c|}{\textbf{Common}}\\ 
       \hline
       \hline 
       \multirow{4}{*}{\textbf{Main Memory}}      & HMC v2.0 Module~\cite{HMC2} 32 vaults, 8 DRAM banks/vault, 256~B row buffer \\
                                                  & 8~GB total size; DRAM@166 MHz; 4-links@8~GHz\\
                                                  & 8~B burst width at 2:1 core-to-bus freq. ratio; 
                                                   Open-page policy; HMC default interleaving~\cite{HMC2, ghose.sigmetrics20}\footnotemark[10]\\
                                                  & Energy: 2~pJ/bit internal, 8~pJ/bit logic layer ~\cite{top-pim,gao2015practical,tsai:micro:2018:ams}, 2~pJ/bit links~\cite{kim2013memory, NDC_ISPASS_2014, tsai:micro:2018:ams}\\
        \hline
        \end{tabular}
    }
\end{table*}

\end{savenotes}

\subsubsection{\gfii{Choosing an NDP Architecture.}}

\geraldo{\sg{We note that across the proposed \gls{NDP} architectures \gfii{in literature}, there is a lack of consensus on whether the architectures should make use of general-purpose \gls{NDP} cores or specialized \gls{NDP} accelerators~\cite{ghose.ibmjrd19, mutlu2020modern}.} In this work, we focus on general-purpose \gls{NDP} cores for \gfiii{two major} reasons. First, many prior works \geraldorevi{(e.g., \gfiii{\cite{tsai:micro:2018:ams,ahn2015scalable,boroumand2018google,lazypim, smc_sim,drumond2017mondrian,NDC_ISPASS_2014, IBM_ActiveCube, gao2015practical,lim2017triple, lockerman2020livia,fernandez2020natsa, syncron,de2018design,singh2019napel}})} suggest that general-purpose cores (especially simple in-order cores) can successfully accelerate memory-bound applications in \gls{NDP} architectures. In fact, UPMEM~\cite{devaux2019true}, \sg{a start-up building some of the first commercial \gfiii{in-DRAM} NDP \gfiii{systems}}, utilizes simple in-order cores \gfii{in} their \gls{NDP} units \gfiii{inside DRAM chips}~\cite{devaux2019true, juansigmetrics21}. Therefore, we believe that general-purpose \gls{NDP} cores are a promising candidate for future \gls{NDP} \sg{architectures}. Second, \sg{the goal of our work} is not to perform a design space exploration of \sg{different \gls{NDP} architectures,} but rather to understand the key properties \gfii{of} applications that lead to memory bottlenecks that can be mitigated by a simple \gls{NDP} engine. \sg{While we expect that each application could potentially benefit further from an \gls{NDP}} accelerator tailored to its computational \geraldorevi{and memory} requirements, \sg{such customized architectures open many challenges for a methodical characterization, such as the need for significant code refactoring, change\geraldorevi{s} in data mapping, and code partitioning between \gls{NDP} accelerators and host CPUs.\footnote{\gfiii{We show in Section~\ref{sec:case_study_2} that our \bench benchmark suite is \gfiv{useful} to \gfiv{rigorously} study \gls{NDP} accelerators.}}$^{,}$\footnote{\gfiii{The development of a \gfiv{new} methodology \gfiv{or extension of our methodology} to perform analysis targeting function-specific\gfiv{, customized, or reconfigurable} NDP accelerators is a good \gfiv{direction} for future work.}}}}

%% file: 04_sec_scalability_analysis.tex
\section{Characterizing Memory Bottlenecks}
\label{sec:characterization}

\geraldorevi{In this section, we apply our three-step \gfii{workload characterization methodology} to characterize the sources of memory bottlenecks across a wide range of applications. First, we apply \emph{Step~1} to identify memory-bound functions within an application (Section~\ref{sec:vtune}). 
Second, we apply \emph{Step~2} and cluster \juan{the identified functions using \gfiii{two} architecture-independent metrics (spatial and temporal locality)} 
(Section~\ref{sec:locality}). Third, we apply \emph{Step~3} and combine the architecture-dependent and architecture-independent metrics to classify the different sources of memory bottlenecks we observe (Section~\ref{sec:scalability}).}

\geraldorevi{We \gfii{also} evaluate \gfii{various other} aspects of our three-step \gfii{workload characterization methodology}. We investigate the effect of increasing the last-level cache on our memory bottleneck classification in Section~\ref{sec_scalability_nuca}. We provide a \geraldorevi{validation} of our memory bottleneck classification in Section~\ref{sec_summary}. \gfiii{W}e discuss  the limitations of our proposed methodology in Section~\ref{sec_scalability_limitations}}.

\subsection{\geraldorevi{Step 1: Memory-Bound Function Identification}}
\label{sec:vtune}
\label{sec_selected}

We first \sg{apply Step~1 of our methodology across 345~applications \gfii{(listed in Appendix~\ref{sec:evaluatedapp}}) to} identify functions whose performance is significantly \geraldorevi{affected} by \sg{data movement.  We} use \gfii{the} previously-proposed top-down analysis \gfii{methodology}~\cite{yasin_ispass2014} that has been used by several recent workload characterization \geraldorevi{studies}~\cite{kanev_isca2015, sirin_damon2017, appuswamy_ipdpsw2018}. \sg{As discussed in Section~\ref{sec:step1}, we use \gfii{the} Intel VTune \gfiii{Profiler}~\cite{vtune}, which we run} on an Intel Xeon E3-1240 processor~\gfiii{\cite{E3-1240}} with \sg{four} cores. We disable \sg{hyper-threading for} more accurate profiling results, as recommended by the VTune documentation~\cite{vtune_HT}. For the applications that we \sg{analyze}, we select functions \geraldorevi{(1)} that take at least 3\% of the clock cycles, and \geraldorevi{(2) that have} \geraldorevi{a} Memory Bound percentage \geraldorevi{that is} greater than 30\%.
We choose 30\% as \geraldorevi{the threshold for} this metric because, in preliminary simulation experiments, we \sg{do not observe significant performance improvement or energy savings with data movement mitigation mechanisms for functions whose Memory Bound \geraldorevi{percentage} is} less than 30\%.

\sg{The applications we analyze come from a \gfii{variety of} sources, such as} popular \gfiii{workload} suites (Chai~\cite{gomezluna_ispass2017}, CORAL~\cite{coral}, Parboil~\cite{stratton2012parboil}, PARSEC~\cite{bienia2008parsec}, Rodinia~\cite{che_iiswc2009}, SD-VBS~\cite{Venkata_iiswc2009}, SPLASH-2~\cite{woo_isca1995}), benchmarking (STREAM~\cite{mccalpin_stream1995}, HPCC~\cite{luszczek_hpcc2006}, HPCG~\cite{dongarra_hpcg2015}), 
bioinformatics~\cite{ahmed2016comparison}, databases~\cite{balkesen_TKDE2015, gomezluna_icpp2015}, graph processing frameworks (GraphMat~\cite{sundaram_vldbendow2015}, Ligra~\cite{shun_ppopp2013}), a map-reduce framework (Phoenix~\cite{yoo_iiswc2009}), \sg{and} neural networks (AlexNet \cite{alexnet2012}, Darknet~\cite{redmon_darknet2013}). We explore different input \gfiii{dataset} sizes for the applications and choose real-world input datasets that impose high pressure on the memory subsystem (as we expect that such \gfiii{real-world} inputs are best suited for stressing the memory hierarchy). We also use different inputs for \gfii{applications} whose performance is tightly related to the input \gfiii{data}set properties. For example, we use two different graphs with varying connectivity degrees (rMat~\cite{rMat} and USA~\cite{dimacs}) to evaluate graph processing applications and two different read sequences to evaluate read alignment algorithms\gfiii{\cite{cali2020genasm,alser2017gatekeeper,alser2020accelerating}}. \addtocounter{footnote}{1}\footnotetext{\geraldorevi{The default HMC interleaving scheme (Row:Column:Bank:Vault~\cite{HMC2}) interleaves consecutive cache lines across vaults, and then across banks~\cite{hadidi2017demystifying}.}}

\begin{figure*}[!t]
    \centering
    \includegraphics[width=0.9\linewidth]{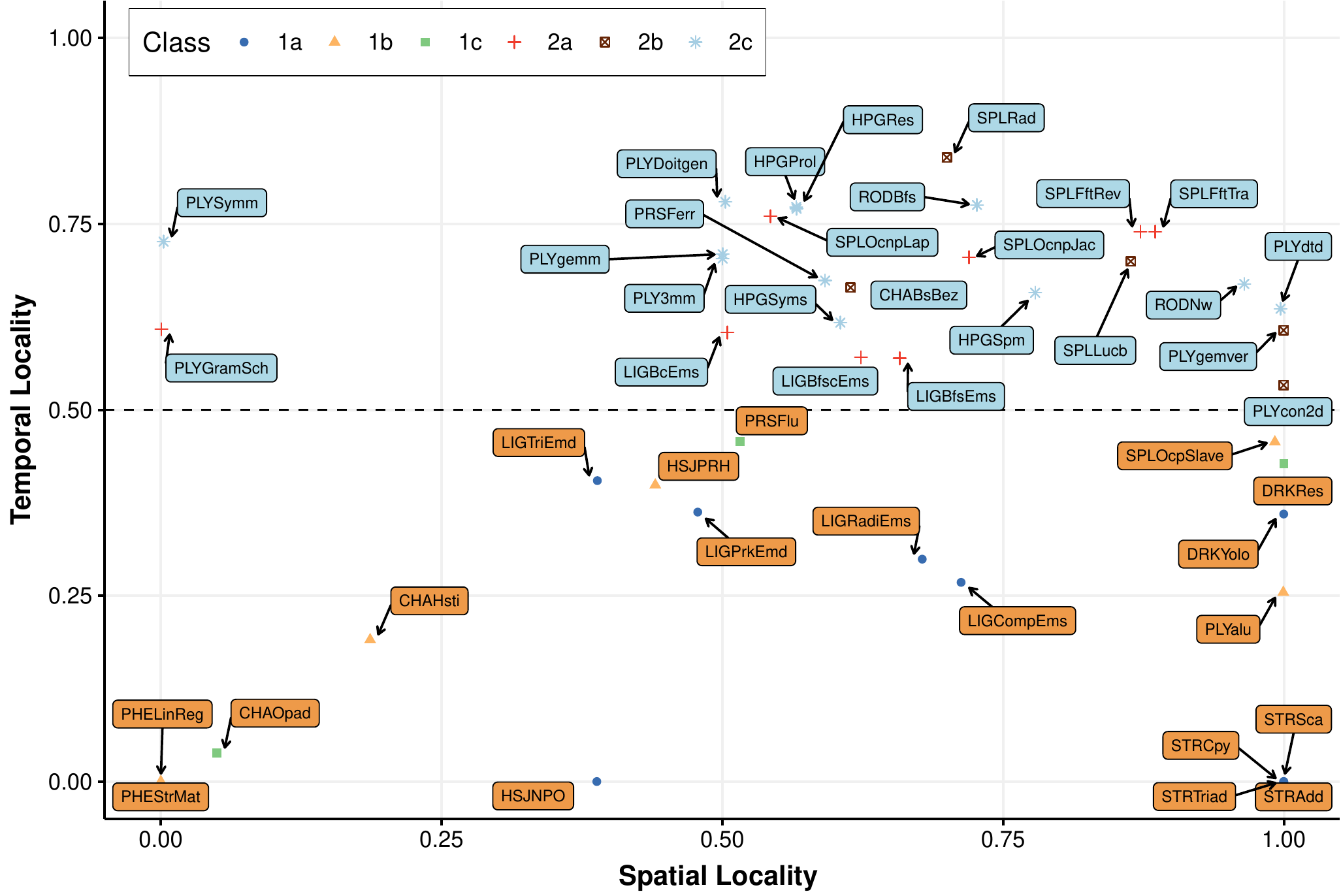}
     \caption{Locality-based clustering \gfii{of 44 representative functions}.}
  \label{fig:locality_chart}
\end{figure*}

{In total, our application analysis covers more than 77K functions.} To date, this is the most extensive analysis of data movement bottlenecks in real-world applications. We find a set of \geraldo{144} functions that take at least 3\% of the total clock cycles and have a value of the Memory Bound metric greater or equal to 30\%, \sg{which forms the basis of \gfiii{\bench,} our new data movement benchmark suite. \geraldorevi{We provide a} list of all 144 functions selected \geraldorevi{based on our analysis} \gfii{and their major characteristics} in Appendix~\ref{sec:benchlist}.} 

After identifying memory-bound functions over a wide range of applications, we apply Steps 2 and 3 of our methodology to classify the primary sources of memory bottlenecks for \gfii{our selected functions}. We evaluate a total of 144 functions out of the \gfiii{77K functions we analyze} in Step 1. \geraldorevi{These functions} \sg{span} across 74 different applications, belonging to 16 different widely-used benchmark suites \sg{or frameworks}.

From the 144~\sg{functions that} we analyze \gfiii{further}, we select a subset of 44 \sg{representative functions to explore in-depth in \gfiii{Sections~\ref{sec:locality} and \ref{sec:scalability}} \geraldorevi{and to drive our bottleneck classification analysis.} \gfiii{We use the 44 representative functions to ease our explanations and make figures more easily readable.} Table~\ref{tab:benchmarks} in Appendix~\ref{sec:benchlist} lists the \geraldorevi{44} representative functions that we select.} The table includes one column that indicates the class \sg{of data movement bottleneck experienced by each function (we discuss the classes \geraldorevi{in Section~\ref{sec:scalability}})}, and another column representing the percentage of clock cycles of the selected function in the whole application. \geraldorevi{We} select \gfiii{representative} \gfii{functions} that belong to a variety of domains: benchmarking, bioinformatics, data analytics, databases, data mining, data reorganization, graph processing, neural networks, physics, and signal processing. \geraldorevi{In Section~\ref{sec_summary}, we validate our classification using the remaining 100 functions \gfiv{and provide a summary of the results of our methodology when applied to all 144 functions}.}

\subsection{Step 2: Locality-Based Clustering}
\label{sec:locality}

We cluster the \gfii{44 representative functions} across both spatial and temporal locality using \gfii{the} K-means \gfii{clustering algorithm}~\cite{hartigan1979algorithm}. \geraldorevi{Figure~\ref{fig:locality_chart} shows how each \geraldorevi{\gfii{function}} is grouped.} \geraldorevi{We} find that two groups emerge from the clustering:
(1)~low temporal locality \gfii{functions} (orange \geraldorevi{boxes in Figure~\ref{fig:locality_chart}}), and
(2)~high temporal locality \gfii{functions} (blue \geraldorevi{boxes in Figure~\ref{fig:locality_chart}}). Intuitively, the closer a \geraldorevi{\gfii{function}} is to the bottom-left corner of the figure, the less likely it is to take advantage of a multi-level cache hierarchy. These \geraldorevi{\gfii{functions}} are more likely to be good candidates for NDP.  However, as we see \geraldorevi{in Section~\ref{sec:scalability}}, the \gfii{\gls{NDP}} suitability of a \gfii{function} also depends on a number of other factors.

\begin{figure*}[!t]
    \centering
  \includegraphics[width=\linewidth]{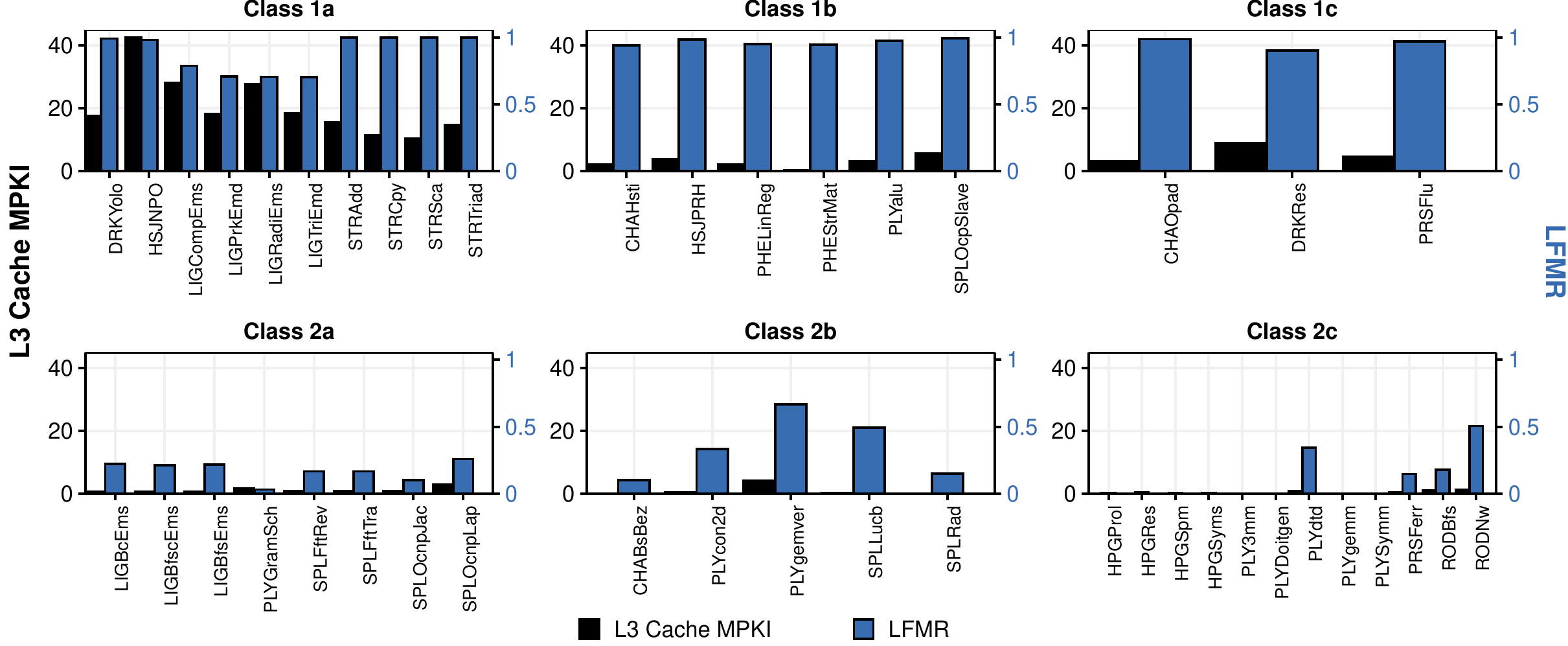}
  \caption{\gfii{L3 Cache \gls{MPKI} and} Last-to-First Miss Ratio (LFMR) \geraldorevi{for 44 representative \gfii{functions}. 
  }}
  \label{fig:lfmr}
\end{figure*}

\subsection{Step 3: Bottleneck Classification}
\label{sec:scalability}

Within the two groups of \gfii{functions} identified in Section~\ref{sec:locality}, we use three key metrics (\gls{AI}, \gls{MPKI}, and \gls{LFMR}) to classify the memory bottlenecks.
We observe that the \gls{AI} of the analyzed low temporal locality \gfii{functions} is low (i.e., \gfiii{always} less than 2.2 op\geraldorevi{s}/cache line\gfiii{, with an average of 1.3 ops/cache line}). Among the high temporal locality \gfii{functions}, there are some with low \gls{AI} (minimum of 0.3 op\geraldorevi{s}/cache line) and others with high \gls{AI} (maximum of 44 op\geraldorevi{s}/cache line). \gls{LFMR} indicates whether a \gfii{function} benefits from a deeper cache hierarchy.  When \gls{LFMR} is low (i.e., less than 0.1), then a \gfii{function} benefits significantly from a deeper cache hierarchy, as most misses from the L1 cache hit in either the L2 or L3 caches.  When \gls{LFMR} is high (i.e., greater than 0.7), then most L1 misses are not serviced by the the L2 or L3 caches, and must go to memory.  A medium \gls{LFMR} (0.1--0.7) indicates that a deeper cache hierarchy can mitigate some, but not \gfiii{a very large fraction of} L1 cache misses. \gls{MPKI} indicates the memory intensity of a \gfii{function} (i.e., the rate at which requests are issued to DRAM). We say that a \gfii{function} is memory-intensive (i.e., it has a high \gls{MPKI}) when the \gls{MPKI} is greater than 10, which is the same threshold used by prior works~\cite{hashemi2016accelerating,chou2015reducing,kim2010atlas,kim2010thread,muralidhara2011reducing, subramanian2016bliss,usui2016dash}.

We find that six classes of \gfii{functions} emerge, based on their temporal locality, \gls{AI}, \gls{MPKI}, and \gls{LFMR} values\geraldorevi{, as we observe from Figures~\ref{fig:locality_chart} and \ref{fig:lfmr}}. We observe that spatial locality is not a key metric for our classification (i.e., it does not define a bottleneck class) \geraldorevi{because} the L1 cache, \juan{which is \gfii{present} in both host \gfii{CPU} and NDP \gfii{system} configurations,} can capture most of the spatial locality for a \gfii{function}. Figure~\ref{fig:lfmr} shows the \gls{LFMR} and \gls{MPKI} values for each class.  Note that we do not have classes of \gfii{functions} for all possible combinations of metrics. In our analysis, we obtain the \geraldorevi{temporal locality, \gls{AI}, \gls{MPKI}, and \gls{LFMR} values} and their combinations empirically. \gfii{F}undamentally, \geraldorevi{not all value combinations of different metrics are possible.} We list some of the combinations we do \gfiii{\emph{not}} observe in our analysis \geraldorevi{of 144 \gfii{functions}}:
\begin{itemize}[noitemsep, leftmargin=*, topsep=0pt]
    \item A \gfii{function} with high \gfii{LLC} \gls{MPKI} does \emph{not} display low \gls{LFMR}\geraldorevi{.} \gfii{This is because} \gfii{a low LFMR happens when most L1 misses hit the L2/L3 caches. Thus,} \geraldorevi{it becomes highly unlikely for the L3 cache to suffer many misses when the L2/L3 caches do a good job \juan{in} fulfilling L1 cache misses.}
    
    \item  A \gfii{function} with high temporal locality does \emph{not} display \geraldorevi{both} high \gls{LFMR} and high \gls{MPKI}. \gfii{This is because} \geraldorevi{a \gfii{function} with high temporal locality will likely issue repeated memory requests to few memory addresses, which will likely \gfiii{be} serviced by the cache hierarchy.}
    \item A \gfii{function} with low temporal locality does \emph{not} display low \gls{LFMR} since there is little data locality to be captured by the cache hierarchy. 
\end{itemize}

\begin{figure*}[h]
    \centering
    \includegraphics[width=\linewidth]{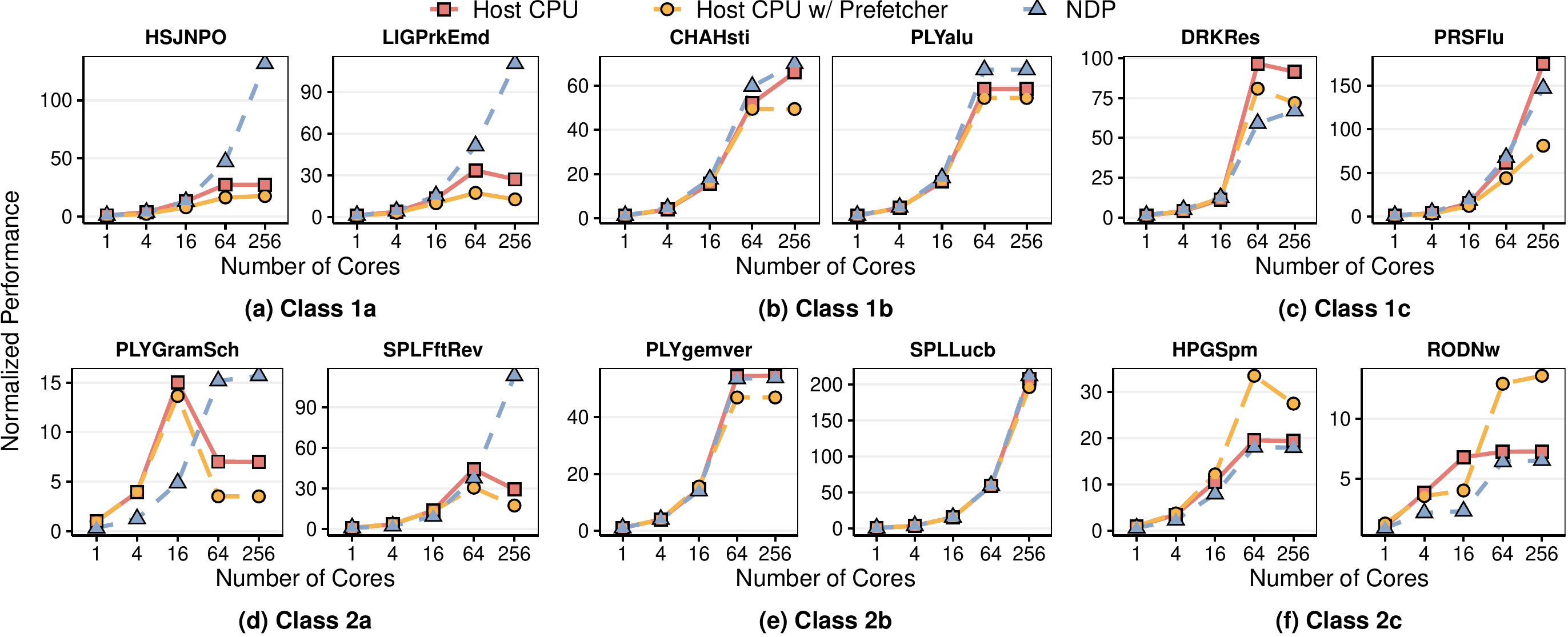}
    \caption{Performance of \gfiii{12} representative  \gfii{functions} on \geraldorevi{three systems:} host \gfii{CPU}\geraldorevi{, host \gfii{CPU} with prefetcher,} and NDP, normalized to one host \gfii{CPU} core.}
    \label{figure_performance}
\end{figure*}

We discuss each class in detail below, identifying the memory bottlenecks for each class and whether \gfii{the} NDP \gfii{system} can alleviate these bottlenecks. To simplify our explanations, we focus on a \gfiii{smaller} set of \geraldorevi{12} representative \gfii{functions} \gfii{(out of the 44 representative functions)} for this part of the analysis. \geraldo{Figure~\ref{figure_performance} shows how each \gfiii{of the 12} \gfii{function\gfiii{s}} scales in terms of performance for the \geraldorevi{\textit{host \gfii{CPU}}, \textit{host \gfii{CPU} with prefetcher}, and \textit{NDP}} \gfii{system} configurations.}

\subsubsection{Class~1a: Low Temporal Locality, Low \gls{AI}, High \gls{LFMR}, and High \gls{MPKI} \gfii{(\textit{DRAM Bandwidth-Bound Functions})}}
\label{sec_scalability_class1a}

\gfii{Functions} in this class exert high \geraldorevi{main} memory pressure \geraldorevi{since they are} highly memory intensive \geraldorevi{and have low data reuse}.  To understand how this affects \geraldorevi{a} \gfii{function}\geraldorevi{'s} suitability for \gls{NDP}, we study how performance scales as we increase the number of cores available to a \gfii{function}, for \geraldorevi{the host \gfii{CPU}, host \gfii{CPU} with prefetcher, and NDP \gfii{system} configurations}.  Figure~\ref{figure_performance}(a) \geraldorevi{depicts} performance\footnote{\geraldorevi{P}erformance \geraldorevi{is} the inverse of \geraldorevi{application} execution time.} as we increase the core count, normalized to the performance of one host \gfii{CPU} core, for two representative \gfii{functions} from \geraldorevi{Class~1a} (\texttt{HSJNPO} and \texttt{LIGPrkEmd}; we see similar trends for all \gfii{functions} in the class).

\begin{figure}[ht]
\centering
  \centering
   \includegraphics[width=\linewidth]{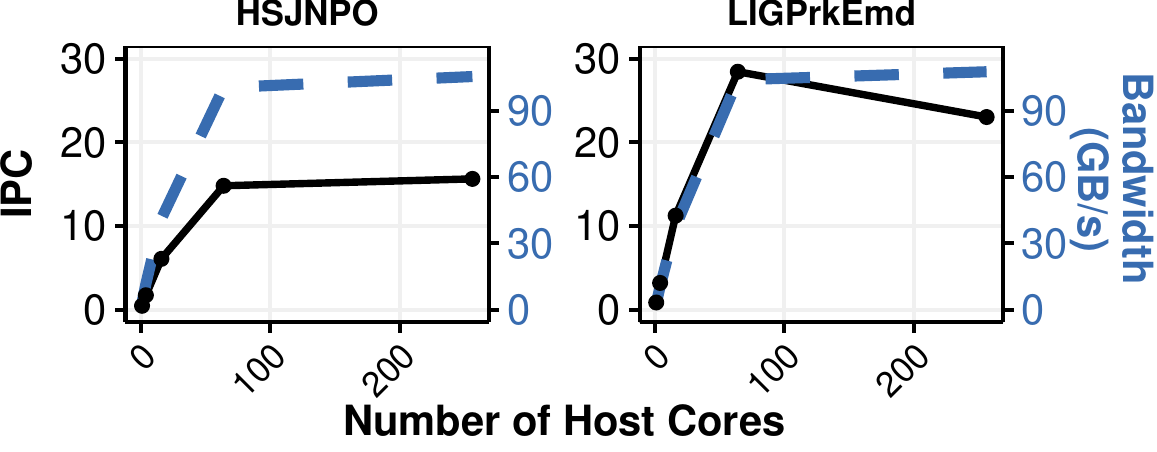}
   \caption{Host \gfii{CPU} \geraldorevi{system} IPC vs. \gfiii{utilized} \geraldorevi{\gfii{DRAM}} Bandwidth for \geraldorevi{representative} Class~1a \geraldorevi{\gfii{functions}}.}
   \label{fig:bw_group_1}
\end{figure}

We make \geraldorevi{three} observations from the figure. First, as the number of host \gfii{CPU} cores increases, performance eventually stops increasing significantly.  For \texttt{HSJNPO}, \gfii{host CPU} performance increases by \geraldorevi{27.5$\times$ going from 1 to 64 host \gfii{CPU} cores but} only 27\% going from 64 host \gfii{CPU} cores to 256 host \gfii{CPU} cores\geraldorevi{. Fo}r \texttt{LIGPrkEmd}, \gfii{host CPU} performance \geraldorevi{increases by 33$\times$ going from 1 to 64 host \gfii{CPU} cores but} \emph{decreases} by 20\% \geraldorevi{going from 64 to 256 host \gfii{CPU} cores}.
We find that the lack of performance improvement \geraldorevi{at large host \gfii{CPU} core counts} is due to \gfii{main} memory bandwidth saturation, as shown in Figure~\ref{fig:bw_group_1}. 
Given the limited \gfii{DRAM} bandwidth available across the off-chip memory channel, we find that Class~1a \gfii{functions} saturate the \gfii{DRAM} bandwidth once enough \gfii{host CPU} cores (e.g., 64) are used,
\gfiii{and thus} these \gfii{functions} \gfiii{are} \emph{bottlenecked by the DRAM bandwidth}.
\geraldorevi{Second, the host \gfii{CPU system} with prefetcher slows down the execution of the \gfiii{\texttt{HSJNPO} (\texttt{LIGPrkEmd})}  \gfii{function} compared with the host \gfii{CPU system without prefetcher} by 43\% (38\%)\gfii{, on average across all core counts}. The prefetcher is \gfii{ineffective} since these \gfii{functions} have low temporal and spatial locality.} \geraldorevi{Third}, when running on \gfii{the} \gls{NDP} \geraldorevi{\gfii{system}}, the \gfii{functions} see \geraldorevi{continued performance improvements} as the number of \gfii{NDP} cores increase\gfii{s}.  By providing the \gfii{functions} with access to the much higher bandwidth available inside memory, the \gls{NDP} \gfii{system} can \geraldorevi{greatly} outperform the host \gfii{CPU system} at a high enough core count. \geraldorevi{For example,} at \geraldorevi{64/}256~cores, \gfii{the} \gls{NDP} \gfii{system} outperform\gfiii{s} \gfii{the} host \gfii{CPU system} by \geraldorevi{1.7$\times$/}4.8$\times$ for \texttt{HSJNPO}, and by \geraldorevi{1.5$\times$/}4.\gfii{1}$\times$ for \texttt{LIGPrkEmd}.

\begin{figure}[ht]
  \centering
   \includegraphics[width=\linewidth]{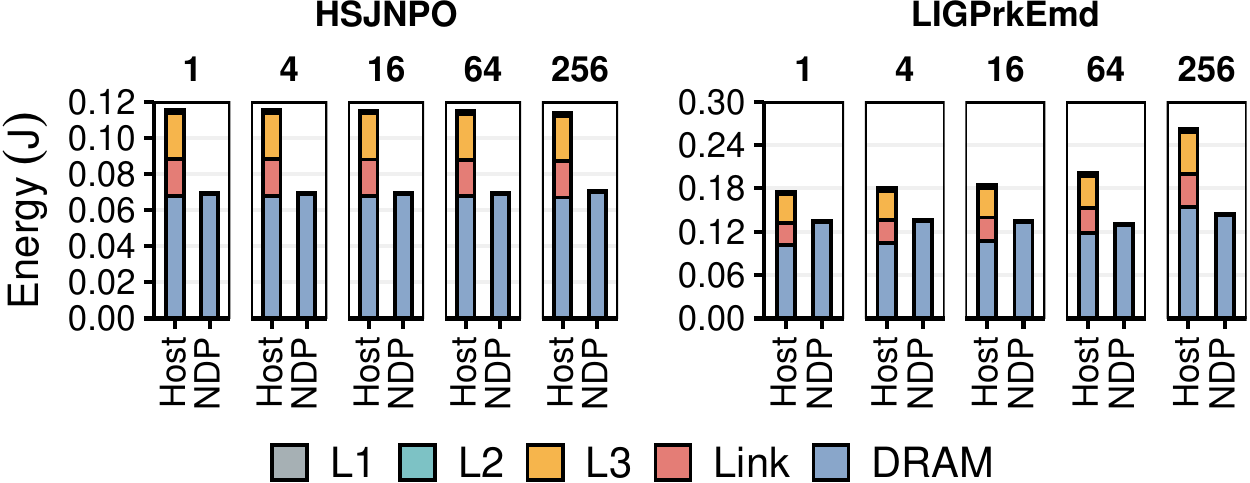}
  \caption{Cache and DRAM energy \geraldorevi{breakdown} for \geraldorevi{representative} Class~1a \geraldorevi{\gfii{functions}} \gfii{at 1, 4, 16, 64, and 256 cores}.}
  \label{fig:energy_group_1}
\end{figure}

Figure~\ref{fig:energy_group_1} depicts the energy breakdown for our two representative \gfii{functions}. We make two observations from the figure. First, for \texttt{HSJNPO}, the energy spent on DRAM for both host \gfii{CPU} \gfii{system} and \gls{NDP} \gfii{system} are similar. This is due to the \gfii{function}'s poor locality, as 98\% of its memory requests miss in the L1 cache.  \geraldorevi{Since} \gls{LFMR} \geraldorevi{is} near 1, \gfii{L1 miss requests} almost always miss in the L2 and L3 caches and go to DRAM \geraldorevi{in the host \gfii{CPU} system} \gfiii{for all core counts we evaluate}, which requires significant energy to query the large cache\geraldorevi{s} and then to \geraldorevi{perform} off-chip \geraldorevi{data transfer\gfii{s}}. 
\geraldorevi{The} NDP \geraldorevi{system} does not \geraldorevi{access} L2, L3, and off-chip link\geraldorevi{s}, \geraldorevi{leading to large system} energy reduction. Second, for \texttt{\texttt{LIGPrkEmd}}, the DRAM energy is higher in \gfii{the} \gls{NDP} \gfii{system} than in the host \gfii{CPU system}. Since the \gfii{function\gfiii{'s}} \gls{LFMR} is 0.7, some memory requests that would be cache hits in the host \gfii{CPU}'s L2 \geraldorevi{and} L3 caches are instead sent directly to DRAM \geraldorevi{in the} \gls{NDP} \geraldorevi{system}. However, the total energy consumption on the host \gfii{CPU} \geraldorevi{system} is still larger than \gfii{that} on \geraldorevi{the} \gls{NDP} \geraldorevi{system}, again because \gfii{the} \gls{NDP} \gfii{system} eliminates the L2\geraldorevi{,} L3 and off-chip link energy.

\geraldorevi{\gfii{DRAM}} bandwidth-bound applications such as those in Class~1a have been the primary focus of a large number of proposed NDP architectures (e.g., \cite{IBM_ActiveCube,azarkhish2018neurostream,ahn2015scalable,azarkhish2016memory,gao2017tetris, nai2017graphpim, boroumand2018google,ke2019recnmp, NDC_ISPASS_2014, pugsley2014comparing}), as they benefit from increased \gfii{main memory} bandwidth and do not have high \gls{AI} (and, thus, do not benefit from complex cores \geraldorevi{on} the host \gfii{CPU} \geraldorevi{system}).  
An NDP architecture for \geraldorevi{a} \gfii{function} in Class~1a  \gfii{needs to} extract enough \gls{MLP}~\gfiv{\cite{glew1998mlp, mutlu2003runahead, qureshi2006case, mutlu2008parallelism,mutlu2006efficient,mutlu2005techniques,chou2004microarchitecture,tuck2006scalable,phadke2011mlp,van2009mlp,everman2007memory,patsilaras2012efficiently}} to maximize the usage of the available internal \geraldorevi{memory} bandwidth. \gfii{However,} prior work has shown that this can be challenging due to the area and power constraints in the logic layer of a 3D-stacked DRAM~\cite{boroumand2018google, ahn2015scalable}. To exploit the high \geraldorevi{memory} bandwidth while \gfii{satisfying} these \gfii{area and power} constraints, the NDP architecture should leverage application \gfii{memory} access patterns to efficiently maximize \gfii{main memory bandwidth} utilization. 

We find that there are two dominant types of memory access patterns among our Class~1a \gfii{functions}. First, \gfii{functions} with regular access patterns (\texttt{DRKYolo}, \texttt{STRAdd}, \texttt{STRCpy}, \texttt{STRSca}, \texttt{STRTriad}) can take advantage of specialized accelerators or \gls{SIMD} architectures~\cite{drumond2017mondrian, boroumand2018google}, which can exploit the regular access patterns to issue many memory requests concurrently. Such accelerators or \gls{SIMD} architectures have \geraldorevi{hardware} area and thermal dissipation that fall \gfii{well} within the constraints of 3D-stacked DRAM~\cite{top-pim,eckert2014, boroumand2018google, ahn2015scalable}. Second, \gfii{functions} with irregular access patterns (\texttt{HSJNPO}, \texttt{LIGCompEms}, \texttt{LIGPrkEmd}, \texttt{LIGRadiEms}) require techniques to extract \gls{MLP} while still fitting within the design constraints.  This requires techniques that cater to the irregular memory access patterns, such as prefetching algorithms designed for graph processing~\gfiv{\cite{ahn2015scalable,nilakant2014prefedge,kaushik2021gretch,ainsworth2016graph,basak2019analysis,yan2019alleviating}}, \gfii{pre-execution of difficult access patterns~\gfiv{\cite{mutlu2003runahead, hashemi2016continuous, srinivasan2004continual, mutlu2006efficient, mutlu2005techniques,hashemi2016accelerating,annavaram2001data,collins2001dynamic,dundas1997improving,mutlu2003runaheadmicro,ramirez2008runahead,ramirez2010efficient,zhang2007accelerating}}} or \gfiii{hardware accelerators for} pointer chasing~\gfiv{\cite{hsieh2016accelerating, ebrahimi2009techniques, cooksey2002stateless,roth1999effective,santos2018processing,lockerman2020livia,hong2016accelerating}}. 

\subsubsection{Class~1b: Low Temporal Locality, Low \gls{AI}, High \gls{LFMR}, and Low \gls{MPKI} \gfii{(\textit{DRAM Latency-Bound Functions})}}
\label{sec_scalability_class1b}

While \gfii{functions} in this class do not effectively use the host \gfii{CPU} caches, they do \emph{not} exert high pressure on the \gfii{main }memory due to their low \gls{MPKI}. Across \geraldorevi{all Class~1b} \gfii{functions}, the average \geraldorevi{\gfii{DRAM}} bandwidth consumption is only 0.5 GB/s.  However, all the \gfii{functions} have very high \gls{LFMR} values (the minimum is 0.94 for \texttt{CHAHsti}), indicating that the \geraldorevi{host} \gfii{CPU} L2 and L3 caches are ineffective. Because the \gfii{functions} cannot exploit significant \gls{MLP} but still incur long-latency requests to DRAM, the DRAM requests fall on the critical path of execution and stall forward progress~\gfiv{\cite{mutlu2003runahead, ghose.isca13, mutlu2008parallelism,hashemi2016accelerating,hashemi2016continuous}}. Thus, Class~1b \gfii{functions} are \emph{bottlenecked by \gfii{DRAM} latency}. Figure~\ref{figure_performance}(b) shows performance of both \gfiii{the} host \gfii{CPU system} and \gfiii{the} NDP \gfii{system} for two representative \gfii{functions} from Class~1b (\texttt{CHAHsti} and  \texttt{PLYalu}). We observe that while \gfii{performance of} both \gfiii{the} host \gfii{CPU} \gfii{system} and \geraldorevi{the} NDP \geraldorevi{system} scale well as the core count increases, NDP \geraldorevi{system} performance is always higher than the host \gfii{CPU system} performance for the same core count.  The \geraldorevi{maximum (average)} speedup with NDP over host \gfii{CPU} at the same core count is \geraldorevi{1.15$\times$ (1.12$\times$)} for \texttt{CHAHsti} and \geraldorevi{1.23$\times$ (1.13$\times$)} for \texttt{PLYalu}.

\begin{figure}[ht]
    \centering
     \includegraphics[width=1.0\linewidth]{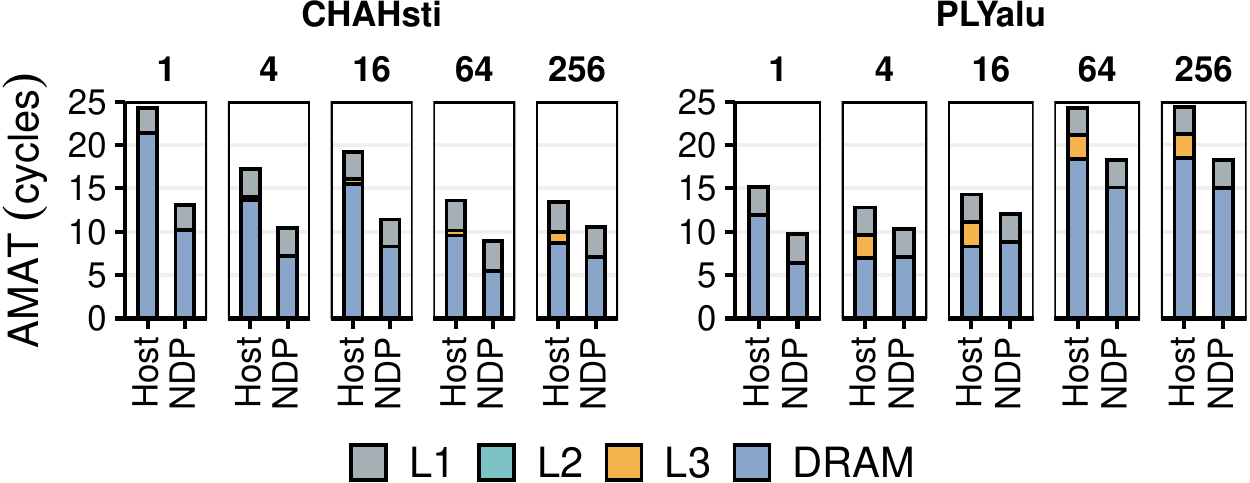}%
      \caption{\gfii{Average Memory Access Time (\gls{AMAT})} for \geraldorevi{representative} Class~1b \geraldorevi{\gfii{functions}}. }
      \label{figure_amat_group_2}
\end{figure}

We find that \gfii{the} NDP \gfii{system}'s improved performance is due to a reduction in the \gls{AMAT}~\cite{amatref}.
Figure~\ref{figure_amat_group_2} shows the \gls{AMAT} for our two representative \gfii{functions}.  Memory accesses take significantly longer in the host \gfii{CPU} \geraldorevi{system} than in \geraldorevi{the} NDP \geraldorevi{system} due to the additional latency of looking up requests in the L2 and L3 caches, \geraldorevi{even though} data \geraldorevi{is rarely present in} those caches\gfiii{, and going through the off-chip links}.

\begin{figure}[ht]
  \centering
   \includegraphics[width=1.0\linewidth]{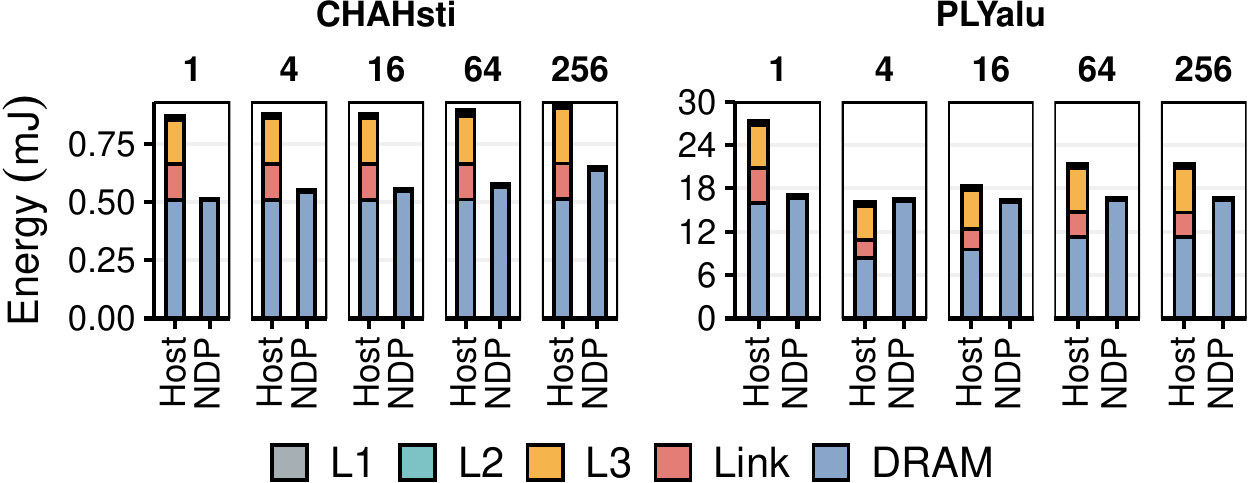}%
   \caption{Energy breakdown for \geraldorevi{representative} Class~1b \geraldorevi{\gfii{functions}}.}
\label{fig:energy_group_2}
\end{figure}

Figure~\ref{fig:energy_group_2} shows the energy breakdown for Class~1b \gfii{representative functions}. Similar to Class~1a, we observe that the L2/L3 caches and off-chip links are a large source of energy usage in the host \gfii{CPU} \geraldorevi{system}. While DRAM energy increases in \gfii{the} NDP \gfii{system}, as L2/L3 hits in the host \gfii{CPU system} become DRAM lookups \gfii{with} NDP, the overall energy consumption \geraldorevi{in the NDP system} is \geraldorevi{greatly smaller (by 69\% {maximum} and 39\% on average)} due to the lack of L2 and L3 caches.

Class~1b \gfii{functions} benefit from \gfii{the} NDP \gfii{system}, but primarily because of the lower memory access latency \geraldorevi{(and energy)} that \gfii{the} NDP \gfii{system} provides for memory requests that need to be serviced by DRAM.  These \gfii{functions} could benefit from other latency \geraldorevi{and energy} reduction techniques, such as L2/L3 cache bypassing~\gfvi{\cite{tsai:micro:2018:ams, johnson1999run, sembrant2014direct,tsai2017jenga, seshadri2014dirty,seshadri2015mitigating,johnson1997run,tyson1995modified,memik2003just,kharbutli2008counter,gupta2013adaptive,li2012optimal,sridharan2016discrete}}, \gfvii{low-latency DRAM~\cite{Tiered-Latency_LEE, lee2015adaptive,lee2017design,chang2017understanding,chang.sigmetrics2016,hassan2016chargecache,hassan2019crow,seshadri2013rowclone,wang2018reducing,kim2018solar,das2018vrl,chang2016low,chang2014improving,chang2017understandingphd,hassan2017softmc,lee2016reducing,luo2020clr,choi2015multiple,son2013reducing,liu2012raidr,rldram,sato1998fast,orosa2021codic,seongil2014row}, and better memory access scheduling~\gfvii{\cite{mutlu2008parallelism, subramanian2016bliss, kim2010atlas,kim2010thread, mutlu2007stall,ausavarungnirun2012staged, usui2016dash, subramanian2014blacklisting, muralidhara2011reducing,subramanian2015application,subramanian2013mise,rixner2004memory,rixner2000memory,zuravleff1997controller,moscibroda2008distributed,ebrahimi2011parallel,ipek2008self,hur2004adaptive,ebrahimi2010fairness,ghose.isca13,xie2014improving,yuan2009complexity,wang2014memory}}}.  However, they generally do \geraldorevi{\emph{not}} benefit significantly from prefetching (as seen in Figure~\ref{figure_performance}(b)), since infrequent memory requests make it difficult for the prefetcher to successfully train on an access pattern. 

\subsubsection{Class~1c: Low Temporal Locality, Low AI, Decreasing LFMR with Core Count, and Low MPKI \gfii{(\textit{L1/L2 Cache Capacity Bottlenecked Functions})}}
\label{sec_scalability_class1c}

We find that the behavior of \gfii{functions} in this class depends on the number of cores they are using. Figure~\ref{figure_performance}(c) shows the host \gfii{CPU system} and \geraldorevi{the} NDP \geraldorevi{system} performance as we increase the core count for two representative \gfii{functions} (\texttt{DRKRes} and \texttt{PRSFlu}). We make two observations from the figure. First, at low core counts, \geraldorevi{the} NDP \geraldorevi{system} outperforms the host \gfii{CPU} \geraldorevi{system}. \gfii{With} a low number of cores, the \gfii{functions} have medium to high LFMR (0.5 for \texttt{DRKRes} \gfii{at 1 and 4 host CPU cores; 0.97 at 1 host CPU core and \gfii{0.91} at 4 host CPU cores for \texttt{PRSFlu})}, and behave like Class~1b \gfii{functions}, where they are \gfiii{DRAM} latency-sensitive. Second, as the core count increases, \geraldorevi{the} host \gfii{CPU} \geraldorevi{system} \gfii{begins} to outperform \geraldorevi{the} NDP \geraldorevi{system}.  For example, beyond 16 \geraldorevi{(64)} cores, the host \gfii{CPU system}  outperforms \gfii{the} NDP \gfii{system} for \texttt{DRKRes} \geraldorevi{(\texttt{PRSFlu})}.  This is because as the core count increases, the \gfii{aggregate} L1 and L2 cache size available at the host \gfii{CPU system} grow\gfii{s}, \geraldorevi{which} \geraldorevi{reduces} the \gfii{miss rates of both L2 and L3} caches. As a result, the LFMR decreases significantly (e.g., at 256 cores, \gfii{LFMR is} 0.09 for \texttt{DRKRes} and 0.35 for \texttt{PRSFlu}). This indicates that the \emph{available \gfiii{L1/L2} cache capacity} bottlenecks Class 1c \gfii{functions}.

\begin{figure}[ht]
    \centering
    \includegraphics[width=\linewidth]{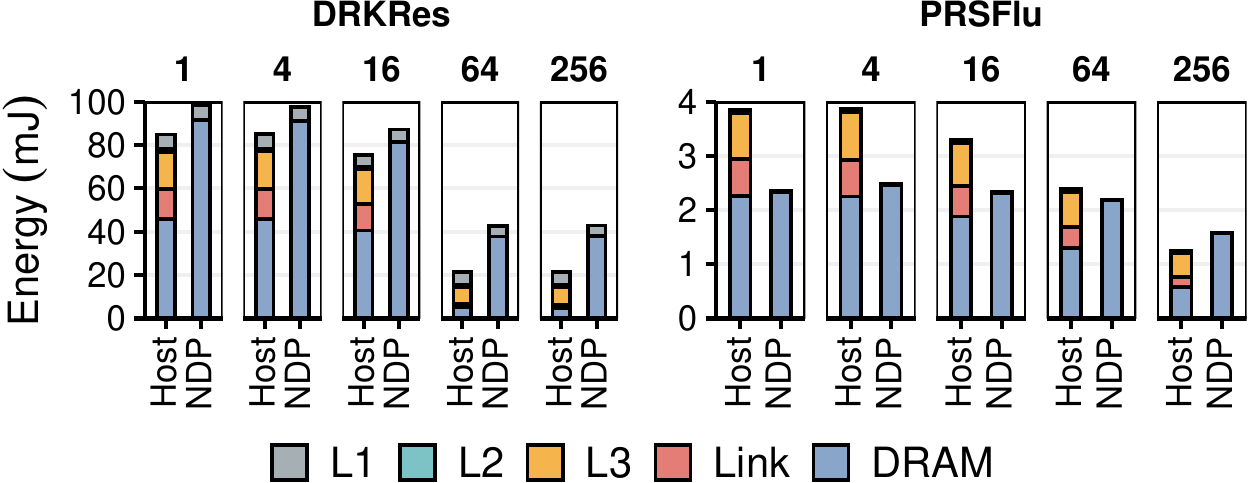}%
    \caption{Energy breakdown for \geraldorevi{representative} Class~1c \geraldorevi{\gfii{functions}}.}
    \label{fig:energy_group_3}
\end{figure}

Figure~\ref{fig:energy_group_3} shows the energy breakdown for Class~1c \gfii{functions}. We make three observations from the figure. First, for \gfii{functions} with larger LFMR values (\texttt{PRSFlu}), \gfii{the} NDP \gfii{system} provides energy savings over the host \gfii{CPU system} at lower core counts, \gfii{since} \gfii{the} NDP \gfii{system} \gfii{eliminates the energy consumed due to L3 and off-chip link accesses}. Second, for \gfii{functions} with smaller LFMR \gfii{values} (\texttt{DRKRes}), \gfii{the} NDP \gfii{system} does not provide energy savings even for \gfii{low} core counts. Due to the medium LFMR, \geraldorevi{enough} requests still hit in the host \gfii{CPU system} L2/L3 caches, and these \geraldorevi{cache} hits become DRAM accesses in \gfii{the} NDP \gfii{system}, which consume more energy than the cache hits. Third, at high\geraldorevi{-enough} core counts, \gfii{the} NDP \gfii{system} consum\gfii{es} more energy than the host \gfii{CPU system} for all Class~1c \gfii{functions}. 
\gfii{As the LFMR decreases, the \gfii{functions} effectively utilize the caches in the host CPU system, reducing the off-chip traffic and, consequently, the energy Class~1c functions spend on accessing DRAM.} \gfiii{The \gls{NDP} system, which does not have L2 and L3 caches, pays the larger energy cost of a DRAM access for all L2/L3 hits in the host CPU system.}

We find that \gfii{the primary source of the memory bottleneck in Class~1c functions is limited \gfiii{L1/L2} cache capacity. Therefore,} while \gfii{the} NDP \gfii{system} \geraldorevi{improve\gfii{s} performance and energy of \gfii{some} Class~1c \gfii{functions} at low core counts \gfii{(with lower associated \gfiii{L1/L2} cache capacity)}}, \gfii{the} NDP \gfii{system} does not provide \geraldorevi{performance and energy} benefits across all core counts for Class~1c \gfii{functions}.

\subsubsection{Class~2a: High Temporal Locality, Low AI, Increasing LFMR with Core Count, and Low MPKI \gfii{(\textit{L3 Cache Contention Bottlenecked Functions})}}
\label{sec_scalability_class2a}

\begin{figure}[h]
  \centering
 \includegraphics[width=1.0\linewidth]{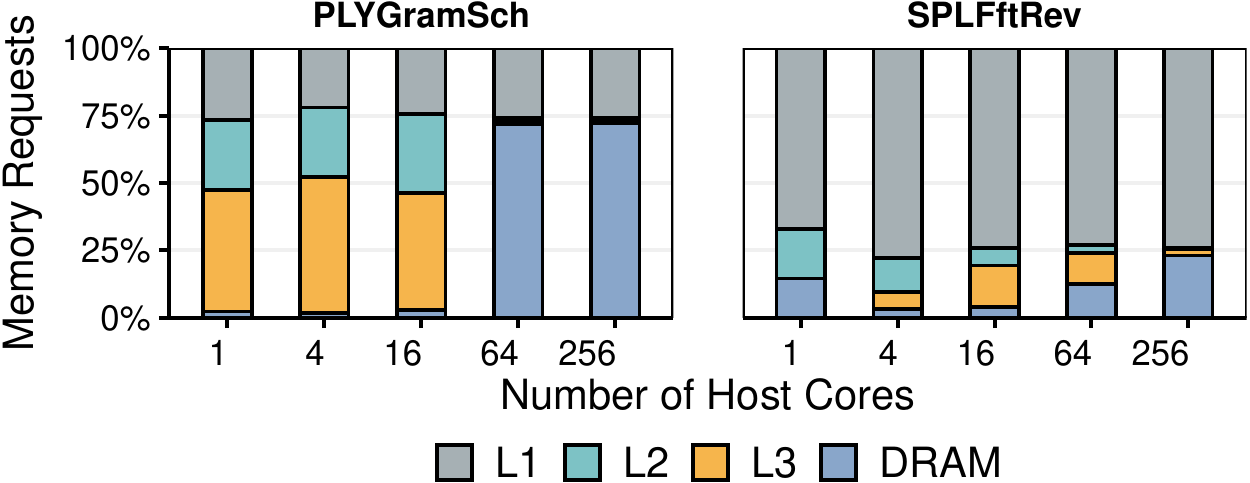}%
 \caption{Memory request breakdown for \geraldorevi{representative} Class~2a \geraldorevi{\gfii{functions}}.}
 \label{fig:latency_class_4}
\end{figure}

Like Class~1c \gfii{functions}, the behavior of \gfii{the} \gfii{functions} in this class depends on the number of cores that they \gfii{use}.  Figure~\ref{figure_performance}(d) shows the host \gfii{CPU system} and \geraldorevi{the} NDP \geraldorevi{system} performance as we increase the core count for two representative \gfii{functions} (\texttt{PLYGramSch} and \texttt{SPLFftRev}).  We make two observations from the figure. First, at low core counts, the \gfii{functions} do \emph{not} benefit from \gfii{the} NDP \gfii{system}. In fact, for a single core \geraldorevi{(16 cores)}, \texttt{PLYGramSch} \emph{slows down} by 67\% \geraldorevi{(3$\times$)} when running on \geraldorevi{the} NDP \geraldorevi{system}, compared to running on the host \gfii{CPU} \geraldorevi{system}. This is because, at low core counts, the\gfiii{se} \gfii{functions} make reasonably good use of the cache hierarchy, with LFMR values of 0.03 for \texttt{PLYGramSch} and \gfii{lower than} 0.44 for \texttt{SPLFftRev} \gfii{until 16 host CPU cores}.
We confirm this in Figure~\ref{fig:latency_class_4}, where we see that very few memory requests for \texttt{PLYGramSch} and \texttt{SPLFftRev} go to DRAM (5\% for \texttt{PLYGramSch}, and at most 13\% for \texttt{SPLFftRev}) \geraldorevi{at \gfiii{core counts} low\gfii{er than 16}}.
Second, at high core counts (\gfii{i.e.,} 64 for \texttt{PLYGramSch} and 256 for \texttt{SPLFftRev}), the host \gfii{CPU} \geraldorevi{system} performance starts to \emph{decrease}.  This is because Class~2a \gfii{functions} are \emph{bottlenecked by cache contention}.  At 256 cores, this contention undermines the \gfii{cache effectiveness} and causes the LFMR to increase to 0.97 for \texttt{PLYGramSch} and 0.93 for \texttt{SPLFftRev}.  With the last-level cache rendered  \geraldorevi{essentially} ineffective, the NDP \geraldorevi{system greatly improves} performance over the host \gfii{CPU} \geraldorevi{system:} by 2.23$\times$ for \texttt{PLYGramSch} and 3.85$\times$ for \texttt{SPLFftRev} \gfii{at 256 cores}.

One impact of the increased cache contention is that it \geraldorevi{converts} these high-\geraldorevi{temporal-}locality \gfii{functions} into \gfii{memory} latency-bound \gfii{functions}.  We find that with the increased number of requests going to DRAM \gfii{due to cache contention}, the \gls{AMAT} increases significantly, in large part due to queuing at the memory controller.  At 256~cores, the queuing becomes so severe that a large fraction of requests (\gfii{24\% for \texttt{PLYGramSch} and 67\% for \texttt{SPLFftRev}}) must be reissued because the \gfiii{memory controller} queues are full.
The increased \gfii{main memory} bandwidth available \gfii{to the NDP cores allows} the NDP \geraldorevi{system} to issue many more requests concurrently, which reduces the average length of the queue and, thus, \gfii{the main memory latency}. \geraldorevi{\gfii{T}he NDP system \gfii{also reduces} memory access latency \gfii{by getting} rid of \gfiii{L2/L3} cache lookup and interconnect latencies.}

\begin{figure}[h]
  \centering
  \includegraphics[width=1.0\linewidth]{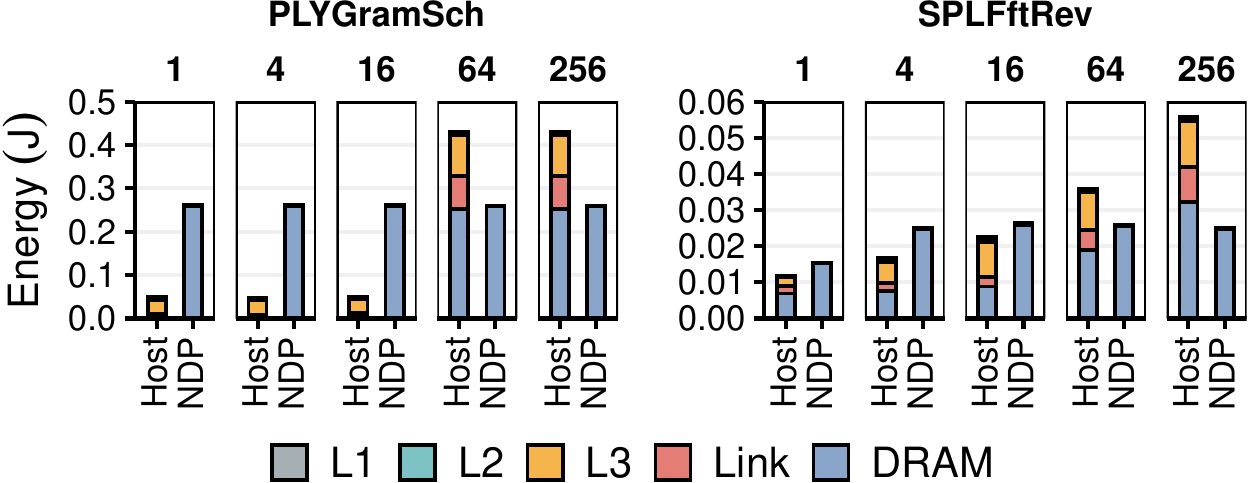}%
    \caption{Energy breakdown for \geraldorevi{representative} Class~2a \geraldorevi{\gfii{functions}}.}
    \label{fig:energy_group_4}
\end{figure}

Figure~\ref{fig:energy_group_4} shows the energy \gfii{breakdown} for \geraldorevi{the two representative} Class~2a \geraldorevi{\gfii{functions}}. We make two observations. First, the host \geraldorevi{\gfii{CPU} system} is more energy-efficient than the NDP \geraldorevi{system} \gfiii{at} low core count\geraldorevi{s}, as most of the memory requests are \geraldorevi{\gfii{served} by} on-chip \geraldorevi{caches in the host \gfii{CPU} system}. Second, \geraldorevi{the} NDP \geraldorevi{system} provides \gfii{large} energy savings over the host \geraldorevi{\gfii {CPU} system} at high core counts.  This is due to the increased cache contention, which increases the number of off-chip requests that the host  \gfii{CPU system} must make, increasing the L3 and off-chip link energy.

We conclude that cache contention is the primary \geraldorevi{scalability} bottleneck \geraldorevi{for Class~2a \gfii{functions}}, \gfii{and}
\gfii{the} NDP \gfii{system} can provide an effective way of mitigating \gfii{this} cache contention \gfii{bottleneck} without incurring the high area and energy overheads of providing additional cache capacity in the host \gfii{CPU system}, \gfii{thereby improving the scalability of these applications to high core counts.}

\subsubsection{Class~2b: High Temporal Locality, Low AI, Low/Medium LFMR, and Low MPKI \gfii{(\textit{L1 Cache Capacity Bottlenecked Functions})}}
\label{sec_scalability_class2b}

\begin{figure}[ht]
  \centering
  \includegraphics[width=1.0\linewidth]{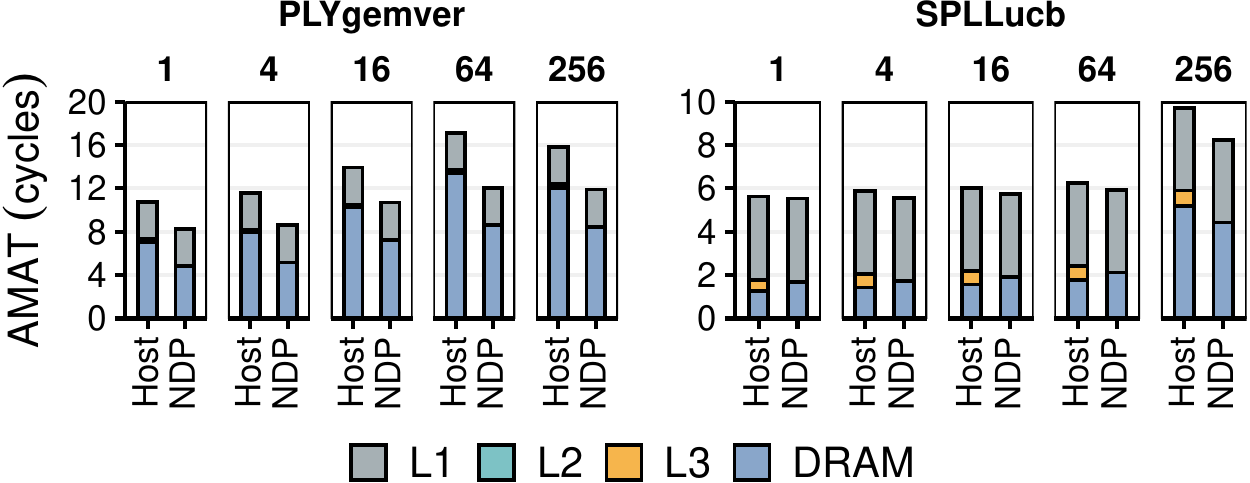}%
  \caption{\gls{AMAT} for \geraldorevi{representative} Class~2b \geraldorevi{\gfii{functions}}.}
   \label{fig_lt_group_5}
\end{figure}

Figure~\ref{figure_performance}(e) shows the host \gfii{CPU system} and \gfii{the} NDP \geraldorevi{system} performance for \texttt{PLYgemver} and \texttt{SPLLucb}. We make two observations from the figure. First, \geraldorevi{as} the number of cores \geraldorevi{increases}, performance \geraldorevi{of} the host \gfii{CPU system} and \geraldorevi{the} NDP \geraldorevi{system} scale in a \geraldorevi{very} similar fashion. The NDP \geraldorevi{system and the host \gfii{CPU} system perform essentially on par with \gfiii{(\gfiii{i.e., }within 1\% of)} each other at all core counts.} Second, even though \gfii{the} NDP \gfii{system} does not \geraldorevi{provide} any performance improvement for \geraldorevi{Class~2b} \gfii{functions}, it also does not hurt performance. Figure~\ref{fig_lt_group_5} shows the \gls{AMAT} for our two representative \gfii{functions}. When \texttt{PLYgemver} executes on the host \gfii{CPU} \geraldorevi{system}, up to 77\% of the memory latency \gfii{comes from} accessing L3 and DRAM, which can be explained by the \gfii{function}\geraldorevi{'s} medium LFMR \gfii{(0.5)}. For \texttt{SPLLucb}, even though up to 73\% of memory latency \gfii{comes from} L1 accesses, some requests still hit in the L3 cache (\geraldorevi{its} LFMR \geraldorevi{is} 0.2), translating to around 10\% of the memory latency. However, the latency \gfii{that comes from} L3 + DRAM for the host \gfii{CPU system} is similar to the latency to access DRAM \gfii{in} the NDP \gfii{system}, resulting in similar performance between the host \gfii{CPU system} and \geraldorevi{the} NDP \geraldorevi{system}.

\begin{figure}[ht]
 \centering
    \includegraphics[width=1.0\linewidth]{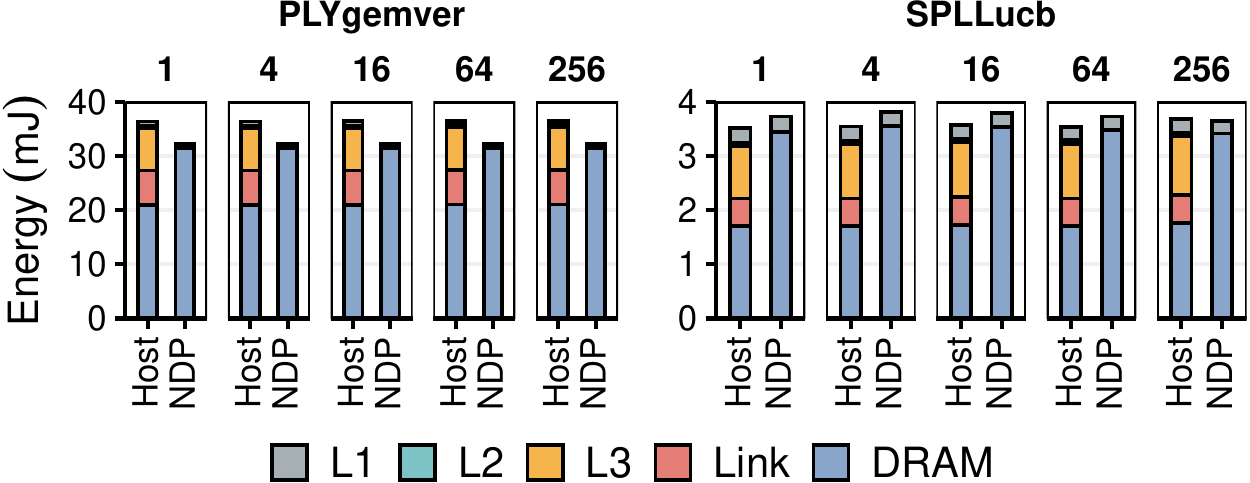}%
    \caption{Energy breakdown for \geraldorevi{representative} Class~2b \geraldorevi{\gfii{functions}}.}
    \label{fig:energy_group_5}
\end{figure}

We make a similar observation for the energy consumption for the host \gfii{CPU system} and \geraldorevi{the} NDP \geraldorevi{system} (Figure~\ref{fig:energy_group_5}). Even though \geraldorevi{a small number of} memory requests hit \gfii{in} L3, the total energy consumption for both \gfii{the} host \gfii{CPU system} and \gfii{the} NDP \gfii{system} is \gfii{similar} due to L3 and off-chip link energy. \gfii{For some functions in Class~2b, we observe that the NDP system slightly reduces energy consumption compared to the host CPU system. For example, the NDP system provides an \gfiii{12\%} average reduction in energy consumption, across all core counts, compared to the host CPU system for \texttt{PLYgemver}}.

We conclude that while \gfii{the} NDP \gfii{system} does not solve any memory bottlenecks for Class~2b \gfii{functions},
it can be used to reduce the overall SRAM area in the system without \gfii{any} performance or energy penalty \gfii{(and sometimes with energy savings)}.

\subsubsection{Class~2c: High Temporal Locality, High AI, Low LFMR, and Low MPKI \gfii{(\textit{Compute-Bound Functions}).}}
\label{sec_scalability_class2c}

Aside from one exception (\texttt{PLYSymm}), all of the \gfii{11} \gfii{functions} in this class exhibit high temporal locality.  When combined with the high AI and low memory intensity, we find that these characteristics significantly impact how \geraldorevi{the} NDP \geraldorevi{system performance} scales for this class. Figure~\ref{figure_performance}(f) shows the host \gfii{CPU system} and \geraldorevi{the} NDP \geraldorevi{system} performance for \texttt{HPGSpm} and \texttt{RODNw}, two representative \gfii{functions} from the class.  We make two observations from the figure. First, the host \geraldorevi{\gfii{CPU} system} performance is \emph{always} greater than the NDP \geraldorevi{system} performance \geraldorevi{(by 44\% for \texttt{HPGSpm} and 54\% for \texttt{RODNw}, on average)}. The high AI (more than 12~ops per cache line), combined with the high temporal locality \geraldorevi{and low MPKI}, \geraldorevi{enables} these \gfii{functions}  \geraldorevi{to make} excellent use of the host \geraldorevi{\gfii{CPU} system} resources. Second, \geraldorevi{both of the} \gfii{functions} benefit \gfii{greatly} from prefetching in the host \gfii{CPU system}. This is a direct result of these \gfii{functions}’ high spatial locality, which allows the prefetcher to be highly accurate \geraldorevi{and effective} in predicting which lines to retrieve \gfii{from main memory}.

\begin{figure}[h]
    \centering
    \includegraphics[width=\linewidth]{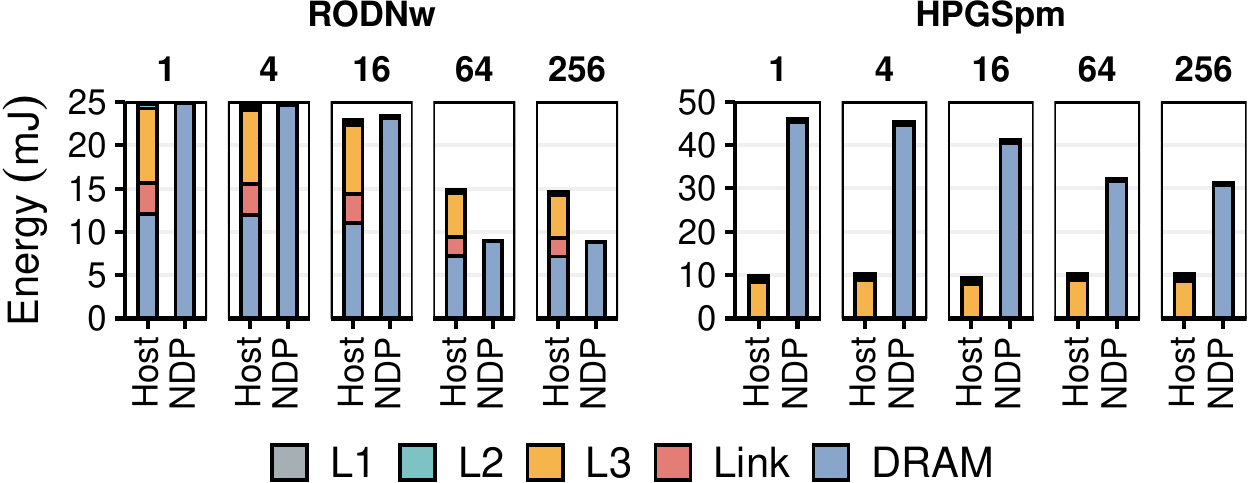}%
    \caption{Energy breakdown for \geraldorevi{representative} Class~2c \geraldorevi{\gfii{functions}}.}
    \label{fig:energy_group_6}
\end{figure}

\geraldorevi{Figure~\ref{fig:energy_group_6} shows the energy \gfii{breakdown} consumption for the two representative Class~2c \gfii{functions}. We make two observations. First, the host \gfii{CPU} system is 77\% more energy-efficient than the NDP system \gfii{for \texttt{HPGSpm}}, on average \gfii{across all core counts}. Second, the NDP system provides energy savings over the host \gfii{CPU} system at high core counts for \texttt{RODNw} (up to 65\% \gfii{at 256 cores}). \sgv{When the core count increases, the aggregate L1 cache capacity across all cores increases as well, which in turn decreases the number of L1 cache misses. Compared to executing on a single core, executing on 256~cores decreases the L1 cache miss count by 43\%, reducing the memory subsystem energy consumption by 40\%.} \gfv{However, due to \texttt{RODNw}'s medium LFMR of 0.5, the host CPU system \gfiv{still suffers from L2 and L3 cache misses at high core counts}, which \gfiii{require} the \gfiii{large} L3 and off-chip link energy. \gfiv{In contrast, the NDP system eliminates the energy of accessing the L3 cache and the off-chip link energy by directly sending L1 cache misses to DRAM, which\gfv{, at high core counts, leads to lower} energy consumption \gfv{than} the host CPU system.} }}

We conclude that Class~2c \gfii{functions} do not experience \gfiii{large} memory bottlenecks and are not a good fit for \gfii{the} NDP \gfii{system} \geraldorevi{in terms of performance. However, \gfii{the} NDP \gfii{system} can \gfii{sometimes} provide energy savings for \gfii{functions} that experience medium LFMR. }

\subsection{\gfvii{Effect of the Last-Level Cache Size}}
\label{sec_scalability_nuca}
 
\geraldorevi{The bottleneck classification we present in Section~\ref{sec:scalability} depends 
\juan{on} two key architecture-dependent metrics \gfii{(\gls{LFMR} and \gls{MPKI})} that are directly \gfii{affected} by the \gfiii{parameters} and the organization of the cache hierarchy. \gfii{O}ur analysis in Section~\ref{sec:scalability} partially evaluate\gfii{s} \gfii{the effect of caching} by scaling the aggregated size of the private \gfiii{(L1/L2)} caches with the number of cores in the system while maintaining the size of the L3 cache fixed at 8~MB for the host \gfii{CPU} system. However, we \gfiii{also} need to understand the impact of the L3 cache size on our bottleneck classification analysis. To this end, t}his section evaluates the effects on our \geraldorevi{bottleneck classification analysis} of using an alternative cache hierarchy configuration, \geraldorevi{where we employ} a \gls{NUCA}~\cite{kim2002adaptive} model \geraldorevi{to scale the size of the L3 cache \gfii{with the number of cores} in the host \gfii{CPU} system.}

 \begin{figure*}[!t]
    \centering
    \includegraphics[width=\linewidth]{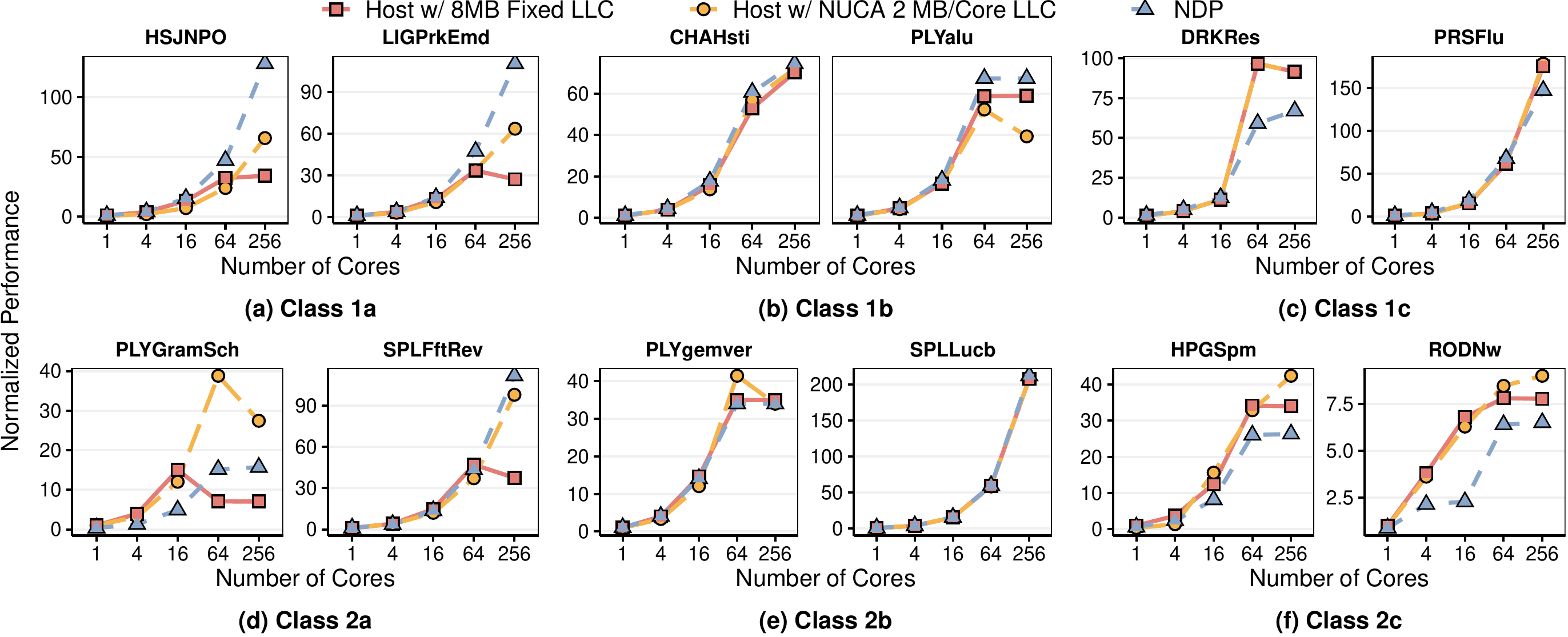}%
    \caption{Performance of \geraldorevi{the} host and \geraldorevi{the} NDP \geraldorevi{system} \geraldorevi{as we vary the} LLC size, normalized to one host core with \geraldorevi{a} fixed 8MB LLC size.}
    \label{fig_nuca_analysis}%
\end{figure*}

In this configuration, we maintain the sizes of the private L1 and L2 caches (32~kB and 256~kB \juan{per core}, respectively) while increasing the shared L3 cache size with the core count (we use 2~MB/core) \gfii{in the host CPU system}. The cores, shared L3 caches\geraldo{, and DRAM memory controller} are interconnected using a 2D-mesh \gls{NoC}~\gfiv{\cite{moscibroda2009case,das2009application,das2010aergia,nychis2012chip,besta2018slim,fallin2011chipper,grot2011kilo,benini2002networks}} of size $(n+1) \times (n+1)$ \geraldorevi{\gfii{(an extra interconnection dimension is added to place the DRAM memory controller\gfiii{s})}}. To faithfully simulate the \gls{NUCA} model (e.g., including network contention in our simulations), we integrate the M/D/1 network model proposed by ZSim++~\cite{zsimplusplus} in our \geraldorevi{\bench} simulator~\cite{damov}. We use a latency of 3 cycles per hop in our analysis, as suggested by prior work~\cite{zhang2018minnow}. \gfii{We adapt our energy model to account for the energy consumption of the \gls{NoC} in the \gls{NUCA} system. We consider router energy consumption of 63 pJ per request and energy consumed per link traversal of 71 pJ, same as previous work~\cite{tsai2017jenga}.}

Figure~\ref{fig_nuca_analysis} shows the \gfii{performance} scalability curves for representative \geraldorevi{\gfii{functions}} from each one of our bottleneck classes \geraldorevi{presented in Section~\ref{sec:scalability}} \gfii{for the baseline host CPU system (\emph{Host with 8MB Fixed LLC}), the host CPU NUCA system (\emph{Host with NUCA 2MB/Core LLC}), and the NDP system}. We make \geraldorevi{two} observations. First, \geraldorevi{the observations we make for our bottleneck classification \juangg{(Section~\ref{sec:scalability})}} \geraldorevi{are} \gfiii{\emph{not}} affected by increasing the L3 cache size for \geraldorevi{C}lasses 1a, 1b, 1c, 2b, and 2c. \geraldorevi{We observe that Class~1a \gfii{functions} benefit from a large L3 cache size (by up to \juangg{1.9$\times$/2.3$\times$} for \juangg{\texttt{HSJNPO}/\texttt{LIGPrkEmd}}
at 256 cores). However, the NDP system still provides performance benefits compared to the \gfii{host CPU} NUCA system. We observe that increasing the L3 size reduces some of the pressure on main memory but cannot fully reduce the \gfii{DRAM} bandwidth bottleneck for \gfii{Class~1a functions}. \gfii{Functions} in Class~1b do \emph{not} benefit from extra L3 capacity} (we do not observe a decrease in \gls{LFMR} 
\juangg{or} \gls{MPKI}). \gfii{Functions} in Class 1c do \gfiii{\emph{not}} benefit from extra \gfiii{L3} cache \geraldorevi{capacity}. \gfii{W}e observe that the private L1 and L2 caches capture most of their data locality, \gfiii{as mentioned \geraldorevi{in Section~\ref{sec_scalability_class1c}}}, \geraldorevi{and} thus, \geraldorevi{these \gfii{functions} do \emph{not}} benefit \geraldorevi{from} \gfii{increasing the} L3 size. \gfiii{Functions in Class~2b do \emph{not} benefit from extra L3 cache capacity, which can even lead to a decrease in performance at high core counts for the host CPU NUCA system in some Class~2b functions} \gfiv{due to long NUCA L3 access latencies}. \gfii{For example, we observe that  \texttt{PLYgemver}'s performance drops 18\% \juangg{when increasing the core count} from 64 to 256 in the host CPU NUCA system. We do \emph{not} observe \juangg{such} a performance drop for the host CPU system with fixed LLC size. \juangg{The} performance drop in the host CPU NUCA system is due to the increase in the number of hops that L3 requests need to travel in the NoC at high core counts, which increase the function's AMAT.} \gfii{Class~2c \gfii{functions} benefit from a \juangg{larger} last-level cache. We observe that \juangg{their} performance improves by 1.3$\times$/1.2$\times$ for \texttt{HPGSpm}/\texttt{RODNw} compared to the host CPU system with 8MB fixed LLC at 256 cores.}

\geraldorevi{Second}, we observe two different \geraldorevi{types of} behavior for \gfii{functions} in Class 2a. Since cache conflicts \geraldorevi{are the major bottleneck for} \gfii{functions} in this class, we observe that increasing the L3 cache size can mitigate \geraldorevi{this} bottleneck. In \gfiii{Figure~\ref{fig_nuca_analysis}}, we observe that for both \texttt{PLYGramSch} and \texttt{SPLFftRev}, the \gfii{host system with} \geraldorevi{NUCA} 2MB/Core LLC provides better performance than the \gfii{host system with 8MB fixed LLC}. However, \gfii{the} \gls{NDP} \gfii{system} can still provide performance benefits in case of contention \gfiii{on the L3 \gls{NoC}} \gfii{(e.g., in \texttt{SPLFftRev})}. \geraldorevi{For \gfii{example}, the NDP system provides 14\% performance improvement \gfii{for \texttt{SPLFftRev}} compared to the NUCA system \gfiii{(with 512~MB L3 cache)} for 256 cores.}

\geraldorevi{In summary, we conclude that the key takeaways and observations we present \gfii{in} our bottleneck classification in Section~\ref{sec:scalability} are also valid for a host system with a shared last-level cache \gfii{whose size scales with core count.}} \gfiii{In particular, different workload classes get affected by an increase in L3 cache size as expected by their characteristics distilled by our classification.}

\begin{figure*}[h]
    \centering
    \includegraphics[width=\linewidth]{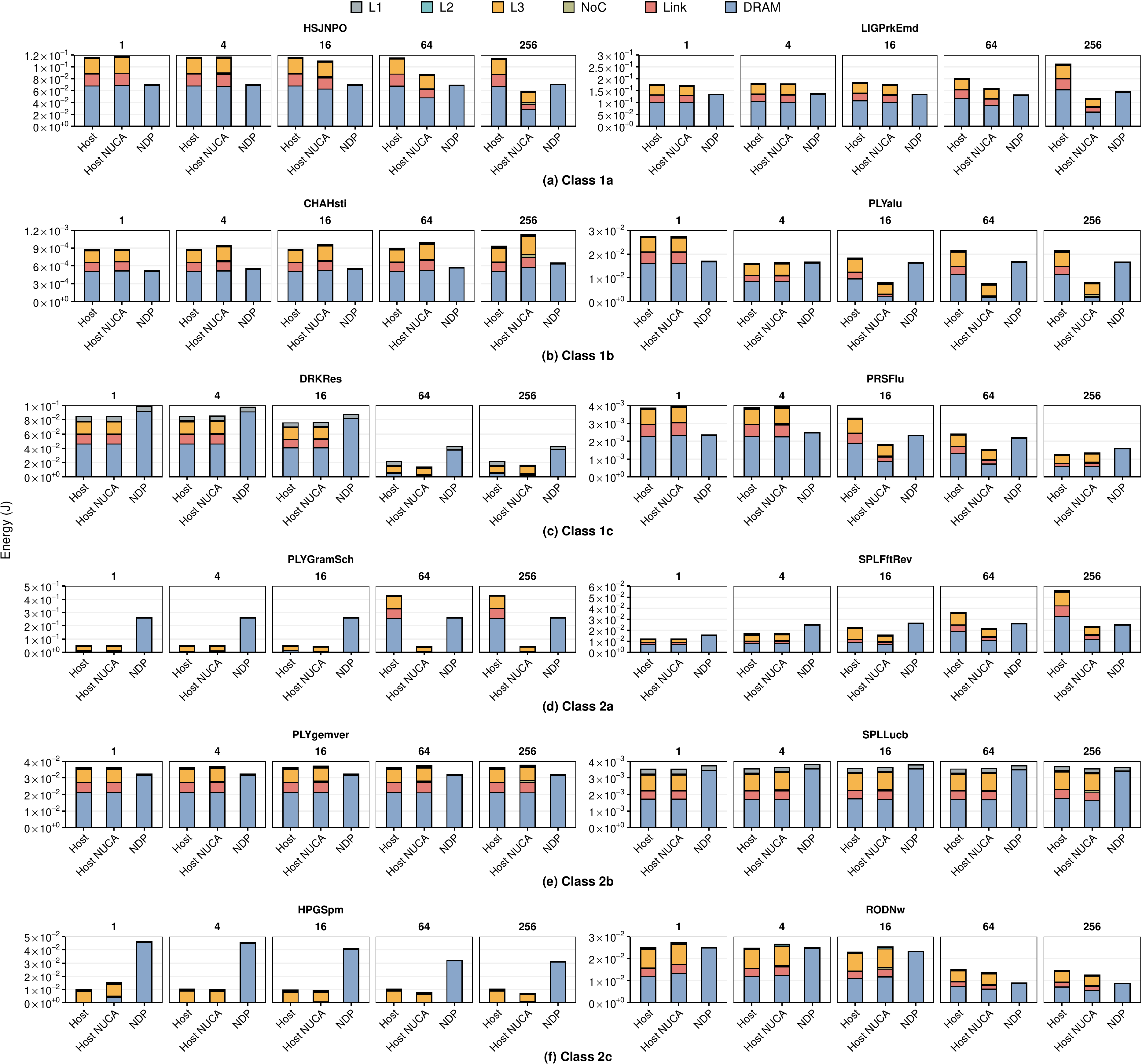}%
    \caption{\gfii{Energy of the host and the NDP system as we vary the LLC size. \emph{Host} refers to the host system with a fixed 8MB LLC size; \emph{Host NUCA} refers to the host system with 2MB/Core LLC.}}
    \label{fig_nuca_analysis_energy}%
\end{figure*}

\gfii{Figure~\ref{fig_nuca_analysis_energy} shows the \gfii{energy} consumption for representative \gfii{functions} from each one of our bottleneck classes  presented in Section~\ref{sec:scalability}. \gfiii{We observe that the NDP system can provide \gfiii{substantial} energy savings for functions in different bottleneck classes, even compared against a system with \gfiii{very} large \gfiii{(e.g., 512~MB)} cache sizes.} We make the following observations for each bottleneck class:}

\begin{itemize}[noitemsep, leftmargin=*, topsep=0pt]
    \item \gfii{\emph{Class~1a}: First, for both representative functions in this bottleneck class, the host CPU NUCA system and the NDP system reduce energy consumption compared to the baseline host CPU system. However, we observe that the NDP system provides \juangg{larger} energy savings than the host CPU NUCA system. On average, across all core counts, the NDP system and the host CPU NUCA system reduce energy consumption compared to the host CPU system for \texttt{HSJNPO}/\texttt{LIGPrkEmd} by  46\%/65\% and 25\%/22\%, respectively. Second, at 256 cores, the host CPU NUCA system provides larger energy savings than the NDP system for both representative functions. This happens because at 256 cores, the large L3 cache \gfiii{(i.e., 512~MB)} capture\gfiii{s a large} portion of the dataset for these functions, reducing costly DRAM traffic. The host CPU NUCA system reduces energy consumption compared to the host CPU system for \texttt{HSJNPO}/\texttt{LIGPrkEmd} \gfiii{at 256 cores} by 2.0$\times$/2.2$\times$ while the NDP system reduces energy consumption by 1.6$\times$/1.8$\times$. \gfiii{The L3 cache capacity needed to make the host CPU NUCA system more energy efficient than the NDP system is \gfv{\emph{very}} large (512~MB SRAM), which is likely not cost-effective.}}
    
    \item \gfii{\emph{Class~1b}: First, for \texttt{CHAHsti}, the host CPU NUCA system \emph{increases} energy consumption compared to the host CPU system by 9\%\gfiii{, on average across all core counts}. In contrast, the NDP system \emph{reduces} energy consumption by 57\%. Due to its low spatial and temporal locality (Figure~\ref{fig:locality_chart}), this function does not benefit from a deep cache hierarchy. In the host CPU NUCA system, the extra energy from the \gfiii{large amount of} \gls{NoC} traffic further increases the cache hierarchy's overall energy consumption. Second, for \texttt{PLYalu}, the host CPU NUCA system and the NDP system reduce energy consumption compared to the host CPU system by 76\% and 23\%, on average across all core counts. Even though the increase in LLC size does not translate to performance improvements, the large LLC sizes in the host CPU NUCA system aid to reduce DRAM traffic, \gfiii{thereby} providing energy savings compared to the baseline host CPU system.} 
    
    \item \gfii{\emph{Class~1c}: First, for \texttt{DRKRes}, the host CPU NUCA system reduces energy consumption compared to the host CPU system by 15\%\gfiii{, on average across all core counts}. In contrast, the NDP system increases energy consumption by 30\%, which is due to the function's medium \gls{LFMR} (Section~\ref{sec_scalability_class1c}). Second, for \texttt{PRSFlu}, we observe that the NDP system provides large energy savings than the host CPU NUCA system. The host CPU NUCA system reduces energy consumption compared to the host CPU system by 21\%, while the NDP system reduces energy consumption by 25\%, on average across all core counts. However, the energy savings of both host CPU NUCA and NDP systems compared to the host CPU system reduces \gfiii{at high-enough core counts} (\gfiii{the energy consumption of the host CPU NUCA system (NDP system) is 0.6$\times$ (0.9$\times$) that of the host CPU system at 64 cores and 1.1$\times$ (1.3$\times$) that of the host CPU system at 256 cores).} This result is expected for Class~1c functions since \gfiii{the functions in this class have decreasing \gls{LFMR}, i.e., the function\gfiii{s} effectively utilize the private L1/L2 caches in the host CPU system at high-enough core counts}.} 
    
    \item \gfii{\emph{Class~2a}: First, for \texttt{PLYGramSch}, \gfiii{compared to the host CPU system} the host CPU NUCA system reduces energy consumption  by 2.53$\times$ and the NDP system increases energy consumption by 55\%\gfiii{, on average across all core counts}. \gfiii{Even though at high core counts (64 and 256 cores) the host CPU NUCA system provides larger energy savings than the NDP system compared to the host CPU system (the host CPU NUCA system and the NDP system reduce energy consumption compare to the host CPU system by 9$\times$ and 65\% respectively, \gfiii{averaged across 64 and 256 cores}), such large energy savings come at the cost of very large (e.g., 512~MB) cache sizes}. Second, for \texttt{SPLFftRev}, the host CPU NUCA system and the NDP system reduce energy consumption compared to the host CPU system by 42\% and 7\%, on average across all core counts. The NDP system increases energy consumption compared to the host CPU system at low core counts (an increase of 33\%, averaged across 1, 4, and 16 cores). However, it provides similar energy savings as the host CPU NUCA system for large core counts (99\% and 75\% energy \gfiii{reduction compare to the host CPU system} for the host CPU NUCA system and the NDP system, respectively, averaged across 64 and 256 cores counts). Since the function suffers from high network contention, the increase in core count increases \gls{NoC} traffic, which \gfiii{in turn} increases energy consumption for the host CPU NUCA system. \gfiii{We conclude that the NDP system provides energy savings for Class~2a applications compared to the host CPU system at lower cost than the host CPU NUCA system.}} 
    
    \item \gfii{\emph{Class~2b}: First, for \texttt{PLYgemver}, the host CPU NUCA system increases energy consumption compared to the host CPU system by 2\%\gfiii{, on average across all core counts}. In contrast, the NDP system reduces energy consumption by 13\%. This function does not benefit from large L3 cache sizes since Class~2b functions are bottlenecked by L1 capacity. Thus, the \gls{NoC} only adds extra \gfiii{static and dynamic} energy consumption. Second, for \texttt{SPLLucb}, the host CPU NUCA system consumes the same energy as the host CPU system while the NDP system increases energy consumption by 5\%, averaged across all core counts.}
    
    \item \gfii{\emph{Class~2c}: For both representative functions in this class, the host CPU NUCA system reduces energy consumption compared to the host CPU system while the NDP system increases energy consumption. For \texttt{HPGSpm}/\texttt{RODNw}, the host CPU NUCA system reduces energy consumption by 6\%/9\% while the NDP system increases energy consumption by 74\%/22\%, averaged across all core counts. This result is expected since Class~2c functions are compute-bound and highly benefit from a deep cache hierarchy.}
\end{itemize}

\gfii{In conclusion, the NDP system can provide \gfiii{substantial} energy savings for functions in different bottleneck classes, even compared against a system with \gfiii{very} large \gfiii{(e.g., 512~MB)} cache sizes.}

\subsection{ \geraldorevi{Validation} \gfiii{and Summary} of Our \geraldorevi{Workload} Characterization \juanggg{Methodology}}
\label{sec_summary}

\gfiii{In this section, we present \gfiv{the} validation \gfiii{and a summary} of our \gfiv{new} workload characterization \gfiv{methodology}. \gfiii{First, we use the remaining 100 memory-bound functions we obtain from \emph{Step~1} \gfiv{(see Section~\ref{sec:vtune})} to \juanggg{validate our workload characterization methodology. To do so, we calculate the accuracy of our workload classification by using the remaining 100 memory-bound functions\gfiv{, which were not used to identify the six classes we found and described in Section~\ref{sec:scalability}}.}
\gfiii{Second}, we present a summary of the key metrics we obtain for all 144 memory-bound functions, including our analysis of the host CPU system and the NDP system using \gfv{two types of cores (in-order and out-of-order)}.}}

\subsubsection{Validation of Our Workload Characterization  \juanggg{Methodology}}
\label{sec::sub::validation}

\gfiii{Our goal is to evaluate the accuracy of our workload characterization \gfv{methodically} on a large set of \gfiv{functions}. To this end, we apply \emph{Step 2} and \emph{Step 3} of our memory bottleneck classification methodology \gfiv{(as described in Sections~\ref{sec:step2} and \ref{sec:step3})} to the remaining 100 memory-bound functions we obtain from \emph{Step~1} (in Section~\ref{sec:vtune}). Then, we \gfiv{perform} a two-phase validation to calculate the accuracy of our workload characterization.}

\gfiii{In \emph{phase 1} of our validation, we calculate the threshold values that define the low/high boundaries of each of the four metrics we use to cluster the initial 44 functions in the six memory bottleneck \gfiv{classes \gfv{in}} Section~\ref{sec:scalability} (i.e., temporal locality, \gls{LFMR}, LLC \gls{MPKI}, and \gls{AI}). We also include the LFMR curve slope to indicate when \gfii{the LFMR} increases, decreases or stays constant \gfiv{as we} scal\gfiv{e} the core count. We calculate the threshold values for a metric \texttt{M} by computing the middle point between (i) the average value of \texttt{M} across the memory bottleneck \gfiv{classes} with \emph{low} values of \texttt{M} and (ii) the average value of \texttt{M}  across the memory bottleneck \gfiv{classes} with \emph{high} values of \texttt{M} values out of the 44 functions. In \emph{phase 2} of our validation, we calculate the accuracy of our workload characterization by classifying the remaining 100 memory-bound functions using the threshold values obtained from \emph{phase 1} and the \gls{LFMR} curve slope. \gfiv{After \emph{phase 2}\gfv{,} a \gfv{function} is \gfv{considered to be \emph{accurately}} classified into a correct memory bottleneck \gfiv{class} if \gfv{and only if} it \gfv{(1)} fits the definition of the assigned class using the threshold values obtained from \emph{phase 1} \gfiv{and \gfv{(2)} follows the expected performance trends \gfv{of the assigned class} when the \gfv{function} is executed in the host CPU system and the NDP system}. For example, a \gfv{function} is correctly classified into Class~1a \emph{if and only if} it \gfv{(1)} displays low temporal locality, low \gls{AI}, high \gls{LFMR}, high \gls{MPKI} \gfiv{and \gfv{(2)} the NDP system outperforms the host CPU system \gfv{as we scale} the core count when executing the \gfv{function}}.} The final \gfv{\emph{accuracy of our workload characterization methodology}} is calculated by \gfiv{computing} the percentage of the functions that are \gfv{\emph{accurately}} classified into one of the six memory bottleneck \gfiv{classes}.
}

\begin{figure*}[!h]
\vspace{-8pt}
\begin{subfigure}{\textwidth}
  \centering
 \includegraphics[width=\textwidth]{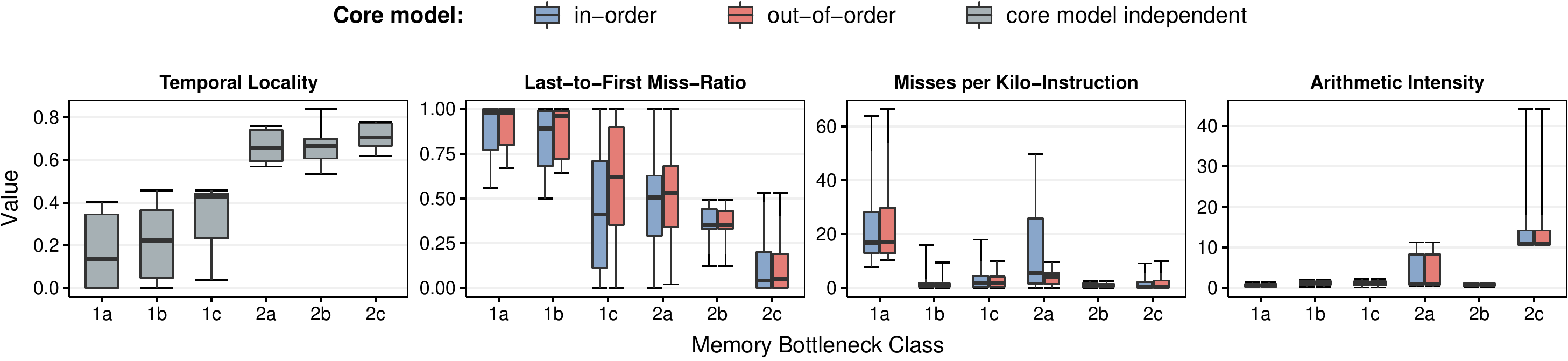}%
 \vspace{-3pt}
  \caption{\gfiv{Summary of the key metrics for each memory bottleneck class. }}
  \label{fig:summary:metrics}
\end{subfigure}
\par\bigskip
\begin{subfigure}{\textwidth}
  \centering
 \includegraphics[width=0.99\textwidth]{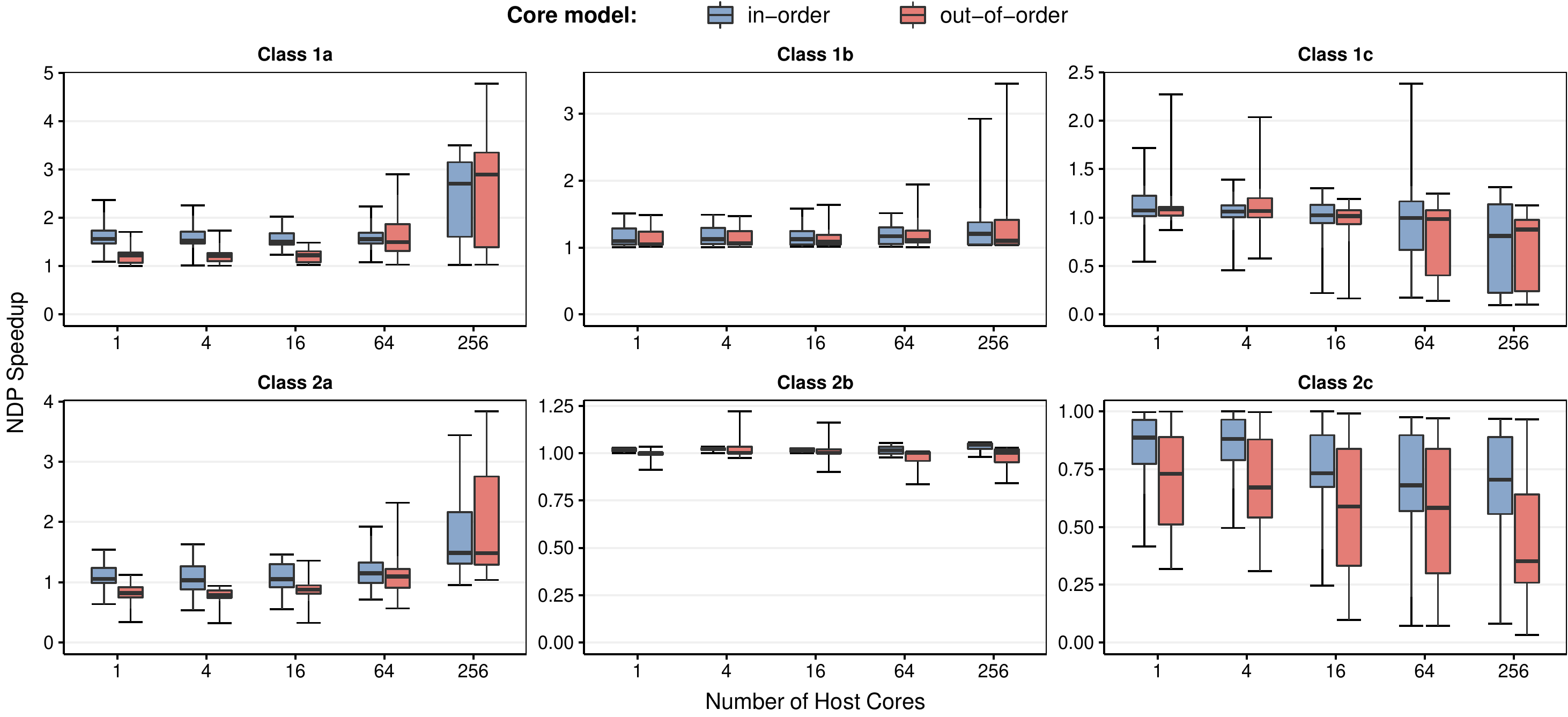}%
 \vspace{-3pt}
  \caption{\gfiv{Summary of NDP speedup for each memory bottleneck class at 1, 4, 16, 64, and 256 cores.}}
  \label{fig:summary:speedup}
\end{subfigure}
     \caption{Summary of our characterization for \gfiv{all} 144 \gfv{memory-bound} functions. \gfv{Each box is lower-bounded by the first quartile and upper-bounded by the third quartile. The median falls within the box. The inter-quartile range (IQR) is the distance between the first and third quartiles (i.e., box size). Whiskers extend to the minimum and maximum data point values on either sides of the box.}}
     \label{fig_summary}
\end{figure*}

\gfiii{First, by applying \emph{phase 1} of our two-phase validation, we obtain that the threshold values are: 0.48 for \emph{temporal locality}, 0.56 for \emph{\gls{LFMR}}, 11.0 for \emph{\gls{MPKI}}, and 8.5 for \emph{\gls{AI}}. Second, by applying \emph{phase 2} of our two-phase validation, we \gfv{find that we can accurately} classify \textit{97\% of the 100 memory-bound functions} into one of our six memory bottleneck \gfiv{classes} \gfiv{(i.e., the accuracy of our workload characterization methodology is 97\%)}. We observe that three functions (\textit{Ligra:ConnectedComponents:compute:rMat}, \textit{Ligra:MaximalIndependentSet:edgeMapDense:USA}, and \textit{SPLASH-2:Oceanncp:relax}) could not be \gfv{accurately} classified into their \gfv{correct} memory bottleneck \gfiv{class} (Class~1a). We observe that these functions have LLC \gls{MPKI} values \emph{lower} than the \gls{MPKI} threshold expected for Class~1a functions. We expect that the accuracy of our \gfv{methodology} can be further improved by incorporating more workloads into our workload suite and fine-tuning each metric to encompass an even large\gfiv{r} set of applications.}

\gfiii{\gfiv{W}e conclude that our workload characterization methodology can accurately classify \gfiv{a given new} application\gfiv{/function} into \gfiv{its} appropriate memory bottleneck \gfiv{class}.}

\subsubsection{Summary of \gfiv{O}ur \gfiv{W}orkload \gfiv{C}haracterization \gfiv{Results}.} 
\label{sec::sub::summary}

\gfiii{Figure~\ref{fig:summary:metrics} summarizes the metrics we collect for all 144 functions across all core counts (i.e., from 1 to 256 cores) and different core microarchitectures \gfiv{(i.e., out-of-order and in-order cores)}. The figure shows the distribution of the key metrics we use during our workload characterization for each memory bottleneck class in Section~\ref{sec:scalability}, including architecture-independent metrics (i.e., temporal locality) and architecture-dependent metrics (i.e., \gls{AI}, \gls{LFMR}, and LLC \gls{MPKI}). We report the architecture-dependent metrics for two core models: (i) in-order and (ii) out-of-order cores.\footnote{\gfiv{In Section~\ref{sec:scalability}, we collect and report the values of the architecture-independent metrics and architecture-dependent metrics for a subset of 44 representative functions out of the 144 memory-bound functions we identify in \emph{Step 1} of our workload characterization methodology. In Section~\ref{sec::sub::summary}, we report values for the \emph{complete set} of 144 memory-bound functions.}} Together with the out-of-order core model that we use in Section~\ref{sec:scalability}, we incorporate an in-order core model to \gfiv{our} analysis, \gfiv{so as} to show that our memory bottleneck classification methodology focuses on data movement requirements and \gfiv{works} independent\gfiv{ly} of the core microarchitecture. \gfiv{Figure~\ref{fig:summary:speedup}} shows the distribution of speedups we observe for when \gfiv {we} offload the function to our general-purpose \gls{NDP} cores, \gfiv{while} employing the same core type as the host CPU system.}

\gfiii{We make two key observations from \gfv{Figure~\ref{fig_summary}}. First, \gfiv{we observe similar values for each \gfv{architecture-dependent} key metric \gfv{(i.e., \gls{LFMR}, \gls{MPKI}, \gls{AI})} regardless of core type for all 144 functions \gfv{(in Figure~\ref{fig:summary:metrics})}.}} \gfiii{Second, we observe that the \gls{NDP} system achieves similar speedup\gfv{s over the host CPU system, when using both} in-order and out-of-order core configurations \gfv{(in Figure~\ref{fig:summary:speedup})}. The speedup provided by the NDP system compared to the host CPU system when both systems use out-of-order (in-order) cores for Classes 1a, 1b, 1c, 2a, 2b, and 2c is 1.59 (1.77), 1.22 (1.15), 0.96 (0.95), 1.04 (1.22), 0.94 (1.01), and 0.56 (0.76), respectively, on average across all core counts and functions within a \gfiv{memory} bottleneck class. \gfiv{The NDP system greatly outperforms the host CPU system across \emph{all core counts} for Class~1a and 1b functions, with a maximum speedup for the out-of-order (in-order) core model of 4.8 (3.5) and 3.4 (2.9), respectively. The NDP system greatly outperforms the host CPU system at \emph{low core counts} for Class~1c functions and at \emph{high core counts} for Class~2a functions, with a maximum speedup for the out-of-order (in-order) core model of 2.3 (2.4) and 3.8 (3.4), respectively. The NDP system provides a modest speedup compared to the host CPU system across \emph{all core counts} for Class~2b functions and slowdown for Class~2c functions, with a maximum speedup for the out-of-order (in-order) core model of 1.2 (1.1) and \gfv{1.0 (1.0)}, respectively.} We observe that\gfv{, averaged across all classes and core types,} the average speedup provided by the NDP system using in-order cores is \gfiv{11\%} higher than the average speedup offered by the NDP system using out-of-order cores. This \gfiv{is} because the host CPU system \gfiv{with} out-of-order cores can hide the \gfiv{performance} impact of memory access latency to some degree (e.g., using \gfiv{dynamic} instruction \gfiv{scheduling})~\gfv{\cite{mutlu2006efficient,mutlu2005techniques,mutlu2003runahead,hashemi2016continuous,oooexec,mutlu2003runaheadmicro}}. On the other hand, the host CPU system using in-order cores has \gfv{little tolerance} to hide memory access latency~\gfv{\cite{mutlu2006efficient,mutlu2005techniques,mutlu2003runahead,hashemi2016continuous,oooexec,mutlu2003runaheadmicro}}.}

\gfiii{We conclude that our methodology to classify memory bottlenecks \gfv{of} application\gfv{s} is \emph{robust} \gfv{and \emph{effective}} \gfiv{since we observe similar trends for the six memory bottleneck classes across a large range of \gfv{(144)} functions and two \gfv{very} different core models.}}

\subsection{Limitations of Our Methodology}
\label{sec_scalability_limitations}

\geraldorevi{We identify three limitations to our workload characterization methodology. We discuss each limitation next.}

\noindent
\textbf{NDP Architecture \geraldorevi{Design Space}.} Our methodology uses the same type and number of cores in \gfii{the} host \gfii{CPU} and \gfiii{the} NDP \gfii{system} configurations \gfiii{for our} scalability analysis (Section~\ref{sec:scalability}) because our main goal is to highlight the performance and energy differences between \gfii{the} host \gfii{CPU system} and \gfii{the} NDP \gfii{system that are} caused by data movement. We do not consider practical limitations related to area or thermal dissipation that could affect the type and the maximum number of cores \gfii{in} the NDP \gfii{system}, because our goal is \textbf{not} \geraldorevi{to propose} NDP architectures but \geraldorevi{to} characteriz\geraldorevi{e data movement \gfiii{and understand the different data movement bottlenecks} in modern} workloads. Proposing NDP architectures for the \geraldorevi{workload classes} that our methodology identifies as \gfiii{suitable for} NDP is a \geraldorevi{\gfii{promising}} topic for future work.

\noindent
\textbf{Function-level Analysis.} We \geraldorevi{choose} to conduct our analysis at a function granularity rather than at the application \geraldorevi{granularity} for two major reasons. First, general-purpose NDP architectures are typically leveraged as accelerator\geraldorevi{s} to which only \emph{parts} of the application or specific \gfiii{functions} are offloaded~\gfv{\cite{boroumand2018google, PEI, hsieh2016transparent, ke2019recnmp, top-pim, kim2018grim, seshadri2017ambit, nai2017graphpim,hadidi2017cairo,hajinazarsimdram,gao2016hrl,lockerman2020livia,boroumand2021polynesia,dualitycache,xin2020elp2im,devaux2019true,IBM_ActiveCube,Asghari-Moghaddam_2016,alves2016large, seshadri2013rowclone,azarkhish2016design,asghari2016chameleon}}, rather than the entire application. Functions typically form natural boundaries for \geraldorevi{parts} of algorithms/applications that can potentially be offloaded. Second, it is well-known that applications go through distinct phases during execution. Each phase may have different characteristics (e.g., a phase might be more compute-bound, while another one might be \gfiii{more} memory-bound) \geraldorevi{and thus fall into different classes in our analysis}.  A fine-grained analysis at the function level enables us to identify each of those phases and hence, identify more fine-grained opportunities for NDP offloading. \geraldorevi{However, the main drawback of function-level analysis is that \gfii{it does not take into account} data movement across function boundaries\gfii{, which affect\gfiii{s}} the performance and energy benefits the NDP system provides \gfiii{over} the host \gfii{CPU} system. For example, the NDP system might hurt overall system performance and energy consumption when a large amount of data needs to be continuously moved between a function executing on the NDP \gfii{cores} and another executing on the host \gfii{CPU cores}~\cite{boroumand2019conda, lazypim}.}

\noindent
\textbf{Overestimating NDP Potential.} Offloading kernels to NDP cores incur\gfiv{s} overheads that \gfii{our analysis do\gfiii{es}} not account for \gfiii{(e.g., maintaining coherence between the host CPU and the NDP cores~\gfiv{\cite{lazypim,boroumand2019conda}}, efficiently synchronizing computation across NDP cores~\gfiv{\cite{syncron,juansigmetrics21}}, providing virtual memory support for the NDP system~\gfiv{\cite{hsieh2016accelerating,PEI,picorel2017near}}, and dynamic offloading support for NDP-friendly functions~\gfiv{\cite{hsieh2016transparent}})}. \gfiii{Such} overheads \geraldorevi{can} impact the performance benefits NDP can provide when considering the end-to-end application. However, deciding \geraldorevi{how to and} whether or not to offload computation to NDP is an open research topic, which involves several architecture-dependent components in the system\gfiii{, such as the following two examples}. \gfiii{First}, maintaining coherence between \gfii{the} host \gfii{CPU} and \geraldorevi{the} NDP cores is a challenging task that \geraldorevi{recent} works tackle~\cite{lazypim,boroumand2019conda}. \gfiii{Second, enabling efficient synchronization across NDP cores is challenging due to the lack of shared caches and hardware cache coherence protocols in NDP systems. Recent works, such as \cite{syncron,liu2017concurrent}, provide solutions to the NDP synchronization problem.}  \gfiii{Therefore, to focus our analysis on the data movement characteristics of workloads and the broad benefits of NDP, we minimize our assumptions about our target NDP architecture, making our evaluation as broad\gfiv{ly} \gfiv{applicable} as possible.}

\geraldorevi{\section{\bench: The Data Movement Benchmark Suite}}
\label{sec:benchmark}

\geraldorevi{In this section, we present \bench, the DAta MOVement Benchmark Suite. \bench is the collection of the 144 functions we use to drive our memory bottleneck classification in Section~\ref{sec:characterization}. The benchmark suite is divided into each one of the six classes of memory bottlenecks presented in Section~\ref{sec:characterization}. \bench is the first benchmark suite that encompasses \emph{real} applications from a diverse set of application domains tailored to stress different memory bottlenecks in a system. We present the complete description of the functions in \bench in Appendix~A. We highlight the benchmark diversity of the function\gfii{s} in \bench in Section~\ref{sec_scalability_benchmark_diversity}.} \jgl{To me, it does not make much sense to have a section with a single subsection. We can just say that in this section we study benchmark diversity.} \gfiii{We open source DAMOV~\cite{damov} to facilitate further rigorous research in mitigating data movement bottlenecks, including in near data processing.}

\begin{figure*}[h]
    \centering
 \includegraphics[width=0.79\linewidth]{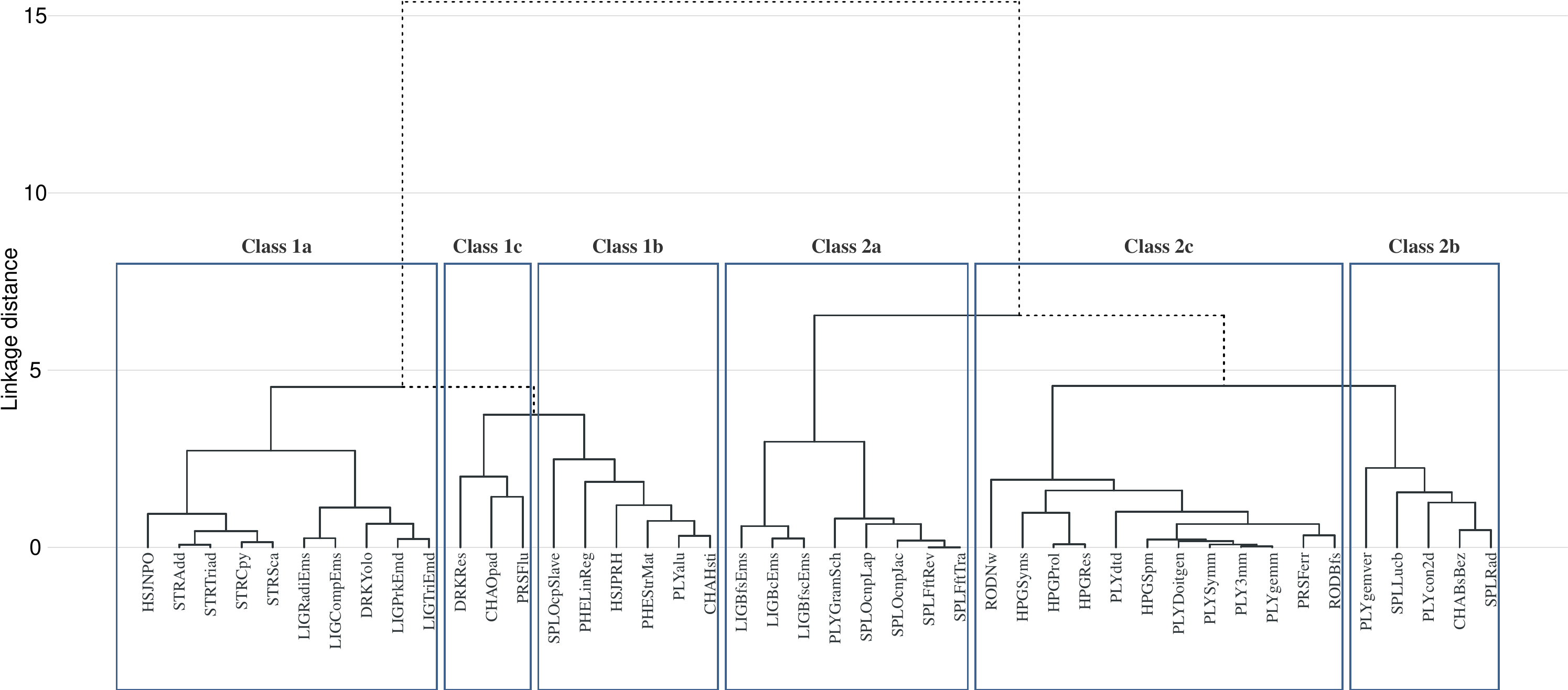}%
    \caption{\geraldorevi{Hierarchical clustering of 44 \gfii{representative} \gfii{functions}.}}
    \label{fig_dendo_validation}
\end{figure*}

\subsection{Benchmark Diversity}
\label{sec_scalability_benchmark_diversity}

\geraldorevi{We perform a hierarchical clustering algorithm with the 44 \gfii{representative} functions we employ in Section~\ref{sec:scalability}.\footnote{\gfii{\gfiii{In Section~\ref{sec_scalability_benchmark_diversity}, w}e use the same 44 representative functions that we use during our bottleneck classification instead of the \gfiii{entire set of} 144 functions in DAMOV, \gfiii{in order} to visualize better the clustering produced by the hierarchical clustering algorithm.}} \gfii{Our goal is to showcase our benchmark suite's diversity and observe whether a clustering algorithm produces a noticeable difference from the application clustering presented Section~\ref{sec:characterization}}. \gfii{The} hierarchical clustering \gfii{algorithm}~\cite{friedman2001elements} \gfii{takes as input a dataset containing features that define each object in the dataset. The algorithm works} by incrementally grouping \gfii{objects} in \gfii{the dataset} \gfii{that are} similar to each other in terms of some distance metric (called \juan{\emph{linkage distance}}), \gfii{which is calculated based on the features' values}. \gfiii{Two} \gfii{objects} with a short linkage distance have more affinity \gfii{to} each other than \gfiii{two} \gfii{objects} with \gfiii{a} large linkage distance. \gfii{To apply the hierarchical clustering algorithm, we create a dataset where each object is one of the 44 representative functions from \bench. We use as features} the same metrics we use for our analysis, i.e., temporal locality, MPKI, LFMR, and AI. We also include the LFMR curve slope to indicate when \gfii{the LFMR} increases, decreases or stays constant when scaling the core count. We use Euclidean distance~\cite{friedman2001elements} to calculate the linkage distance across \gfii{features} in our dataset. We evaluate other linkage distance metrics (\gfiii{such} as Manhattan distance~\cite{friedman2001elements}), and we observe similar clustering results.}

\geraldorevi{Figure~\ref{fig_dendo_validation} shows the dendrogram \juan{that} the hierarchical clustering algorithm produces for our 44 representative functions. \gfii{We indicate in the figure }
the application class each \gfii{function} belongs to, according to our classification. 
We make \gfii{three} observations from the figure.}

\geraldorevi{First, our benchmarks exhibit a wide range of behavior diversity, even among those belonging to the same class. For example, we observe that the functions from Class~1a are divided into two groups, with a linkage distance of 3. Intuitively, \gfii{functions} in the first group (\texttt{HSJNPO}, \texttt{STRAdd}, \texttt{STRCpy}, \texttt{STRSca}, \texttt{STRTriad}) have regular access patterns while \gfii{functions} in the second group (\texttt{DRKYolo}, \texttt{LIGCompEms}, \texttt{LIGPrkEmd}, \texttt{LIGRadiEms}) have  irregular access patterns. We observe a similar clustering in Section~\ref{sec_scalability_class1a}}. 

\geraldorevi{Second, we observe that our application clustering \gfii{(Section~\ref{sec:scalability})} matches the clustering \juan{that the} \gfii{hierarchical clustering} algorithm provides \gfii{(Figure~\ref{fig_dendo_validation})}. From the dendrogram root, we observe that the \juan{right} part of the dendrogram consists of \gfii{functions} with high temporal locality (from \geraldorevi{C}lasses 2a, 2b, and 2c). \juan{Conversely}, the \juan{left} part of the dendrogram consists of \gfii{functions} with low temporal locality (from \geraldorevi{C}lasses 1a, 1b, and 1c). \gfii{The \gfii{functions} in the right and left part of the dendrogram} have a high linkage distance (\gfiii{higher than} 15), which implies that the metrics we use for our clustering are significantly different from each other for these \gfii{functions}. \gfii{Third}, we observe that \gfii{functions} within the same class are clustered into groups with a linkage distance \gfiii{lower than} 5. This grouping matches the six classes of data movement bottlenecks present in \bench. Therefore, we conclude that our methodology can successfully cluster \gfii{functions} into distinct classes, each one representing a different memory bottleneck.}

\geraldorevi{We conclude that (i) \bench provides a heterogeneous \gfiv{and diverse} set of functions \gfiii{to study data movement bottlenecks} and (ii) our memory bottleneck clustering methodology matches the clustering provided by \gfii{a} hierarchical clustering algorithm \gfii{(this section; Figure~\ref{fig_dendo_validation})}. }

%% file: 05_sec_case_studies.tex
\section{Case Studies}

In this section, we \gfii{demonstrate} how our benchmark suite is useful to study open questions \geraldorevi{related to} NDP \geraldorevi{system} designs. We \gfii{provide} four case studies. The first study analyzes the impact of load balance and communication on NDP execution. The second study \juan{assesses} \geraldorevi{the impact of tailored NDP accelerators on our memory bottleneck analysis}. \geraldorevi{T}he third study evaluates the \geraldorevi{effect} \juan{of} \geraldorevi{different core \gfii{designs} on NDP \gfii{system performance}}. The fourth study \juan{analyzes} \geraldorevi{the impact of \gfiv{fine-grained} offloading \gfiv{(i.e., offloading} \gfiii{\gfv{small blocks of} instructions to NDP cores}\gfiv{)} on performance.}

\subsection{Case Study 1: Impact of Load Balance and Inter-Vault Communication on NDP Systems} 
\label{sec:case_study_1}

\begin{figure}[ht]
 \centering
  \includegraphics[width=0.9\linewidth]{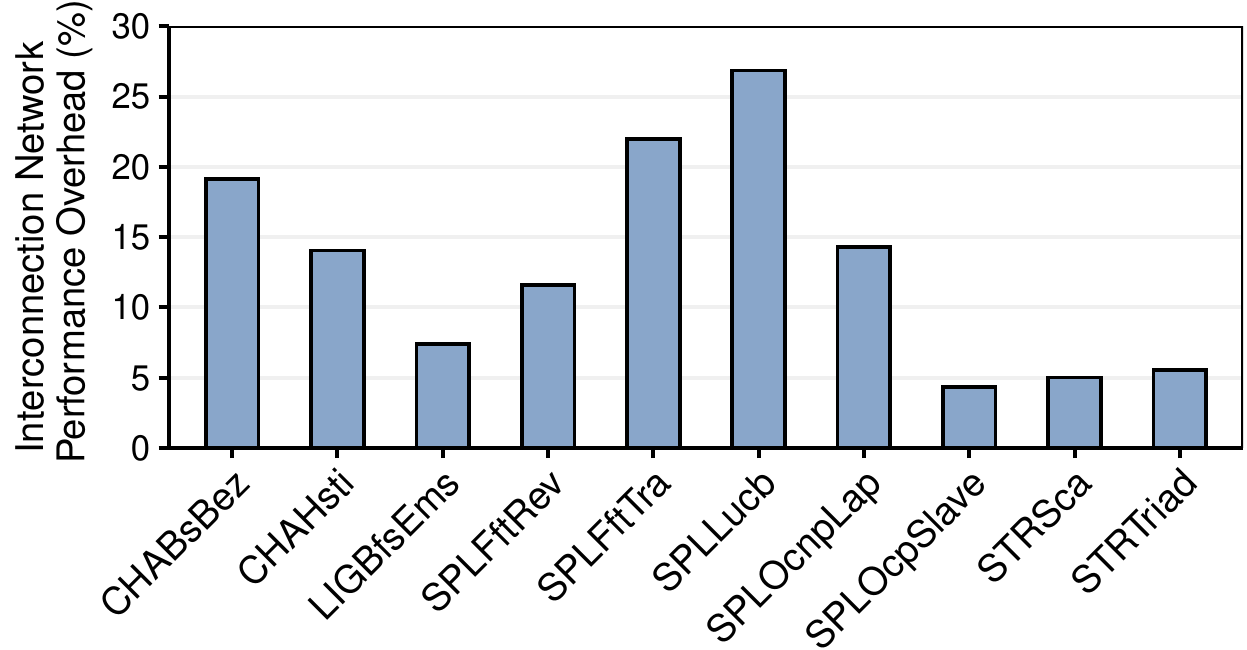}%
  \caption{\geraldorevi{Interconnection network performance overhead \gfii{in our NDP system}.}}
  \label{fig_interconnection_overhead}
\end{figure}

Communication between NDP cores is one of the key challenges for future NDP \geraldorevi{system} designs, especially for NDP architectures based on 3D-stacked memories, where accessing a remote vault incurs extra latency overhead due to network traffic~\cite{ESMC_DATE_2015, ahn2015scalable,syncron}. This case study aims to evaluate the load \geraldorevi{im}balance and inter-vault communication that the NDP cores \geraldorevi{experience} when executing \gfii{functions} from \gfiii{the \bench} benchmark suite. We statically map a \gfii{function} to an NDP core, and  we assume that NDP cores are connected using a 6x6 2D-mesh \geraldorevi{Network-on-Chip (NoC)}, similar to previous works~\gfiii{\cite{drumond2017mondrian,Kim2016,hadidi2018performance,min2019neuralhmc,dai2018graphh}}. \geraldo{Figure~\ref{fig_interconnection_overhead} shows the \geraldorevi{performance} overhead that the \geraldorevi{interconnection network} imposes to NDP cores when running several \gfii{functions} from our benchmark suite.} \geraldorevi{We report performance overheads of \gfii{functions} from different bottleneck classes (i.e., from \geraldorevi{C}lasses 1a, 1b, 2a, and 2b) that experience at least 5\% of performance overhead due to the interconnection network.} We calculate the interconnection \geraldorevi{network performance} overhead by comparing performance \gfii{with} the 2D-mesh versus \gfii{that with} \geraldorevi{an ideal} zero-latency interconnection network. We observe that the interconnection \geraldorevi{network performance} overhead varies across \gfii{functions}, with a minimum overhead of 5\% for \texttt{SPLOcpSlave} and a maximum overhead of 26\% for \texttt{SPLLucb}. 

\begin{figure}[ht]
 \centering
  \includegraphics[width=0.9\linewidth]{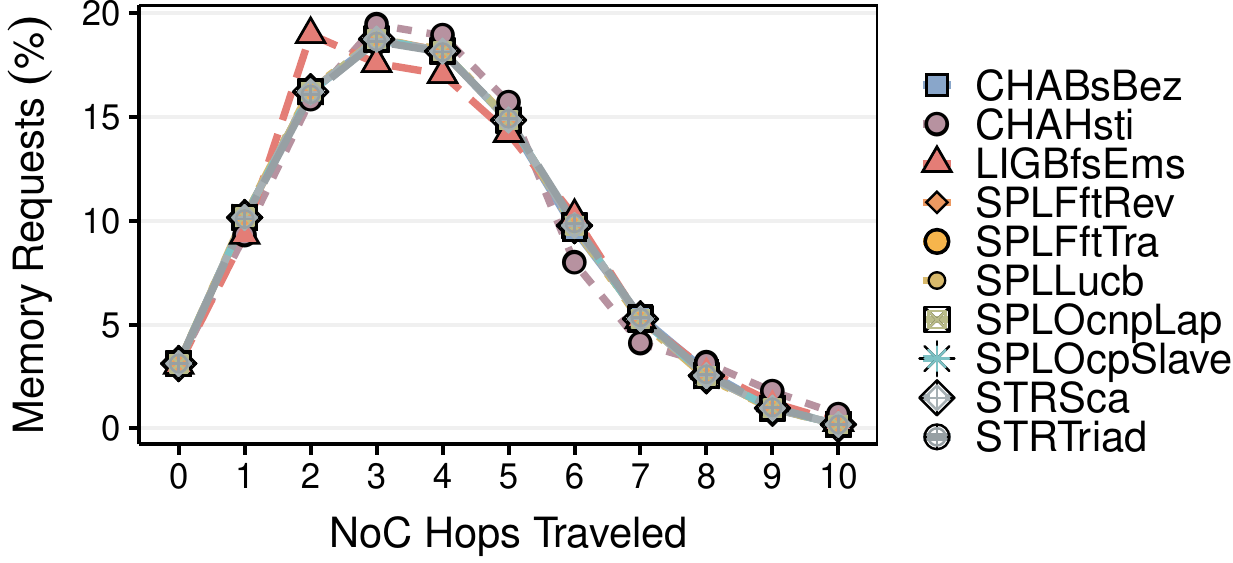}%
  \caption{\geraldorevi{Distribution of NoC hops traveled per memory request.}}
  \label{fig_traffic_distribution}
\end{figure}

\geraldo{We further characterize the \geraldorevi{traffic of memory requests injected into the interconnection network} for these \gfii{functions},  aiming to understand the communication pattern\geraldorevi{s} across NDP cores. Figure~\ref{fig_traffic_distribution} shows \gfii{the distribution of all memory requests (y-axis) in terms of} how many hops \gfii{they} need to travel \geraldorevi{in the NoC between NDP cores \gfiii{(x-axis)} for each \gfii{function}}. We make the following observations.} First, we observe that, on average, 40\% of all memory requests need to travel 3 to 4 hops in the \gls{NoC}, and less than 5\% of all requests are issued to a local vault (0 hops). Even though the \gfii{functions} follow different memory access patterns, they all inject similar network traffic \gfiii{into} the \gls{NoC}.\footnote{\geraldorevi{We use the default HMC data interleaving scheme in our experiments (Table~\ref{table_parameters}).}} Therefore, we conclude that the NDP design can be further optimized by (i) employing \geraldorevi{more intelligent} data mapping and scheduling mechanisms that can efficiently allocate data nearby the NDP core \geraldorevi{that access\juan{es} the data} (\geraldorevi{thereby} reducing inter-vault communication \geraldorevi{and improving data locality}) and (ii) designing interconnection networks that can better fit the traffic \geraldorevi{patterns} that NDP workloads produce. \gfiii{The \bench} \geraldorevi{benchmark suite} can be used to develop \geraldorevi{new ideas as well as evaluate existing ideas in} both directions.

\subsection{\geraldorevi{Case Study 2: Impact of NDP Accelerators on Our Memory Bottleneck Analysis}}
\label{sec:case_study_2}

In \geraldorevi{our second} case study, we \gfii{aim to leverage our memory bottleneck classification to} evaluate the benefits \geraldorevi{an} NDP accelerator provides compared to the same accelerator accessing memory externally. We use the Aladdin accelerator simulator\gfiii{~\cite{shao2014aladdin}} to tailor an accelerator \geraldorevi{for an} application \gfii{function}. \geraldorevi{Aladdin works by estimating the performance of a custom accelerator based on the data-flow graph of the application.} The main difference between \gfii{an} NDP accelerator and \gfii{a} regular accelerator \geraldorevi{(i.e., \gfiii{compute-centric} accelerator)} is that the former \geraldorevi{is placed in the logic layer of a 3D-stacked memory device and thus} can leverage larger memory bandwidth\geraldorevi{,} shorter memory access latency\geraldorevi{, and lower memory access energy, compared to the \gfiii{compute-centric} accelerator \gfii{that is exemplary of existing compute-centric accelerator designs}}.

\gfii{To evaluate the benefits of NDP accelerators, we select \gfii{three} functions from our benchmark suite for this case study: \texttt{DRKYolo} (from Class~1a), \texttt{PLYalu} (from Class~1b), and \texttt{PLY3mm} (from Class~2c). \gfii{We select these functions and memory bottleneck classes \gfii{because we expect them}  \juan{\gfii{to} benefit the most (\gfii{or to show no benefit}) from the near-memory placement of an accelerator.} \gfii{According to our memory bottleneck analysis, we expect that the functions we select to } (i) benefit from NDP due to its high DRAM bandwidth (Class~1a), (ii) benefit from NDP due to its shorter DRAM access latency (Class~1b), or (iii) do \emph{not} benefit from NDP in any way \gfiii{ (Class~2c)}.}}

\begin{figure}[h]
     \centering
     \includegraphics[width=\linewidth]{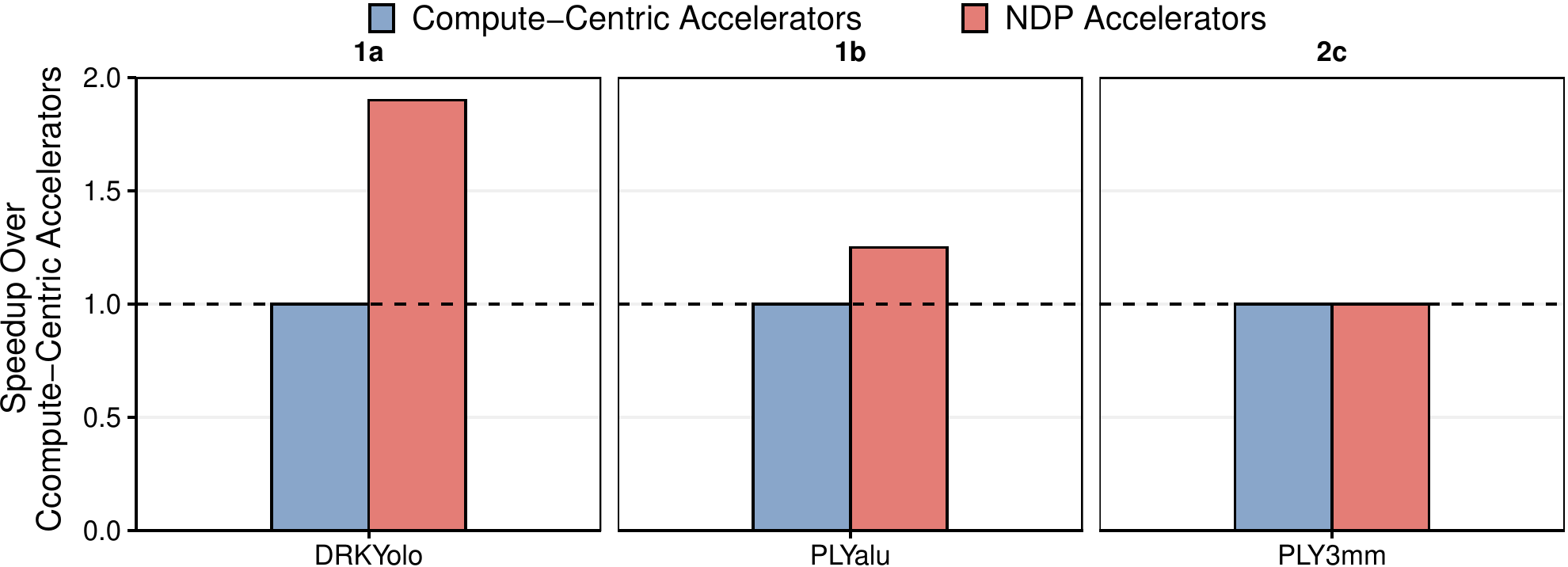}%
     \caption{Speedup \gfii{of the NDP Accelerators over the \gfiii{Compute-Centric} Accelerators for three functions from Classes 1a, 1b, and \gfiii{2}c.}}
     \label{fig:accelerator}
 \end{figure}

Figure~\ref{fig:accelerator} shows the speedup \juan{that} \geraldorevi{\gfii{the} NDP accelerator provides for the different \gfii{functions} compared to \gfii{the} \gfiii{compute-centric} accelerator}. We make \gfii{four} observations. First, \gfiii{as expected based on our classification,} the NDP accelerator provides performance benefits \geraldorevi{compared to the \gfiii{compute-centric} accelerator} for \gfii{functions} in \geraldorevi{C}lasses 1a and 1b. It does not provide performance improvement for the \gfii{function} in Class 2c. \gfii{Second,} the NDP accelerator for \texttt{DRKYolo} shows the largest performance benefits (1.9$\times$ \geraldorevi{performance improvement compared to the \gfiii{compute-centric} accelerator}). Since this \gfii{function} is \gfii{DRAM} bandwidth-bound \geraldorevi{(Class~1a, Section~\ref{sec_scalability_class1a})}, the \geraldorevi{NDP} accelerator can leverage the larger memory bandwidth available \geraldorevi{in the logic layer of the 3D-stacked memory device}. \gfii{Third,} we observe that the NDP accelerator also provides speedup (1.25$\times$) for the \texttt{PLYalu} \gfii{function} \geraldorevi{compared to the \gfiii{compute-centric}  accelerator,} since \geraldorevi{the NDP accelerator provides shorter memory access latency to the \gfii{function}, which is latency-bound (Class~1b, Section~\ref{sec_scalability_class1b}).} \gfii{Fourth,} the NDP accelerator does not provide performance improvement for the \texttt{PLY3mm} \gfii{function} since this \gfii{function} is compute-bound \geraldorevi{(Class~2\gfiii{c}, Section~\ref{sec_scalability_class2c})}. 

\gfiii{In conclusion, our observations for the performance of NDP accelerators are in line with the characteristics of the three memory bottleneck \gfiv{classes} we evaluate in this case study.} Therefore, our memory bottleneck classification can be applied to study other types of system configurations, e.g., the accelerators used in this section. \gfiii{However, since NDP accelerators are often employed under restricted area and power constraints (e.g., limited area available in the logic layer of a 3D-stacked memory~\cite{lazypim, boroumand2019conda}), the core model of the compute-centric and NDP accelerators cannot always be the same}. We leave a thorough analysis that takes area and power constraints \gfiii{in the study of NDP accelerators} into consideration \gfiv{for future research}.

\subsection{\geraldorevi{Case Study 3: Impact of Different Core Models on NDP Architectures}} 
\label{sec:case_study_3}

This case study aims to \gfii{analyze} when a workload can benefit from different core models and numbers of cores while respecting the area and power envelope of the logic layer of a 3D-stacked memory. \gfii{Many prior works employ 3D-stacked memories as the substrate to implement NDP architectures~\gfv{\cite{ahn2015scalable, nai2017graphpim, boroumand2018google, lazypim, top-pim, gao2016hrl, kim2018grim, drumond2017mondrian, RVU, NIM, PEI, gao2017tetris, boroumand2019conda,Kim2016, hsieh2016transparent, cali2020genasm, NDC_ISPASS_2014, farmahini2015nda,pattnaik2016scheduling, hsieh2016accelerating,fernandez2020natsa,syncron,boroumand2021polynesia,boroumand2021mitigating,amiraliphd, IBM_ActiveCube,akin2015data,jang2019charon,nai2015instruction,alves2016large,xie2017processing,lenjani2020fulcrum,kersey2017lightweight,huang2019active,hassan2015near,guo20143d,liu20173d,de2018design,LiM_3D_FFT_MM,Sparse_MM_LiM,glova2019near}}. However, 3D-stacked memories impose severe area and power restrictions on NDP architectures. For example, the area and power budget of the logic layer of a single HMC vault 
\juang{are} 4.4~$mm^2$ and 312~$mW$,
respectively~\cite{lazypim, boroumand2018google}.} 

\gfii{In the case study, we perform an iso-area and iso-power performance evaluation of three functions from our benchmark suite. We configure the host CPU system and the NDP system to guarantee an iso-area and iso-power evaluation, considering the area and power budget for a 32-vault HMC device~\cite{lazypim, boroumand2018google}. \gfiii{W}e use four out-of-order cores with a deep cache hierarchy for the host system configuration and} two \gfiii{different} NDP configurations\gfiii{:} (1) \gfiii{one} using six out-of-order \gfiii{NDP} cores (\textit{NDP+out-of-order}) and (2) using 128 in-order \gfiii{NDP} cores (\textit{NDP+in-order}), without a deep cache hierarchy.  We \geraldorevi{choose} \gfii{functions} from \geraldorevi{C}lasses 1a, 1b, \juan{and 2b} for this case study since the major \gfiii{effects} distinct microarchitectures have on the memory system are: (a) how much DRAM bandwidth they can \juan{sustain,} and (b) how much DRAM latency they can hide. \geraldorevi{\juan{Classes 1a, 1b, and 2b are the most} affected by memory bandwidth and access latency (as shown in Section~\ref{sec:characterization})}. We \geraldorevi{choose} two \gfiii{representative} \gfii{functions} from each of these classes. 

\begin{figure}[ht]
  \vspace{-5pt}
     \centering
     \includegraphics[width=\linewidth]{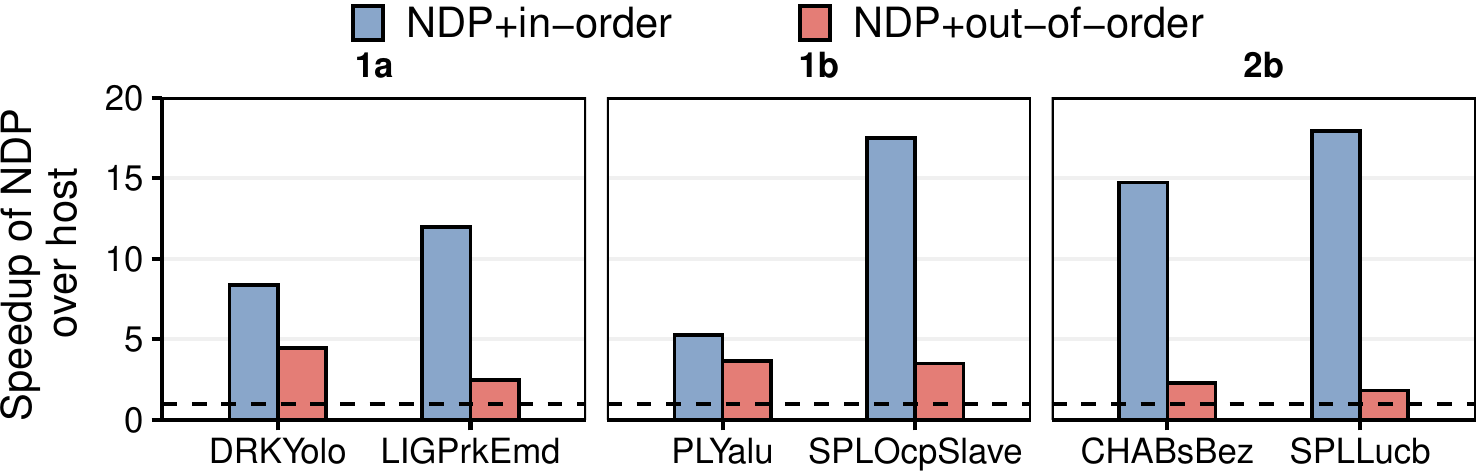}%
    \caption{\geraldorevi{Speedup of NDP architectures \gfiii{over} 4 out-of-order host \gfiii{CPU} cores for two NDP configurations: using \geraldorevi{128 in-order \gfiii{NDP cores} (\emph{NDP+in-order}) and 6 out-of-order \gfiii{NDP cores} (\emph{NDP+out-of-order}) for \gfiii{representative functions} from \geraldorevi{C}lasses 1a, 1b, and \juan{2b.}}}}
     \label{fig:inorder}
 \end{figure}

\geraldorevi{Figure~\ref{fig:inorder} shows the speedup provided by \gfiii{the two} NDP system configurations compared \juan{to} the baseline host system. We make two observations. \gfii{First, in all cases, the \emph{NDP+in-order} system provides higher speedup than the \emph{NDP+out-of-order} system, both compared to the host system. On average across all six functions, the \emph{NDP+in-order} system provides 4$\times$ the speedup of the \emph{NDP+out-of-order} system. The larger speedup the \emph{NDP+in-order} system provides is due to the high number of NDP cores in the \emph{NDP+in-order} system\gfiv{. W}e can fit 128 in-order cores in the logic layer of the 3D-stacked memory \gfiv{as opposed to} \gfiii{only} six out-of-order cores in the same area/power budget.  Second, we observe that the speedup the  \emph{NDP+in-order} system provides compared to the \emph{NDP+out-of-order} system does not scale with the number of cores. For example, the \emph{NDP+in-order} system provides \emph{only} 2$\times$ the performance of the \emph{NDP+out-of-order} system for \texttt{DRKYolo} and \texttt{PLYalu}, even though the \emph{NDP+in-order} system has 21$\times$ the number of NDP cores of the \emph{NDP+out-of-order} system. This implies that even though the \gfii{functions} benefit from \gfii{a large number of NDP cores available in the \emph{NDP+in-order} system}, static instruction scheduling limits performance on the \emph{NDP+in-order} system.}} 

\geraldo{We believe, and \juang{our previous observations suggest,} that a\gfii{n} \gfii{efficient} NDP architecture can be achieved by leveraging \gfii{mechanisms} that can exploit \juang{both} \gfii{dynamic instruction scheduling} and \gfii{many-core design while fitting in the area and power budget \gfiii{of} 3D-stacked memories.} For example, \gfiii{past works~\gfv{\cite{padmanabha2017mirage, fallin2014heterogeneous,villavieja2014yoga,suleman2012morphcore,ipek2007core,suleman2010data,petrica2013flicker,kim2007composable,lukefahr2012composite,mutlu2006efficient,mutlu2005techniques,mutlu2003runahead,suleman2009accelerating,joao2012bottleneck,chaudhry2009simultaneous, hashemi2016continuous, chou2004microarchitecture,joao2013utility,tarjan2008federation,alipour2020delay,kumar2019freeway,alipour2019fiforder,kumar2003single,kumar2004single}} propose techniques that enable the benefits of simple and complex cores at the same time\gfv{, via heterogeneous \gfvi{or adaptive} architectures}. These \gfiv{ideas} can be examined to \gfiv{enable} better core \gfv{and system} design\gfv{s} for NDP systems, and \bench can facilitate their proper \gfv{design, exploration, and} evaluation.} 
}

\subsection{\geraldorevi{Case Study 4: Impact of \gfiv{Fine-Grained} \gfiii{Offloading to NDP} on Performance}}
\label{sec:case_study_4}

\gfv{Several p}rior works on \gls{NDP} \gfv{(e.g., \cite{PEI,nai2017graphpim,nai2015instruction,hadidi2017cairo,azarkhish2016design,hajinazarsimdram, seshadri2017ambit, seshadri2013rowclone,ahmed2019compiler,baskaran2020decentralized,li2019pims})} propose to identify and offload to the \geraldorevi{NDP system} simple primitives (e.g., \geraldorevi{instructions}, atomic operations). \geraldorevi{We refer to this \gls{NDP} offloading scheme as a \emph{fine-grained \gls{NDP} offloading}, in contrast to a \emph{coarse-grain\geraldorevi{ed} \gls{NDP} offloading scheme} that offloads whole functions and applications to \gls{NDP} systems}. \gfiii{A fine-grained \gls{NDP} offloading scheme provides two main benefits compared to a coarse-grained \gls{NDP} offloading scheme. First, a fine-grained \gls{NDP} offloading scheme allows for a reduction in} the complexity of the processing elements used as \gls{NDP} logic\geraldorevi{, since the \gls{NDP} logic can consist of simple processing elements (e.g., arithmetic units, fixed function units) instead of \gfiii{entire} in-order or out-of-order core\gfiii{s often utilized when employing a coarse-grained \gls{NDP} offloading scheme}. \gfiii{Second, a fine-grained \gls{NDP} offloading scheme} can help developing simple} coherence mechanism needed to allow shared host and \gls{NDP} execution~\cite{PEI}. However, identifying arbitrary \gls{NDP} \gfiii{instructions} can be a daunting task since there is no comprehensive methodology that indicates \geraldorevi{what types} of instructions are good offloading candidate\geraldorevi{s}. 

\begin{figure}[ht]
 \centering
  \includegraphics[width=\linewidth]{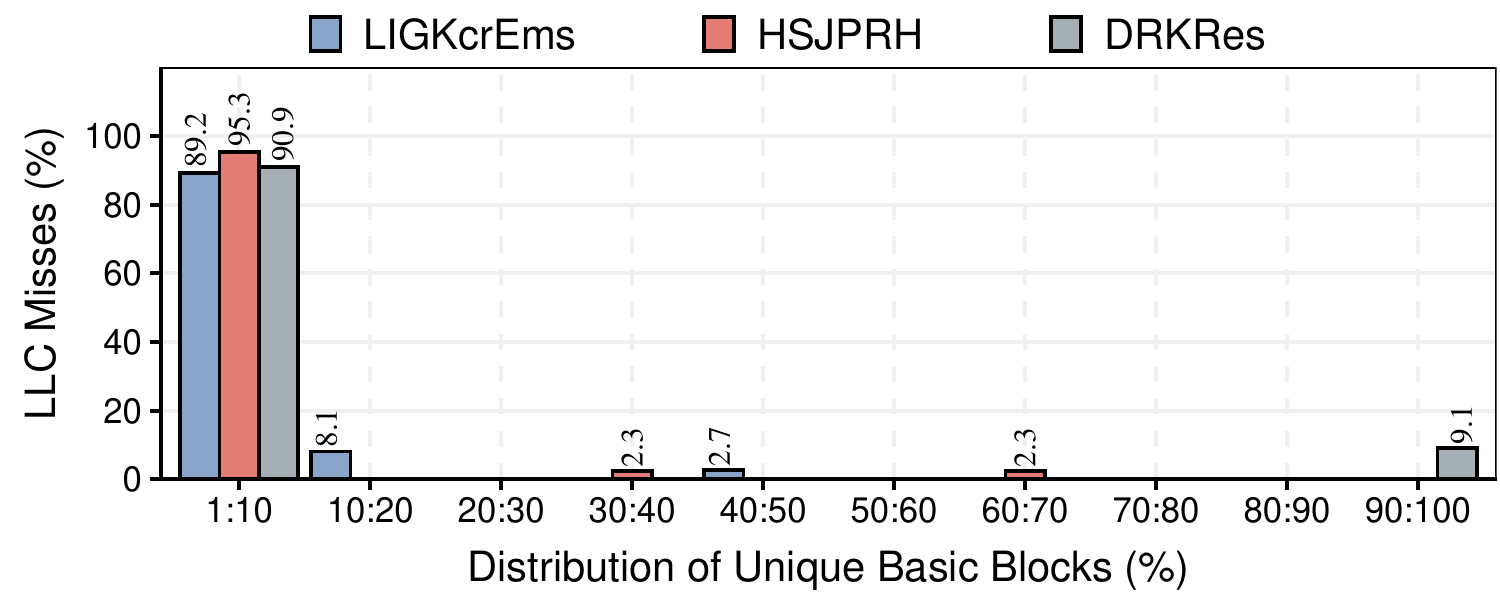}%
  \caption{\geraldorevi{Distribution of unique basic blocks (x-axis) and the percentage of last-level cache misses they produce (y-axis) for three representative functions from Classes 1a (\texttt{LIGKcrEms}), 1b (\texttt{HSJPRH}), and 1c (\texttt{DRKRes}).}}
  \label{fig_bbc_dist}
\end{figure}

\begin{figure}[ht]
  \centering
  \includegraphics[width=0.9\linewidth]{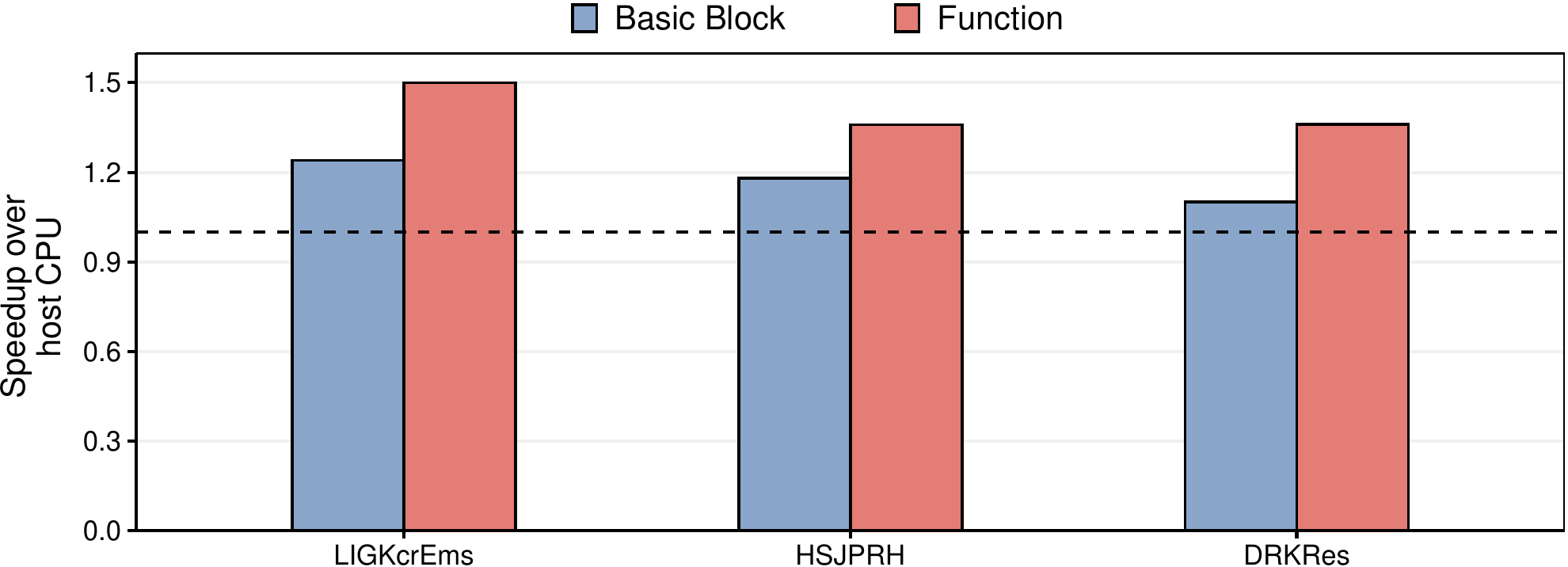}%
  \caption{\geraldorevi{Speedup of offloading to NDP the \emph{hottest} basic block in each function \gfiii{versus} the entire function.}}
  \label{fig_bbc_speedup}
 \end{figure}

As the first step in this direction, we exploit the key insight provided by~\cite{ayers2020classifying, hashemi2016accelerating} to identify potential regions of code that can be candidates for fine-grain\geraldorevi{ed} \gls{NDP} offloading. \cite{ayers2020classifying, collins2001speculative, hashemi2016accelerating} show that few instructions are responsible for generating most of the cache misses during program execution in memory-intensive applications. Thus, these instructions are naturally good candidates for fine-grain\geraldorevi{ed} \gls{NDP} offloading. \geraldorevi{Figure~\ref{fig_bbc_dist} shows the distribution of unique basic blocks (x-axis) and the percentage of last-level cache misses (y-axis) the basic block produces for three representative functions from our benchmark suite. We select functions from Classes 1a (\texttt{LIGKcrEms}), 1b (\texttt{HSJPRH}), and 1c (\texttt{DRKRes}) since \gfii{functions} in these classes have higher L3 MPKI than \gfii{functions} in Classes 2a, 2b, and 2c. We observe from the figure that 1\% to 10\% of the basic blocks in each function are responsible for up to 95.3\% of the LLC misses. \geraldorevi{We call these basic blocks the} \emph{hot\geraldorevi{test}} basic block\gfiii{s}.}\footnote{We observe for \geraldorevi{the} 44 \gfii{functions} \geraldorevi{we evaluate in Section~\ref{sec:characterization}} that in many cases (for 65\% of the evaluated workloads), a single basic block is responsible for 90\% to 100\% of the \gls{LLC} misses during the \gfii{function}'s execution.} \geraldorevi{W}e investigate the data-flow of each basic block and observe that \geraldorevi{these basic blocks} often execute simple read-modify-write operations, with few arithmetic operations. Therefore, we believe that such basic blocks are good candidates for fine-grain\juan{ed} offloading. \geraldorevi{Figure~\ref{fig_bbc_speedup} shows the speedup obtained by offloading \gfii{(i)} the hottest basic block we identified for the three representative functions \gfii{and (ii) the entire function} to the \gls{NDP} system, compared to the host system.} Our initial evaluations show that offloading the hot\geraldorevi{test} basic block \geraldorevi{\gfii{of} each function} to \geraldorevi{the} \gls{NDP} \geraldorevi{system} can provide up to 1.25$\times$ speedup compared to the host \gfiii{CPU}\gfii{, which is \gfiii{half of} the 1.5$\times$ speedup achieved when offloading the entire function}. \geraldorevi{Therefore, we believe that \gfiii{methodically} identifying simple NDP \gfiii{instructions} can be a promising research direction for future NDP system designs, which \gfii{our} DAMOV \gfii{Benchmark Suite} can help with.} 

\section{Key Takeaways}

\gfiii{W}e summarize the key takeaways \geraldorevi{from} our \geraldorevi{extensive characterization of 144 functions using our new three-step methodology to identify data movement bottlenecks.} We also highlight when NDP is a good architectural choice to mitigate a particular memory bottleneck. 

\begin{figure*}[h]
  \centering
  \includegraphics[width=0.90\textwidth]{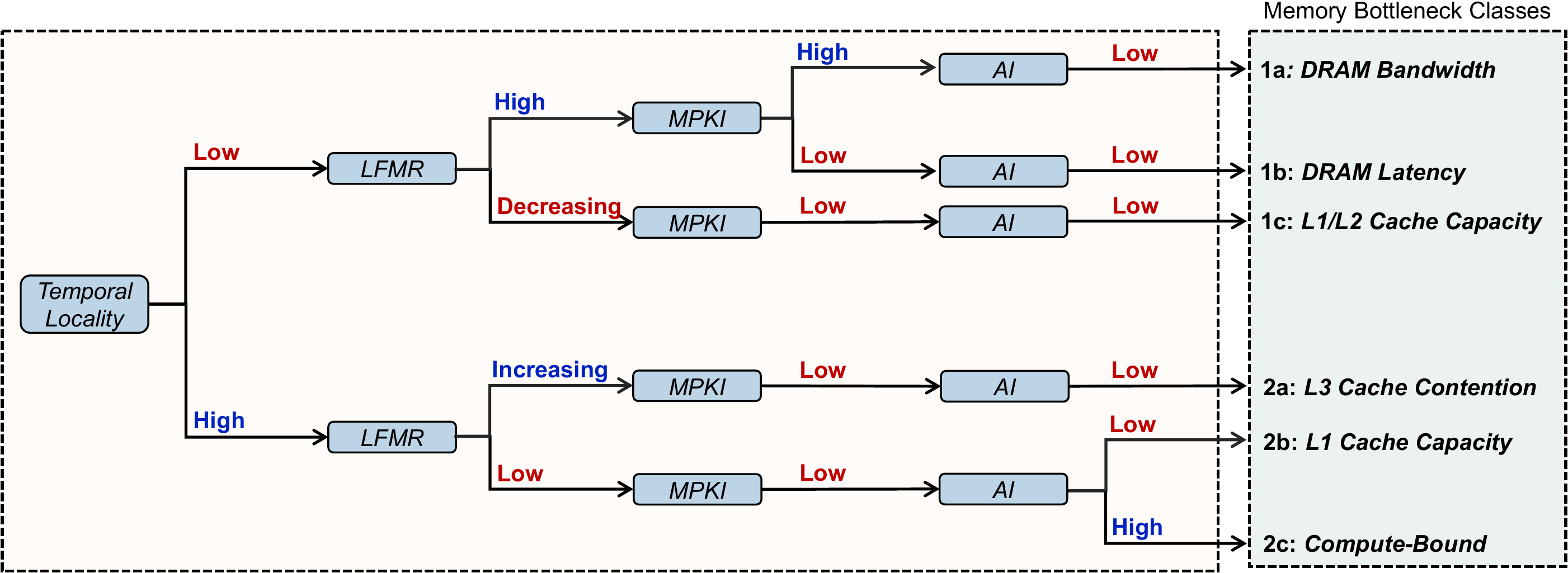}
  \caption{\gfii{Summary of our memory bottleneck classification}.} 
  \label{fig:decision}
\end{figure*}

\gfii{Figure~\ref{fig:decision} pictorially represents the key takeaways we obtain from our memory bottleneck classification. Based on four key metrics, we classify workloads into six classes of memory bottlenecks. We provide the following key takeaways:}

\begin{enumerate}
\item Applications with low temporal locality, high LFMR, high MPKI, \gfii{and low AI} are \geraldorevi{\gfii{DRAM}} \emph{bandwidth-bound} \geraldorevi{(Class~1a, Section~\ref{sec_scalability_class1a})}. They are bottlenecked by the limited off-chip memory bandwidth as they exert high pressure on main memory. \gfii{We make three observations for Class~1a applications. First,} \geraldorevi{these} applications do benefit from prefetching since they display \geraldorevi{a low} degree of spatial locality.  \gfii{Second,} these applications highly benefit from NDP architectures because they take advantage of the high memory bandwidth available within the memory device. \geraldorevi{\gfii{Third}, NDP architectures significantly improve energy for these applications since they eliminate the off-chip I/O traffic between \juan{the} CPU and \juan{the} main memory.}

\item Applications with low temporal locality, high LFMR, low MPKI, and low AI are \geraldorevi{DRAM} \emph{latency-bound} \geraldorevi{(Class~1b, Section~\ref{sec_scalability_class1b})}. \gfii{We make three observations for Class~1b applications. First,} these applications do not significantly benefit from prefetching since infrequent memory requests make it difficult for the prefetcher to train \geraldorevi{successfully} on an access pattern. \gfii{Second, t}hese applications benefit from NDP architectures since they take advantage of NDP's lower memory access latency \geraldorevi{and the elimination of deep \gfiii{L2/L3} cache hierarchies\gfiii{, which fail to capture data locality} for these workloads}. \geraldorevi{\gfii{Third}, NDP architectures significantly improve energy for these applications since they eliminate costly (and unnecessary) L3 cache look-ups and the off-chip I/O traffic between \juan{the} CPU and \juan{the} main memory.}

\item Applications with low temporal locality, decreasing LFMR with core count, low MPKI, and low AI are \emph{bottlenecked by the available \gfii{L1/L2} cache capacity} \geraldorevi{(Class~1c, Section~\ref{sec_scalability_class1c})}. \gfii{We make three observations for Class 1c applications. First, these} applications are \gfii{DRAM} latency-bound at low core counts, thus taking advantage of NDP architectures\geraldorevi{, both \gfiii{in terms of} performance improvement and energy \gfiii{reduction}}. \gfii{Second,} NDP's benefits reduce when core count \geraldorevi{becomes larger}, which consequently allows the working set\juan{s} of such applications \gfii{to} fit inside the cache hierarchy at high core counts. \juan{\gfii{Third}, NDP architectures can be a good design choice for such workloads in systems with limited area budget \gfiii{since NDP architectures do not require large L2/L3 caches to \gfiv{outperform or perform similarly to} the host CPU \gfiv{(in terms of both \gfv{system throughput} and energy)} for these workloads}.}

\item Applications with high temporal locality, increasing LFMR with core count, low MPKI, and low AI  are \emph{bottlenecked by \gfii{L3} cache contention} \geraldorevi{(Class~2a, Section~\ref{sec_scalability_class2a})}. \gfii{We make three observations for Class~2a applications. First, these} applications benefit from a deep cache hierarchy and do not take advantage of NDP architectures at low core counts. \gfii{Second}, the number of cache conflicts increases when the number of cores in the system \geraldorevi{increases}, \geraldorevi{leading to more} pressure on main memory. We observe that NDP can effectively mitigate such cache contention for these applications without incurring the high area and energy overheads of providing additional cache capacity in the host. \geraldorevi{\gfii{Third,} NDP can improve energy for these workloads at high core counts, since it eliminates the costly data movement between the last-level cache and \juan{the} main memory.}
 
\item Applications with high temporal locality, low LFMR, low MPKI, and low AI are \gfii{bottlenecked by} \emph{L1 cache capacity} (Class~2b, Section~\ref{sec_scalability_class2b}). \gfii{We make two observation for Class~2b applications. First,} NDP can provide similar performance \geraldorevi{and energy consumption than the host system by leveraging lower memory access latency and avoiding off-chip energy consumption for these applications}. \geraldorevi{\gfii{Second}, NDP can be used to reduce the overall SRAM area \gfiii{(by eliminating L2/L3 cache\gfiv{s})} in the system without a performance or energy penalty}. 

\item Applications with high temporal locality, low LFMR, low MPKI, and high AI are \emph{compute-bound} (Class~2c, Section~\ref{sec_scalability_class2c}). \gfii{We make three observation\gfiii{s} for Class~2c applications. First, these applications} suffer performance and energy penalties due to the lack of a deep \gfiii{L2/L3} cache hierarchy when executed on the NDP architecture. \geraldorevi{Second, these applications highly benefit from prefetching due to their high temporal \gfiii{and spatial} locality.} \geraldorevi{\gfii{Third}, these applications are not good candidate\gfiii{s} to execute on NDP architectures.}

\end{enumerate}

\subsection{Shaping Future Research with \bench}

\gfii{A key contribution of our work is \bench, the first benchmark suite for \gfiii{main memory} data movement studies. \bench is the collection of 144 functions from 74 different applications, belonging to 16 different benchmark suites or frameworks, classified into six different classes of data movement bottlenecks.}

\geraldorevi{We believe that \bench can be used to explore a wide range of research directions on the study of data movement bottlenecks,  appropriate mitigation mechanisms, \gfii{and open research topics on NDP architectures.}
\gfii{We highlight \bench's usability \gfiii{and potential benefits} with four \gfiii{brief} case studies, which we summarize below:}
}

\begin{itemize}
\item \gfii{In the first case study (Section~\ref{sec:case_study_1}), we use \bench to evaluate the interconnection network overheads that NDP cores placed in different vaults of a 3D-stacked memory suffer \gfiii{from}. We observe that a large portion of the memory requests \gfiii{an} NDP core issues go to 
\juang{remote} vaults, which increases the memory access latency for the NDP core. We believe that \bench can be employed to study better data mapping \gfiii{techniques} and interconnection network \gfiii{designs} that \gfiii{aim} to minimize (i) the number of remote memory accesses the NDP cores execute and (ii) the interconnection \gfiii{network} latency overheads.}
    
\item \gfii{In the second case study (Section~\ref{sec:case_study_2}), we evaluate the benefits that NDP accelerators can provide for three applications \juang{from} our benchmark suite. We compare the performance improvements an NDP accelerator provides against the \gfiii{compute-centric version of the} same accelerator. We observe that the NDP accelerator provides significant performance benefits compared to the \gfiii{compute-centric} accelerator for applications in Classes 1a and 1b. At the same time, it does not improve performance for an application in Class 2c. We believe that \bench can \gfiii{aid the design of} NDP accelerators that target different memory bottlenecks in the system.} 
    
\item \gfii{In the third case study (Section~\ref{sec:case_study_3}), we perform an iso-area/-power performance evaluation to compare NDP systems using in-order and out-of-order cores. We observe that the in-order cores' performance benefits for some applications are limited by the cores' static instruction scheduling mechanism. We believe that better NDP systems can be built by leveraging techniques that enable dynamic instruction scheduling without incurring the large area and power overheads of out-of-order cores. \bench can help in the analysis and development of such NDP architectures. }
    
\item \gfii{In the fourth case study (Section~\ref{sec:case_study_4}), we evaluate the benefits of offloading small portions of code (i.e., a basic block) to NDP, which simplifies the design of NDP systems. We observe that for many applications, a small percentage of basic blocks is responsible for most of the last-level cache misses. By offloading these basic blocks to an NDP core, we observe a performance improvement of up to 1.25$\times$. We believe that \bench can be used to identify simple NDP \gfiii{instructions} that enable building efficient NDP systems in the future.} 
     
\end{itemize}

%% file: 06_sec_related_work.tex
\section{Related Work}

\geraldorevi{To our knowledge, this is the first work that methodically characterizes data movement bottlenecks and evaluates the benefits of different data movement mitigation mechanisms, with a focus on Near-Data Processing (NDP), for a broad range of applications. This is also the first work that provides an extensive \gfii{open-source} benchmark suite, with a diverse range of real world applications, tailored to stress different \gfii{memory-related} data movement bottlenecks in a system.}

Many past works investigate how to reduce data movement cost using a range of different compute-centric (e.g., \geraldorevi{prefetchers~\gfvii{\cite{
gonzalez1997speculative,
chen1995effective, 
bera2019dspatch, 
bakhshalipour2018domino, 
fu1992stride, 
ishii2009access, 
hashemi2018learning,
orosa2018avpp,
jog2013orchestrated,
lee2011prefetch,
ebrahimi2009techniques,
srinath2007feedback,
austin1995zero,
ceze2006cava,
kadjo2014b,
kirman2005checkpointed,
cooksey2002stateless,
joseph1997prefetching,
ebrahimi2011prefetch,
ebrahimi2009coordinated,
lee2009improving,
ChangJooLee,charney1995,charneyphd}}, speculative execution~\gfiii{\cite{gonzalez1997speculative, mutlu2003runahead, hashemi2016continuous, mutlu2005techniques,mutlu2006efficient, mutlu2005using}},} value-prediction~\gfiii{\cite{orosa2018avpp, yazdanbakhsh2016rfvp, eickemeyer1993load, endo2017interactions, gonzalez1997speculative, lipasti1996exceeding,lipasti1996value,calder1999selective,wang1997highly,gabbay1996speculative,burtscher1999exploring,fu1998value,goeman2001differential,nakra1999global,sato2002low,sazeides1997predictability,tuck2005multithreaded,tullsen1999storageless,tune2002quantifying}}, data compression~\gfvii{\cite{pekhimenko2012base, 
dusser2009zero, 
yang2000frequent,
alameldeen2004adaptive, 
zhang2000frequent,
pekhimenko2016case,pekhimenko2015toggle, pekhimenkoenergy, pekhimenko2015exploiting, pekhimenko2013linearly,chen2009c,hallnor2005unified,hammerstrom1977information,islam2009zero,arelakis2014sc,ekman2005robust,vijaykumar2015case,gaur2016base}}, approximate computing~\gfiii{\cite{yazdanbakhsh2016rfvp,koppula2019eden,miguel2015doppelganger,oliveira2018employing}}) and memory-centric techniques~\gfvii{\cite{vijaykumar2018case, yazdanbakhsh2016mitigating, joao2012bottleneck, tsai2018rethinking, sembrant2014direct, PEI,lazypim,MEMSYS_MVX,top-pim,acm,New_PIM_2013,kim2017heterogeneous,pugsley2014comparing,boroumand2018google,azarkhish2016memory,nai2017graphpim,gao2016hrl,de2018design,babarinsa2015jafar,tsai2017jenga,tsai2019compress,vijaykumar2018locality,mutlu2020intelligent,mutlu2020intelligentdate}}. \gfii{These works} evaluate the impact of data movement in different systems\geraldorevi{, including} mobile systems~\gfiii{\cite{boroumand2018google, pandiyan2014quantifying, narancic2014evaluating, yan2015characterizing,shingari2015characterization}}, data centers~\gfiii{\cite{kanev_isca2015, ferdman2012clearing, kozyrakis2010server,yasin2014deep, hashemi2018learning, gupta2019architectural,gan2018architectural,ayers2018memory}}, accelerators\gfiii{-}\geraldorevi{based systems}~\gfiii{\cite{boroumand2018google, gupta2019architectural, singh2019napel, murphy2001characterization, cali2020genasm, kim2018grim,alser2020accelerating}}, and \geraldorevi{desktop} computers~\cite{limaye2018workload, bienia2008parsec, corda2019memory}. They use \gfii{very} different profiling frameworks and methodologies to identify the root cause of data movement for a \geraldorevi{small} set of applications. Thus, it is not possible to generalize \gfii{prior works'} findings \gfii{to} other applications \gfii{than the limited set they analyze}.

\geraldorevi{We highlight two of these prior works, \cite{murphy2001characterization} and \cite{boroumand2018google}, since they also} focus on characterizing applications \gfii{for} \gls{NDP} architectures. In \cite{murphy2001characterization}, the authors provide the first work that characterizes workloads for \gls{NDP}. They analyze \geraldorevi{five} applications \geraldorevi{(FFT, ray tracing, method of moments, image understanding, data management)}. The \gls{NDP} organization \cite{murphy2001characterization} targets is similar to \cite{IRAM_Micro_1997}, where \geraldorevi{vector processing compute} units \geraldorevi{are} integrate\geraldorevi{d into} the DDRx memor\geraldorevi{y modules}. Even though \cite{murphy2001characterization} has a similar goal to our work, it understandably does not provide insights \gfii{into} modern data-intensive applications and \gls{NDP} architectures \gfii{as it dates from 2001}. \gfiii{Also, \cite{murphy2001characterization} focus\gfiv{es its} analysis only on a few workloads, whereas we conduct a broader workload analysis starting from 345 applications.} Therefore, a new\gfiii{, more comprehensive and rigorous} analysis methodology of data movement bottlenecks \gfii{in modern workloads and modern NDP systems} is necessary. A more recent work investigates the memory bottlenecks in \gfii{widely-used} \geraldorevi{consumer} workloads \geraldorevi{from Google} and how \gls{NDP} can mitigate \geraldorevi{such bottlenecks} \cite{boroumand2018google}. \geraldorevi{This work} \geraldorevi{focuses its} analysis on a \geraldorevi{small} number of consumer workloads. Our work presents a comprehensive analysis of a much broader set of applications \geraldorevi{(\gfiii{345} different applications, and a total of \gfiii{77K} application functions)}, which allows us to provide \gfii{a general methodology, a comprehensive workload suite, and} general takeaways and guidelines for future \gls{NDP} research. \geraldorevi{\gfii{With our comprehensive analysis, this work is the first to} develop a rigorous methodology to classify applications into six groups, which have different characteristics with respect to how they benefit from NDP \gfii{systems} \gfiii{as well as other data movement bottleneck mitigation techniques}.}

%% file: 07_sec_conclusion.tex
\section{Conclusion}

\geraldorevi{This paper introduces the first \gfii{rigorous} methodology to characterize \gfii{memory-related} data movement bottlenecks \juan{in modern workloads} and the first data movement benchmark suite, called \bench. We perform the first large-scale characterization of applications to develop a \gfii{three-step workload characterization methodology that introduces and evaluates four key metrics} 
\gfii{to identify} the sources of data movement bottlenecks \gfiii{in real applications}. We \gfii{use our new methodology to} classify \gfii{the primary sources of memory bottlenecks of a broad range of applications into six different classes of memory bottlenecks.} We highlight the benefits of our benchmark suite with four case studies, which showcase how \gfii{representative} \gfii{workloads} in \bench can be used to explore open-research topics on NDP systems \gfii{and reach architectural as well as workload-level \gfiii{insights and} conclusions}. We open-source our benchmark suite and our bottleneck analysis toolchain~\cite{damov}\gfii{. \gfiii{We hope that our work enables} further studies \gfii{and research} \gfiii{on} \gfii{hardware and software solutions for} data movement \gfii{bottlenecks}\gfiv{, including near-data processing}.}}

%% file: 08_appendixA.tex
\geraldorevi{\gfiii{W}e present the list of application functions in each one of the six classes of data movement bottlenecks we identify using our new methodology.}

\geraldorevi{Our benchmark suite is composed of 144 different application functions, collected from 74 different applications. These applications belong to a different set of previously published and widely used benchmark suites. In total, \gfiii{we} collect applications from 16 benchmark suites, including: BWA~\gfiii{\cite{li2009fast}}, Chai~\gfiii{\cite{gomezluna_ispass2017}}, Darknet~\gfiii{\cite{redmon_darknet2013}}, GASE~\gfiii{\cite{ahmed2016comparison}}, Hardware Effects~\gfiii{\cite{hardwareeffects}}, Hashjoin~\gfiii{\cite{balkesen_TKDE2015}}, HPCC~\gfiii{\cite{luszczek_hpcc2006}}, HPCG~\gfiii{\cite{dongarra_hpcg2015}}, Ligra~\gfiii{\cite{shun_ppopp2013}}, PARSEC~\gfiii{\cite{bienia2008parsec}}, Parboil~\gfiii{\cite{stratton2012parboil}}, PolyBench~\gfiii{\cite{pouchet2012polybench}}, Phoenix~\gfiii{\cite{yoo_iiswc2009}}, Rodinia~\gfiii{\cite{che_iiswc2009}}, SPLASH-2~\gfiii{\cite{woo_isca1995}}, STREAM~\gfiii{\cite{mccalpin_stream1995}}. The 144 application functions that are part of \bench are listed across six tables\gfiii{, each designating one of the six classes we identify in Section~\ref{sec:scalability}}:}

\begin{itemize}
    \item Table~\ref{table_1a} lists application functions \gfiv{in Class~1a, i.e.,} that are \gfiii{DRAM} bandwidth\gfiii{-}bound (characterized in Section~\ref{sec_scalability_class1a});
    
    \item Table~\ref{table_1b} lists application functions \gfiv{in Class~1b, i.e.,} that are \gfiii{DRAM} latency\gfiii{-}bound (characterized in Section~\ref{sec_scalability_class1b}); 
    
    \item Table~\ref{table_1c} lists application functions \gfiv{in Class~1c, i.e.,} that are bottlenecked by the available \gfiii{L1/L2} cache capacity (characterized in Section~\ref{sec_scalability_class1c}); 
    
    \item Table~\ref{table_2a} lists application functions \gfiv{in Class~2a, i.e.,} that are bottlenecked by \gfiii{L3} cache contention (characterized in Section~\ref{sec_scalability_class2a}); 
    
    \item Table~\ref{table_2b} lists application functions \gfiv{in Class~2b, i.e.,} that are bottlenecked by L1 cache size (characterized in Section~\ref{sec_scalability_class2b}); 
    
    \item Table~\ref{table_2c} lists application functions \gfiv{in Class~2c, i.e.,} that are compute\gfiii{-}bound (characterized in Section~\ref{sec_scalability_class2c}).
\end{itemize}

In each table we list the benchmark suite, the application name, and the function name. We also list the input size/problem size we use to evaluate each application function.


\begin{table*}[t]
\tempcommand{1.2}
\centering
\caption{List of application functions in Class~1a.}
\label{table_1a}
\resizebox{\textwidth}{!}{%
\begin{tabular}{|c|c|c|c|c|c|}
\hline
\thead{\textbf{Class}} & \thead{\textbf{Suite}} & \thead{\textbf{Benchmark}} & \thead{\textbf{Function}} & \thead{\textbf{Input Set/ } \\ \textbf{Problem Size}} & \thead{\textbf{Representative}\\ \textbf{Function?}} \\ \hline \hline
1a & Chai~\cite{gomezluna_ispass2017} & Transpose & cpu & -m 1024 -n 524288  & No\\ \hline
1a & Chai~\cite{gomezluna_ispass2017} & Vector Pack & run\_cpu\_threads  & -m 268435456 -n 16777216 & No \\ \hline
1a & Chai~\cite{gomezluna_ispass2017} & Vector Unpack & run\_cpu\_threads  & -m 268435456 -n 16777216 & No \\ \hline
1a & Darknet~\cite{redmon_darknet2013} & Yolo & gemm  & ref & Yes \\ \hline
1a & Hardware Effects~\cite{hardwareeffects} & Bandwidth Saturation - Non Temporal & main  & ref &  No \\ \hline
1a & Hardware Effects~\cite{hardwareeffects} & Bandwidth Saturation - Temporal & main  & ref &  No \\ \hline
1a & Hashjoin~\cite{balkesen_TKDE2015} & NPO & knuth  & -r 12800000 -s 12000000 -x 12345 -y 54321 & No \\ \hline
1a & Hashjoin~\cite{balkesen_TKDE2015} & NPO & ProbeHashTable  & -r 12800000 -s 12000000 -x 12345 -y 54321 &Yes \\ \hline
1a & Hashjoin~\cite{balkesen_TKDE2015} & PRH & knuth  & -r 12800000 -s 12000000 -x 12345 -y 54321 & No \\ \hline
1a & Hashjoin~\cite{balkesen_TKDE2015} & PRH & lock  & -r 12800000 -s 12000000 -x 12345 -y 54321 & No\\ \hline
1a & Hashjoin~\cite{balkesen_TKDE2015} & PRHO & knuth  & -r 12800000 -s 12000000 -x 12345 -y 54321 & No\\ \hline
1a & Hashjoin~\cite{balkesen_TKDE2015} & PRHO & radix  & -r 12800000 -s 12000000 -x 12345 -y 54321 & No \\ \hline
1a & Hashjoin~\cite{balkesen_TKDE2015} & PRO & knuth  & -r 12800000 -s 12000000 -x 12345 -y 54321 & No \\ \hline
1a & Hashjoin~\cite{balkesen_TKDE2015} & PRO & parallel  & -r 12800000 -s 12000000 -x 12345 -y 54321 & No \\ \hline
1a & Hashjoin~\cite{balkesen_TKDE2015} & PRO & radix  & -r 12800000 -s 12000000 -x 12345 -y 54321& No \\ \hline
1a & Hashjoin~\cite{balkesen_TKDE2015} & RJ & knuth  & -r 12800000 -s 12000000 -x 12345 -y 54321 & No \\ \hline
1a & Ligra~\cite{shun_ppopp2013} & Betweenness Centrality & edgeMapSparse  & rMat & No \\ \hline
1a & Ligra~\cite{shun_ppopp2013} & Breadth-First Search & edgeMapSparse & rMat & No \\ \hline
1a & Ligra~\cite{shun_ppopp2013} & Connected Components & compute  & rMat & No \\ \hline
1a & Ligra~\cite{shun_ppopp2013} & Connected Components & compute  & USA & No \\ \hline
1a & Ligra~\cite{shun_ppopp2013} & Connected Components & edgeMapDense  & USA & No \\ \hline
1a & Ligra~\cite{shun_ppopp2013} & Connected Components & edgeMapSparse  & USA & Yes \\ \hline
1a & Ligra~\cite{shun_ppopp2013} & K-Core Decomposition & compute  & rMat & No \\ \hline
1a & Ligra~\cite{shun_ppopp2013} & K-Core Decomposition & compute & USA & No \\ \hline
1a & Ligra~\cite{shun_ppopp2013} & K-Core Decomposition & edgeMapDense & USA & No \\ \hline
1a & Ligra~\cite{shun_ppopp2013} & K-Core Decomposition & edgeMapSparse & rMat & No \\ \hline
1a & Ligra~\cite{shun_ppopp2013} & Maximal Independent Set & compute & rMat & No \\ \hline
1a & Ligra~\cite{shun_ppopp2013} & Maximal Independent Set & compute  & USA & No \\ \hline
1a & Ligra~\cite{shun_ppopp2013} & Maximal Independent Set & edgeMapDense  & USA & No \\ \hline
1a & Ligra~\cite{shun_ppopp2013} & Maximal Independent Set & edgeMapSparse  & rMat & No \\ \hline
1a & Ligra~\cite{shun_ppopp2013} & Maximal Independent Set & edgeMapSparse  & USA & No \\ \hline
1a & Ligra~\cite{shun_ppopp2013} & PageRank & compute & rMat & No \\ \hline
1a & Ligra~\cite{shun_ppopp2013} & PageRank & compute & USA & No \\ \hline
1a & Ligra~\cite{shun_ppopp2013} & PageRank & edgeMapDense  & USA & Yes \\ \hline
1a & Ligra~\cite{shun_ppopp2013} & Radii & compute  & rMat & No \\ \hline
1a & Ligra~\cite{shun_ppopp2013} & Radii & compute & USA & No \\ \hline
1a & Ligra~\cite{shun_ppopp2013} & Radii & edgeMapSparse  & USA & No \\ \hline
1a & Ligra~\cite{shun_ppopp2013} & Triangle Count & edgeMapDense  & rMat & Yes \\ \hline
1a & SPLASH-2~\cite{woo_isca1995} & Oceancp & relax  & simlarge & No \\ \hline
1a & SPLASH-2~\cite{woo_isca1995} & Oceanncp & relax  & simlarge & No \\ \hline
1a & STREAM~\cite{mccalpin_stream1995} & Add & Add  & 50000000 & Yes \\ \hline
1a & STREAM~\cite{mccalpin_stream1995} & Copy & Copy  & 50000000 & Yes \\ \hline
1a & STREAM~\cite{mccalpin_stream1995} & Scale & Scale  & 50000000 & Yes \\ \hline
1a & STREAM~\cite{mccalpin_stream1995} & Triad & Triad  &  50000000 &  Yes \\ \hline
\end{tabular}%
}
\end{table*}

\begin{table*}[t]
\tempcommand{1.2}
\centering
\caption{List of application functions in Class~1b.}
\label{table_1b}
\resizebox{\textwidth}{!}{%
\begin{tabular}{|c|c|c|c|c|c|}
\hline
\thead{\textbf{Class}} & \thead{\textbf{Suite}} & \thead{\textbf{Benchmark}} & \thead{\textbf{Function}} & \thead{\textbf{Input Set/ } \\ \textbf{Problem Size}} & \thead{\textbf{Representative}\\ \textbf{Function?}} \\ \hline \hline
1b & Chai~\cite{gomezluna_ispass2017} & Canny Edge Detection & gaussian & ref & No \\ \hline
1b & Chai~\cite{gomezluna_ispass2017} & Canny Edge Detection & supression & ref & No \\ \hline
1b & Chai~\cite{gomezluna_ispass2017} & Histogram - input partition & run\_cpu\_threads & ref & Yes \\ \hline
1b & Chai~\cite{gomezluna_ispass2017} & Select & run\_cpu\_threads & -n 67108864 & No \\ \hline
1b & GASE~\cite{ahmed2016comparison} & FastMap & 2occ4 & Wg2 & No \\ \hline
1b & GASE~\cite{ahmed2016comparison} & FastMap & occ4 & Wg2 & No \\ \hline
1b & Hashjoin~\cite{balkesen_TKDE2015} & PRH & HistogramJoin & -r 12800000 -s 12000000 -x 12345 -y 54321 & Yes \\ \hline
1b & Phoenix~\cite{yoo_iiswc2009} & Linear Regression & linear\_regression\_map & key\_file\_500MB & No \\ \hline
1b & Phoenix~\cite{yoo_iiswc2009}  & PCA & main & ref & No \\ \hline
1b & Phoenix~\cite{yoo_iiswc2009}  & String Match & string\_match\_map & key\_file\_500MB & Yes \\ \hline
1b & PolyBench~\cite{pouchet2012polybench} & linear-algebra & lu & LARGE\_DATASET & Yes \\ \hline
1b & Rodinia~\cite{che_iiswc2009} & Kmeans & euclidDist & 819200.txt & No \\ \hline
1b & Rodinia~\cite{che_iiswc2009} & Kmeans & find & 819200.txt & No \\ \hline
1b & Rodinia~\cite{che_iiswc2009} & Kmeans & main & 819200.txt & No \\ \hline
1b & Rodinia~\cite{che_iiswc2009} & Streamcluster & pengain & ref & No \\ \hline
1b & SPLASH-2~\cite{woo_isca1995} & Oceancp & slave2 & simlarge & Yes \\ \hline
\end{tabular}%
}
\end{table*}

\begin{table*}[t]
\tempcommand{1.2}
\centering
\caption{List of application functions in Class~1c.}
\label{table_1c}
\resizebox{\textwidth}{!}{%
\begin{tabular}{|c|c|c|c|c|c|}
\hline
\thead{\textbf{Class}} & \thead{\textbf{Suite}} & \thead{\textbf{Benchmark}} & \thead{\textbf{Function}} & \thead{\textbf{Input Set/ } \\ \textbf{Problem Size}} & \thead{\textbf{Representative}\\ \textbf{Function?}} \\ \hline \hline
1c & BWA~\cite{li2009fast} & Align & bwa\_aln\_core & Wg1 & No \\ \hline
1c & Chai~\cite{gomezluna_ispass2017} & Breadth-First Search & comp  & USA-road-d & No \\ \hline
1c & Chai~\cite{gomezluna_ispass2017} & Breadth-First Search & fetch  & USA-road-d & No \\ \hline
1c & Chai~\cite{gomezluna_ispass2017} & Breadth-First Search & load  & USA-road-d & No \\ \hline
1c & Chai~\cite{gomezluna_ispass2017} & Breadth-First Search & run\_cpu\_threads  & USA-road-d & No \\ \hline
1c & Chai~\cite{gomezluna_ispass2017} & Canny Edge Detection & hystresis  & ref & No \\ \hline
1c & Chai~\cite{gomezluna_ispass2017} & Canny Edge Detection & sobel  & ref & No \\ \hline
1c & Chai~\cite{gomezluna_ispass2017} & Histogram - output partition & run\_cpu\_threads  & ref & No \\ \hline
1c & Chai~\cite{gomezluna_ispass2017} & Padding & run\_cpu\_threads  & -m 10000 -n 9999 & Yes \\ \hline
1c & Chai~\cite{gomezluna_ispass2017} & Select & fetch  & -n 67108864 & No \\ \hline
1c & Chai~\cite{gomezluna_ispass2017} & Stream Compaction & run\_cpu\_threads  & ref & No \\ \hline
1c & Darknet~\cite{redmon_darknet2013} & Resnet & gemm  & ref & Yes \\ \hline
1c & Hashjoin~\cite{balkesen_TKDE2015} & NPO & lock  & -r 12800000 -s 12000000 -x 12345 -y 54321 & No \\ \hline
1c & Ligra~\cite{shun_ppopp2013} & BFS-Connected Components & edgeMapSparse  & rMat & No \\ \hline
1c & Ligra~\cite{shun_ppopp2013} & Triangle Count & compute & rMat & No \\ \hline
1c & Ligra~\cite{shun_ppopp2013} & Triangle Count & compute & USA & No \\ \hline
1c & Ligra~\cite{shun_ppopp2013} & Triangle Count & edgeMapDense  & USA & No \\ \hline
1c & PARSEC~\cite{bienia2008parsec} & Blackscholes & BlkSchlsEqEuroNoDiv  & simlarge & No \\ \hline
1c & PARSEC~\cite{bienia2008parsec} & Fluidaminate & ProcessCollision2MT  & simlarge & Yes \\ \hline
1c & PARSEC~\cite{bienia2008parsec} & Streamcluster & DistL2Float  & simlarge & No \\ \hline
1c & Rodinia~\cite{che_iiswc2009} & Myocyte & find  & 1000000 & No \\ \hline
1c & Rodinia~\cite{che_iiswc2009} & Myocyte & master  & 1000000 &  No \\ \hline
\end{tabular}%
}
\end{table*}

\begin{table*}[t]
\tempcommand{1.2}
\centering
\caption{List of application functions in Class~2a.}
\label{table_2a}
\resizebox{\textwidth}{!}{%
\begin{tabular}{|c|c|c|c|c|c|}
\hline
\thead{\textbf{Class}} & \thead{\textbf{Suite}} & \thead{\textbf{Benchmark}} & \thead{\textbf{Function}} & \thead{\textbf{Input Set/ } \\ \textbf{Problem Size}} & \thead{\textbf{Representative}\\ \textbf{Function?}} \\ \hline \hline
2a & HPCC~\cite{luszczek_hpcc2006} & RandomAccess & main & ref &  No \\ \hline
2a & HPCC~\cite{luszczek_hpcc2006} & RandomAccess & update & ref &  No \\ \hline
2a & Ligra~\cite{shun_ppopp2013} & Betweenness Centrality & Compute & rMat & No \\ \hline
2a & Ligra~\cite{shun_ppopp2013} & Betweenness Centrality & Compute  & USA & No \\ \hline
2a & Ligra~\cite{shun_ppopp2013} & Betweenness Centrality & edgeMapDense  & rMat & No \\ \hline
2a & Ligra~\cite{shun_ppopp2013} & Betweenness Centrality & *edgeMapSparse  &  USA & Yes \\ \hline
2a & Ligra~\cite{shun_ppopp2013} & BFS-Connected Components & Compute  & rMat & No \\ \hline
2a & Ligra~\cite{shun_ppopp2013} & BFS-Connected Components & Compute & USA & No \\ \hline
2a & Ligra~\cite{shun_ppopp2013} & BFS-Connected Components & edgeMapSparse  & USA & Yes \\ \hline
2a & Ligra~\cite{shun_ppopp2013} & Breadth-First Search & compute  & rMat & No \\ \hline
2a & Ligra~\cite{shun_ppopp2013} & Breadth-First Search & compute & USA & No  \\ \hline
2a & Ligra~\cite{shun_ppopp2013} & Breadth-First Search & edgeMapDense  & rMat & No \\ \hline
2a & Ligra~\cite{shun_ppopp2013} & Breadth-First Search & edgeMapSparse  & USA & Yes \\ \hline
2a & Ligra~\cite{shun_ppopp2013} & Connected Components & edgeMapDense  & rMat  & No \\ \hline
2a & Ligra~\cite{shun_ppopp2013} & Maximal Independent Set & edgeMapDense & rMat & No \\ \hline
2a & Ligra~\cite{shun_ppopp2013} & PageRank & edgeMapDense(Rmat)  & rMat & No \\ \hline
2a & Phoenix~\cite{yoo_iiswc2009}  & WordCount & main & word\_100MB & No \\ \hline
2a & PolyBench~\cite{pouchet2012polybench} & linear-algebra & gramschmidt  & LARGE\_DATASET & Yes \\ \hline
2a & Rodinia~\cite{che_iiswc2009} & CFD Solver & main  & fvcorr.domn.193K  & No \\ \hline
2a & SPLASH-2~\cite{woo_isca1995} & FFT2 & Reverse  & simlarge & Yes \\ \hline
2a & SPLASH-2~\cite{woo_isca1995} & FFT2 & Transpose  & simlarge & Yes \\ \hline
2a & SPLASH-2~\cite{woo_isca1995} & Oceancp & jacobcalc  &  simlarge & No \\ \hline
2a & SPLASH-2~\cite{woo_isca1995} & Oceancp & laplaccalc  & simlarge  & No \\ \hline
2a & SPLASH-2~\cite{woo_isca1995} & Oceanncp & jacobcalc  & simlarge &  Yes \\ \hline
2a & SPLASH-2~\cite{woo_isca1995} & Oceanncp & laplaccalc  & simlarge  & Yes \\ \hline
2a & SPLASH-2~\cite{woo_isca1995} & Oceanncp & slave2  & simlarge & No \\ \hline
\end{tabular}%
}
\end{table*}

\begin{table*}[t]
\tempcommand{1.2}

\centering
\caption{List of application functions in Class~2b.}
\label{table_2b}
\resizebox{\textwidth}{!}{%
\begin{tabular}{|c|c|c|c|c|c|}
\hline
\thead{\textbf{Class}} & \thead{\textbf{Suite}} & \thead{\textbf{Benchmark}} & \thead{\textbf{Function}} & \thead{\textbf{Input Set/ } \\ \textbf{Problem Size}} & \thead{\textbf{Representative}\\ \textbf{Function?}} \\ \hline \hline
2b & Chai~\cite{gomezluna_ispass2017} & Bezier Surface & main\_thread  &  ref &  Yes \\ \hline
2b & Hardware Effects~\cite{hardwareeffects} & False Sharing - Isolated & main  & ref & No \\ \hline
2b & PolyBench~\cite{pouchet2012polybench} & convolution & convolution-2d  & LARGE\_DATASET & No \\ \hline
2b & PolyBench~\cite{pouchet2012polybench} & linear-algebra & gemver  & LARGE\_DATASET &  Yes \\ \hline
2b & SPLASH-2~\cite{woo_isca1995} & Lucb & Bmod  & simlarge &  Yes \\ \hline
2b & SPLASH-2~\cite{woo_isca1995} & Radix & slave2  & simlarge &  Yes \\ \hline
\end{tabular}%
}
\end{table*}


\begin{table*}[!t]
\tempcommand{1.2}
\caption{List of application functions in Class~2c.}
\label{table_2c}
\resizebox{\textwidth}{!}{%
\begin{tabular}{|c|c|c|c|c|c|}
\hline
\thead{\textbf{Class}} & \thead{\textbf{Suite}} & \thead{\textbf{Benchmark}} & \thead{\textbf{Function}} & \thead{\textbf{Input Set/ } \\ \textbf{Problem Size}} & \thead{\textbf{Representative}\\ \textbf{Function?}} \\ \hline \hline
2c & BWA~\cite{li2009fast} & Align & bwa\_aln\_core & Wg2 &  No \\ \hline
2c & Chai~\cite{gomezluna_ispass2017} & Transpose & run\_cpu\_threads  & -m 1024 -n 524288 &  No \\ \hline
2c & Darknet~\cite{redmon_darknet2013} & Alexnet & gemm  & ref & No\\ \hline
2c & Darknet~\cite{redmon_darknet2013} & vgg16 & gemm  & ref & No \\ \hline
2c & Hardware Effects~\cite{hardwareeffects} & False Sharing - Shared & main  & ref & No\\ \hline
2c & HPCG~\cite{dongarra_hpcg2015} & HPCG & ComputePrologation  & ref & Yes \\ \hline
2c & HPCG~\cite{dongarra_hpcg2015} & HPCG & ComputeRestriction  & ref & Yes \\ \hline
2c & HPCG~\cite{dongarra_hpcg2015} & HPCG & ComputeSPMV  & ref & Yes \\ \hline
2c & HPCG~\cite{dongarra_hpcg2015} & HPCG & ComputeSYMGS  & ref & Yes \\ \hline
2c & Ligra~\cite{shun_ppopp2013} & K-Core Decomposition & edgeMapDense & rMat &  No\\ \hline
2c & Ligra~\cite{shun_ppopp2013} & Radii & edgeMapSparse & rMat & No \\ \hline
2c & Parboil~\cite{stratton2012parboil} & Breadth-First Search & BFS\_CPU  & ref & No\\ \hline
2c & Parboil~\cite{stratton2012parboil}  & MRI-Gridding & CPU\_kernels  & ref & No\\ \hline
2c & Parboil~\cite{stratton2012parboil}  & Stencil & cpu\_stencil  & ref & No \\ \hline
2c & Parboil~\cite{stratton2012parboil}  & Two Point Angular Correlation Function & doCompute  & ref & No \\ \hline
2c & PARSEC~\cite{bienia2008parsec} & Bodytrack & FilterRow & ref & No \\ \hline
2c & PARSEC~\cite{bienia2008parsec} & Ferret & DistL2Float & ref & Yes \\ \hline
2c & Phoenix~\cite{yoo_iiswc2009}  & Kmeans & main & ref  & No \\ \hline
2c & PolyBench~\cite{pouchet2012polybench} & linear-algebra & 3mm & LARGE\_DATASET & Yes \\ \hline
2c & PolyBench~\cite{pouchet2012polybench} & linear-algebra & doitgen & LARGE\_DATASET & Yes \\ \hline
2c & PolyBench~\cite{pouchet2012polybench} & linear-algebra & gemm & LARGE\_DATASET  & Yes \\ \hline
2c & PolyBench~\cite{pouchet2012polybench} & linear-algebra & symm & LARGE\_DATASET  & Yes \\ \hline
2c & PolyBench~\cite{pouchet2012polybench} & stencil & fdtd-apml &  LARGE\_DATASET & Yes \\ \hline
2c & Rodinia~\cite{che_iiswc2009} & Back Propagation & adjustweights & 134217728 & No \\ \hline
2c & Rodinia~\cite{che_iiswc2009} & Back Propagation & layerfoward & 134217728 & No \\ \hline
2c & Rodinia~\cite{che_iiswc2009} & Breadth-First Search & main & graph1M\_6 & Yes \\ \hline
2c & Rodinia~\cite{che_iiswc2009} & Needleman-Wunsch & main & 32768 & Yes \\ \hline
2c & Rodinia~\cite{che_iiswc2009}a & Srad & FIN & ref &  No \\ \hline
2c & SPLASH-2~\cite{woo_isca1995} & Barnes & computeForces & simlarge & No \\ \hline
2c & SPLASH-2~\cite{woo_isca1995} & Barnes & gravsub & simlarge & No \\ \hline
\end{tabular}%
}
\vspace{15pt}
\end{table*}


%% file: 08_appendixB.tex
\vspace{10pt}
\begin{table}[!b]
\begin{minipage}{\textwidth}
\caption{\gfiv{44 r}epresentative \gfv{application} functions studied in detail in this work.$^*$}
\label{tab:benchmarks}
 \tempcommand{1.2}
\centering
\resizebox{0.84 \linewidth}{!}{
\begin{tabular}{llllll}
\toprule
\textbf{Suite} & \textbf{Benchmark} & \textbf{Function} & \textbf{Short Name} & \textbf{Class}  & \textbf{\% } \\
\midrule
\multirow{3}{*}{\parbox{1.5cm}{Chai \cite{gomezluna_ispass2017}}} & Bezier Surface & Bezier & CHABsBez & 2b & 100\\
                                                  & Histogram & Histogram & CHAHsti & 1b & 100\\
                                                  & Padding & Padding & CHAOpad & 1c & 75.1\\
\hline
\multirow{2}{*}{\parbox{1.7cm}{Darknet~\cite{redmon_darknet2013}}} & Resnet 152 & gemm\_nn & DRKRes & 1c & 95.2\\
                                                  & Yolo & gemm\_nn & DRKYolo & 1a & 97.1\\
\hline
\multirow{2}{*}{\parbox{1.9cm}{Hashjoin \cite{balkesen_TKDE2015}}} & NPO & ProbeHashTable & HSJNPO & 1a & 47.8 \\
                                                  & PRH & HistogramJoin & HSJPRH & 1b & 53.1\\
\hline
\multirow{4}{*}{\parbox{1.7cm}{HPCG~\cite{dongarra_hpcg2015}}} & HPCG & ComputeProlongation & HPGProl & 2c & 34.3\\
                                               & HPCG & ComputeRestriction & HPGRes & 2c & 42.1\\
                                               & HPCG & ComputeSPMV & HPGSpm & 2c & 30.5\\
                                               & HPCG & ComputeSYMGS & HPGSyms & 2c & 63.6\\
\hline
\multirow{4}{*}{Ligra \cite{shun_ppopp2013}} & Betweenness Centrality & EdgeMapSparse (USA~\cite{dimacs}) & LIGBcEms & 2a & 78.9\\
                                             & Breadth-First Search & EdgeMapSparse (USA) & LIGBfsEms & 2a & 67.0\\
                                             & BFS-Connected Components &  EdgeMapSparse (USA) & LIGBfscEms & 2a & 68.3\\
                                             & Connected Components & EdgeMapSparse (USA) & LIGCompEms & 1a & 25.6\\
                                             & PageRank & EdgeMapDense (USA~\cite{dimacs}) & LIGPrkEmd & 1a & 57.2 \\
                                             & Radii & EdgeMapSparse (USA)& LIGRadiEms & 1a & 67.0\\
                                             & Triangle & EdgeMapDense (Rmat) & LIGTriEmd & 1a & 26.7\\
\hline
\multirow{2}{*}{\parbox{1.7cm}{PARSEC~\cite{bienia2008parsec}}} & Ferret & DistL2Float & PRSFerr  & 2c & 18.6\\
                                               & Fluidaminate & ProcessCollision2MT & PRSFlu & 1c & 23.9 \\
\hline
\multirow{2}{*}{\parbox{1.7cm}{Phoenix~\cite{yoo_iiswc2009}}} & Linear Regression & linear\_regression\_map & PHELinReg & 1b& 76.2 \\
                                              & String Matching & string\_match\_map & PHEStrMat  & 1b & 38.3\\
\hline
\multirow{9}{*}{PolyBench \cite{pouchet2012polybench}} & Linear Algebra & 3 Matrix Multiplications & PLY3mm & 2c & 100.0\\
                                             & Linear Algebra  & Multi-resolution analysis kernel & PLYDoitgen & 2c & 98.3\\
                                             & Linear Algebra  &  Matrix-multiply C=alpha.A.B+beta.C & PLYgemm & 2c & 99.7\\
                                             & Linear Algebra  & Vector Mult. and Matrix Addition & PLYgemver & 2b & 44.4\\
                                             & Linear Algebra  & Gram-Schmidt decomposition & PLYGramSch & 2a & 100.0\\
                                             & Linear Algebra  & LU decomposition & PLYalu & 1b & 100.0\\
                                             & Linear Algebra & Symmetric matrix-multiply & PLYSymm & 2c & 99.9 \\
                                             & Stencil  & 2D Convolution & PLYcon2d & 2b & 100.0 \\
                                             & Stencil & 2-D Finite Different Time Domain & PLYdtd & 2c & 39.8\\
\hline
\multirow{2}{*}{\parbox{1.8cm}{Rodinia~\cite{chen2014big}}} & BFS & BFSGraph & RODBfs  & 2c & 100.0\\
                                           & Needleman-Wunsch & runTest & RODNw  & 2c & 84.9\\
\hline
\multirow{7}{*}{\parbox{1.9cm}{SPLASH-2~\cite{woo_isca1995}}} & FFT & Reverse & SPLFftRev & 2a &12.7 \\
                                              & FFT & Transpose & SPLFftTra & 2a & 8.0\\
                                              & Lucb & Bmod & SPLLucb & 2b & 77.6\\
                                              & Oceanncp & jacobcalc & SPLOcnpJac & 2a &30.7 \\
                                              & Oceanncp & laplaccalc & SPLOcnpLap & 2a & 23.4\\
                                              & Oceancp & slave2 & SPLOcpSlave & 1b & 24.4\\
                                              & Radix & slave\_sort & SPLRad & 2b & 41.1\\
\hline
\multirow{4}{*}{\parbox{1.7cm}{STREAM~\cite{mccalpin_stream1995}}} & Add & Add & STRAdd & 1a & 98.4 \\
                                                   & Copy & Copy & STRCpy & 1a & 98.3\\
                                                   & Scale & Scale & STRSca & 1a & 97.5\\
                                                   & Triad & Triad & STRTriad & 1a & 99.1 \\
\bottomrule
\end{tabular}
}
\newline
\scriptsize{$^*$ Short names are encoded as XXXYyyZzz, where XXX is the source application suite, Yyy is the application name, and Zzz is the function (if more than one per benchmark). For graph processing applications from Ligra, we test two different input graphs, so we append the graph name to the short benchmark name as well. The \% column indicates the percentage of clock cycles that the function consumes as a fraction of the execution time of the entire benchmark}.
\end{minipage}

\end{table}

%% file: 08_appendixC.tex

\vspace{5pt}

\begin{table}[!b]
\begin{minipage}{0.89\textwidth}
\centering
\caption{List of \gfv{the} \gfiv{e}valuated \gfiii{345} \gfiv{a}pplications.}
\label{tab:allapps}
\tempcommand{0.905}
\resizebox{\linewidth}{!}{
\begin{tabular}{|l|l || l|l || l |l|}
\hline
\textbf{Benchmark Suite} &	\textbf{Application} 	&	\textbf{Benchmark Suite} &	\textbf{Application} 	&	\textbf{Benchmark Suite} &	\textbf{Application} \\ \hline \hline						
ArtraCFD~\cite{mo2019mesoscale} &	ArtraCFD	&	\multirow{5}{*}{HPCG~\cite{dongarra_hpcg2015}} & 	Global Dot Product	&	\multirow{10}{*}{SD-VBS - Vision~\cite{Thomas_CortexSuite_IISWC_2014}} & 	disparity	\\	\cline{1-1}	\cline{2-2}		\cline{4-4}		\cline{6-6}
blasr~\cite{chaisson2012mapping} &	Long read aligner	&	&	Multigrid preconditione	&	&	localization	\\	\cline{1-1}	\cline{2-2}		\cline{4-4}		\cline{6-6}
BWA~\cite{li2013aligning} &	aln	&	&	Sparse Matrix Vector Multiplication (SpMV)	&	&	mser	\\		\cline{2-2}		\cline{4-4}		\cline{6-6}
&	fastmap	&	&	Symmetric Gauss-Seidel smoother (SymGS)	&	&	multi\_ncut	\\	\cline{1-1}	\cline{2-2}		\cline{4-4}		\cline{6-6}
\multirow{12}{*}{Chai~\cite{gomezluna_ispass2017}} & 	BFS	&	&	Vector Update	&	&	pca	\\		\cline{2-2}	\cline{3-3}	\cline{4-4}		\cline{6-6}
&	BS	&	\multirow{3}{*}{IMPICA Workloads~\cite{hsieh2016accelerating}} & 	btree	&	&	sift	\\		\cline{2-2}		\cline{4-4}		\cline{6-6}
&	CEDD	&	&	hashtable	&	&	stitch	\\		\cline{2-2}		\cline{4-4}		\cline{6-6}
&	HSTI	&	&	llubenchmark	&	&	svm	\\		\cline{2-2}	\cline{3-3}	\cline{4-4}		\cline{6-6}
&	HSTO	&	\multirow{2}{*}{libvpx~\cite{libvpx}} & 	VP8	&	&	texture\_synthesis	\\		\cline{2-2}		\cline{4-4}		\cline{6-6}
&	OOPPAD	&	&	VP9	&	&	tracking	\\		\cline{2-2}	\cline{3-3}	\cline{4-4}	\cline{5-5}	\cline{6-6}
&	OOPTRNS	&	\multirow{13}{*}{Ligra~\cite{shun_ppopp2013}} & 	BC	&	\multirow{2}{*}{sort-merge-joins~\cite{balkesen2013multi}} & 	m-pass	\\		\cline{2-2}		\cline{4-4}		\cline{6-6}
&	SC	&	&	BellmanFord	&	&	m-way	\\		\cline{2-2}		\cline{4-4}	\cline{5-5}	\cline{6-6}
&	SELECT	&	&	BFS	&	\multirow{29}{*}{SPEC CPU2006~\cite{spec2006}} & 	400.perlbench	\\		\cline{2-2}		\cline{4-4}		\cline{6-6}
&	TRNS	&	&	BFS-Bitvector	&	&	401.bzip2	\\		\cline{2-2}		\cline{4-4}		\cline{6-6}
&	VPACK	&	&	BFS-CC	&	&	403.gcc	\\		\cline{2-2}		\cline{4-4}		\cline{6-6}
&	VUPACK	&	&	CF	&	&	410.bwaves	\\	\cline{1-1}	\cline{2-2}		\cline{4-4}		\cline{6-6}
clstm~\cite{ghosh2016contextual} &	clstm	&	&	Components	&	&	416.gamess	\\	\cline{1-1}	\cline{2-2}		\cline{4-4}		\cline{6-6}
\multirow{9}{*}{CombBLAS~\cite{bulucc2011combinatorial}} & 	BetwCent	&	&	KCore	&	&	429.mcf	\\		\cline{2-2}		\cline{4-4}		\cline{6-6}
&	BipartiteMatchings	&	&	MIS	&	&	433.milc	\\		\cline{2-2}		\cline{4-4}		\cline{6-6}
&	CC	&	&	PageRank	&	&	434.zeusmp	\\		\cline{2-2}		\cline{4-4}		\cline{6-6}
&	DirOptBFS	&	&	PageRankDelta	&	&	435.gromacs	\\		\cline{2-2}		\cline{4-4}		\cline{6-6}
&	FilteredBFS	&	&	Radii	&	&	436.cactusADM	\\		\cline{2-2}		\cline{4-4}		\cline{6-6}
&	FilteredMIS	&	&	Triangle	&	&	437.leslie3d	\\		\cline{2-2}	\cline{3-3}	\cline{4-4}		\cline{6-6}
&	MCL3D	&	\multirow{2}{*}{Metagraph~\cite{karasikov2020metagraph}} & 	annotate	&	&	444.namd	\\		\cline{2-2}		\cline{4-4}		\cline{6-6}
&	Ordering/RCM	&	&	classify	&	&	445.gobmk	\\		\cline{2-2}	\cline{3-3}	\cline{4-4}		\cline{6-6}
&	TopDownBFS	&	\multirow{5}{*}{MKL~\cite{mkl}} & 	ASUM	&	&	447.dealII	\\	\cline{1-1}	\cline{2-2}		\cline{4-4}		\cline{6-6}
\multirow{17}{*}{CORAL~\cite{coral}} &	AMG2013	&	&	AXPY	&	&	450.soplex	\\		\cline{2-2}		\cline{4-4}		\cline{6-6}
&	CAM-SE	&	&	DOT	&	&	453.povray	\\		\cline{2-2}		\cline{4-4}		\cline{6-6}
&	Graph500	&	&	GEMM	&	&	454.calculix	\\		\cline{2-2}		\cline{4-4}		\cline{6-6}
&	HACC	&	&	GEMV	&	&	456.hmmer	\\		\cline{2-2}	\cline{3-3}	\cline{4-4}		\cline{6-6}
&	Hash	&	\multirow{11}{*}{Parboil~\cite{stratton2012parboil}} & 	mri-q	&	&	458.sjeng	\\		\cline{2-2}		\cline{4-4}		\cline{6-6}
&	homme1\_3\_6	&	&	BFS	&	&	459.GemsFDTD	\\		\cline{2-2}		\cline{4-4}		\cline{6-6}
&	Integer Sort	&	&	cutcp	&	&	462.libquantum	\\		\cline{2-2}		\cline{4-4}		\cline{6-6}
&	KMI	&	&	histo	&	&	464.h264ref	\\		\cline{2-2}		\cline{4-4}		\cline{6-6}
&	LSMS	&	&	lbm	&	&	465.tonto	\\		\cline{2-2}		\cline{4-4}		\cline{6-6}
&	LULESH	&	&	mri-gridding	&	&	470.lbm	\\		\cline{2-2}		\cline{4-4}		\cline{6-6}
&	MCB	&	&	sad	&	&	471.omnetpp	\\		\cline{2-2}		\cline{4-4}		\cline{6-6}
&	miniFE	&	&	sgemm	&	&	473.astar	\\		\cline{2-2}		\cline{4-4}		\cline{6-6}
&	Nekbone	&	&	spmv	&	&	481.wrf	\\		\cline{2-2}		\cline{4-4}		\cline{6-6}
&	QBOX	&	&	stencil	&	&	482.sphinx3	\\		\cline{2-2}		\cline{4-4}		\cline{6-6}
&	SNAP	&	&	tpacf	&	&	483.xalancbmk	\\		\cline{2-2}	\cline{3-3}	\cline{4-4}	\cline{5-5}	\cline{6-6}
&	SPECint2006"peak"	&	\multirow{12}{*}{PARSEC~\cite{bienia2008parsec}} &	blackscholes	&	\multirow{43}{*}{SPEC CPU2017~\cite{spec2017}} & 	500.perlbench\_r	\\		\cline{2-2}		\cline{4-4}		\cline{6-6}
&	UMT2013	&	&	bodytrack	&	&	502.gcc\_r	\\	\cline{1-1}	\cline{2-2}		\cline{4-4}		\cline{6-6}
\multirow{15}{*}{Darknet~\cite{redmon_darknet2013}} & 	AlexNet	&	&	canneal	&	&	503.bwaves\_r	\\		\cline{2-2}		\cline{4-4}		\cline{6-6}
&	Darknet19	&	&	dedup	&	&	505.mcf\_r	\\		\cline{2-2}		\cline{4-4}		\cline{6-6}
&	Darknet53	&	&	facesim	&	&	507.cactuBSSN\_r	\\		\cline{2-2}		\cline{4-4}		\cline{6-6}
&	Densenet 201	&	&	ferret	&	&	508.namd\_r	\\		\cline{2-2}		\cline{4-4}		\cline{6-6}
&	Extraction	&	&	fluidanimate	&	&	510.parest\_r	\\		\cline{2-2}		\cline{4-4}		\cline{6-6}
&	Resnet 101	&	&	freqmine	&	&	511.povray\_r	\\		\cline{2-2}		\cline{4-4}		\cline{6-6}
&	Resnet 152	&	&	raytrace	&	&	519.lbm\_r	\\		\cline{2-2}		\cline{4-4}		\cline{6-6}
&	Resnet 18	&	&	streamcluster	&	&	520.omnetpp\_r	\\		\cline{2-2}		\cline{4-4}		\cline{6-6}
&	Resnet 34	&	&	swaptions	&	&	521.wrf\_r	\\		\cline{2-2}		\cline{4-4}		\cline{6-6}
&	Resnet 50	&	&	vips	&	&	523.xalancbmk\_r	\\		\cline{2-2}		\cline{4-4}		\cline{6-6}
&	ResNeXt 101	&	&	x264	&	&	525.x264\_r	\\		\cline{2-2}	\cline{3-3}	\cline{4-4}		\cline{6-6}
&	ResNext 152	&	\multirow{7}{*}{Phoenix~\cite{yoo_iiswc2009}} & 	histogram	&	&	526.blender\_r	\\		\cline{2-2}		\cline{4-4}		\cline{6-6}
&	ResNeXt50	&	&	kmeans	&	&	527.cam4\_r	\\		\cline{2-2}		\cline{4-4}		\cline{6-6}
&	VGG-16	&	&	linear-regression	&	&	531.deepsjeng\_r	\\		\cline{2-2}		\cline{4-4}		\cline{6-6}
&	Yolo	&	&	matrix multiply	&	&	538.imagick\_r	\\	\cline{1-1}	\cline{2-2}		\cline{4-4}		\cline{6-6}
DBT-5~\cite{nascimento2010dbt} & 	TPC-E 	&	&	pca	&	&	541.leela\_r	\\	\cline{1-1}	\cline{2-2}		\cline{4-4}		\cline{6-6}
\multirow{10}{*}{DBx1000~\cite{yu2014staring}} & 	TPCC DL\_DETECT	&	&	string\_match	&	&	544.nab\_r	\\		\cline{2-2}		\cline{4-4}		\cline{6-6}
&	TPCC HEKATON	&	&	word\_count	&	&	548.exchange2\_r	\\		\cline{2-2}	\cline{3-3}	\cline{4-4}		\cline{6-6}
&	TPCC NO\_WAIT	&	\multirow{23}{*}{PolyBench~\cite{pouchet2012polybench}} & 	2mm	&	&	549.fotonik3d\_r	\\		\cline{2-2}		\cline{4-4}		\cline{6-6}
&	TPCC SILO	&	&	3mm	&	&	554.roms\_r	\\		\cline{2-2}		\cline{4-4}		\cline{6-6}
&	TPCC TICTOC	&	&	atax	&	&	557.xz\_r	\\		\cline{2-2}		\cline{4-4}		\cline{6-6}
&	YCSB DL\_DETECT	&	&	bicg	&	&	600.perlbench\_s	\\		\cline{2-2}		\cline{4-4}		\cline{6-6}
&	YCSB HEKATON	&	&	cholesky	&	&	602.gcc\_s	\\		\cline{2-2}		\cline{4-4}		\cline{6-6}
&	YCSB NO\_WAIT	&	&	convolution-2d	&	&	603.bwaves\_s	\\		\cline{2-2}		\cline{4-4}		\cline{6-6}
&	YCSB SILO	&	&	convolution-3d	&	&	605.mcf\_s	\\		\cline{2-2}		\cline{4-4}		\cline{6-6}
&	YCSB TICTOC	&	&	correlation	&	&	607.cactuBSSN\_s	\\	\cline{1-1}	\cline{2-2}		\cline{4-4}		\cline{6-6}
\multirow{4}{*}{DLRM~\cite{DLRM19}} & 	RM1-large~\cite{ke2019recnmp}	&	&	covariance	&	&	619.lbm\_s	\\		\cline{2-2}		\cline{4-4}		\cline{6-6}
&	RM1-small~\cite{ke2019recnmp}	&	&	doitgen	&	&	620.omnetpp\_s	\\		\cline{2-2}		\cline{4-4}		\cline{6-6}
&	RM2-large~\cite{ke2019recnmp}	&	&	durbin	&	&	621.wrf\_s	\\		\cline{2-2}		\cline{4-4}		\cline{6-6}
&	RM2-small~\cite{ke2019recnmp}	&	&	fdtd-apm	&	&	623.xalancbmk\_s	\\	\cline{1-1}	\cline{2-2}		\cline{4-4}		\cline{6-6}
\multirow{2}{*}{GASE~\cite{ahmed2016comparison}} &	FastMap	&	&	gemm	&	&	625.x264\_s	\\		\cline{2-2}		\cline{4-4}		\cline{6-6}
&	gale\_aln	&	&	gemver	&	&	627.cam4\_s	\\	\cline{1-1}	\cline{2-2}		\cline{4-4}		\cline{6-6}
\multirow{9}{*}{GraphMat~\cite{sundaram_vldbendow2015}} & 	BFS	&	&	gramschmidt	&	&	628.pop2\_s	\\		\cline{2-2}		\cline{4-4}		\cline{6-6}
&	DeltaStepping	&	&	gramschmidt	&	&	631.deepsjeng\_s	\\		\cline{2-2}		\cline{4-4}		\cline{6-6}
&	Incremental PageRank	&	&	lu	&	&	638.imagick\_s	\\		\cline{2-2}		\cline{4-4}		\cline{6-6}
&	LDA	&	&	lu	&	&	641.leela\_s	\\		\cline{2-2}		\cline{4-4}		\cline{6-6}
&	PageRank	&	&	mvt	&	&	644.nab\_s	\\		\cline{2-2}		\cline{4-4}		\cline{6-6}
&	SDG	&	&	symm	&	&	648.exchange2\_s	\\		\cline{2-2}		\cline{4-4}		\cline{6-6}
&	SSSP	&	&	syr2k	&	&	649.fotonik3d\_s	\\		\cline{2-2}		\cline{4-4}		\cline{6-6}
&	Topological Sort	&	&	syrk	&	&	654.roms\_s	\\		\cline{2-2}		\cline{4-4}		\cline{6-6}
&	Triangle Counting	&	&	trmm	&	&	657.xz\_s	\\	\cline{1-1}	\cline{2-2}	\cline{3-3}	\cline{4-4}	\cline{5-5}	\cline{6-6}
\end{tabular}%
}

\end{minipage}
\end{table}

\begin{table*}[h]
\centering
\label{table_aps}
\resizebox{\linewidth}{!}{
\begin{tabular}{|l|l || l|l || l |l|}
\hline
\textbf{Benchmark Suite} &	\textbf{Application} 	&	\textbf{Benchmark Suite} &	\textbf{Application} 	&	\textbf{Benchmark Suite} &	\textbf{Application}	\\ \hline \hline
\multirow{22}{*}{Hardware Effects~\cite{hardwareeffects}} & 	4k aliasing	&	resectionvolume~\cite{palomar2018high} & 	resectionvolume	&	\multirow{14}{*}{SPLASH-2~\cite{woo_isca1995}} & 	barnes	\\		\cline{2-2}	\cline{3-3}	\cline{4-4}		\cline{6-6}
&	bandwidth saturation non-temporal	&	\multirow{19}{*}{Rodinia~\cite{che_iiswc2009}} & 	b+tree	&	&	cholesky	\\		\cline{2-2}		\cline{4-4}		\cline{6-6}
&	bandwidth saturation temporal	&	&	backprop	&	&	fft	\\		\cline{2-2}		\cline{4-4}		\cline{6-6}
&	branch misprediction sort	&	&	bfs	&	&	fmm	\\		\cline{2-2}		\cline{4-4}		\cline{6-6}
&	branch misprediction unsort	&	&	cfd	&	&	lu\_cb	\\		\cline{2-2}		\cline{4-4}		\cline{6-6}
&	branch target misprediction	&	&	heartwall	&	&	lu\_ncb	\\		\cline{2-2}		\cline{4-4}		\cline{6-6}
&	cache conflicts	&	&	hotspot	&	&	ocean\_cp	\\		\cline{2-2}		\cline{4-4}		\cline{6-6}
&	cache/memory hierarchy bandwidth	&	&	hotspot3D	&	&	ocean\_ncp	\\		\cline{2-2}		\cline{4-4}		\cline{6-6}
&	data dependencies	&	&	kmeans	&	&	radiosity	\\		\cline{2-2}		\cline{4-4}		\cline{6-6}
&	denormal floating point numbers	&	&	lavaMD	&	&	radix	\\		\cline{2-2}		\cline{4-4}		\cline{6-6}
&	denormal floating point numbers flush	&	&	leukocyte	&	&	raytrace	\\		\cline{2-2}		\cline{4-4}		\cline{6-6}
&	DRAM refresh interval	&	&	lud	&	&	volrend	\\		\cline{2-2}		\cline{4-4}		\cline{6-6}
&	false sharing	&	&	mummergpu	&	&	water\_nsquared	\\		\cline{2-2}		\cline{4-4}		\cline{6-6}
&	hardware prefetching	&	&	myocyte	&	&	water\_spatial	\\		\cline{2-2}		\cline{4-4}	\cline{5-5}	\cline{6-6}
&	hardware prefetching shuffle	&	&	nn	&	\multirow{8}{*}{Tailbench~\cite{kasture2016tailbench}} & 	img-dnn	\\		\cline{2-2}		\cline{4-4}		\cline{6-6}
&	hardware store elimination	&	&	nw	&	&	masstree	\\		\cline{2-2}		\cline{4-4}		\cline{6-6}
&	memory-bound program	&	&	particlefilter	&	&	moses	\\		\cline{2-2}		\cline{4-4}		\cline{6-6}
&	misaligned accesses	&	&	pathfinder	&	&	shore	\\		\cline{2-2}		\cline{4-4}		\cline{6-6}
&	non-temporal stores	&	&	srad	&	&	silo	\\		\cline{2-2}		\cline{4-4}		\cline{6-6}
&	software prefetching	&	&	streamcluster	&	&	specjbb	\\		\cline{2-2}		\cline{4-4}		\cline{6-6}
&	store buffer capacity	&	\multirow{8}{*}{SD-VBS- Cortex~\cite{Thomas_CortexSuite_IISWC_2014}} & 	lda	&	&	sphinx	\\		\cline{2-2}	\cline{3-3}	\cline{4-4}		\cline{6-6}
&	write combining	&	&	libl	&	&	xapian	\\	\cline{1-1}	\cline{2-2}		\cline{4-4}	\cline{5-5}	\cline{6-6}
\multirow{5}{*}{Hashjoin~\cite{balkesen_TKDE2015}} & 	NPO	&	&	me	&	\multirow{11}{*}{WHISPER~\cite{nalli2017analysis}} & 	ctree	\\		\cline{2-2}		\cline{4-4}		\cline{6-6}
&	PRH	&	&	pca	&	&	echo	\\		\cline{2-2}		\cline{4-4}		\cline{6-6}
&	PRHO	&	&	rbm	&	&	exim	\\		\cline{2-2}		\cline{4-4}		\cline{6-6}
&	PRO	&	&	sphinix	&	&	hashmap	\\		\cline{2-2}		\cline{4-4}		\cline{6-6}
&	RJ	&	&	srr	&	&	memcached	\\	\cline{1-1}	\cline{2-2}		\cline{4-4}		\cline{6-6}
HPCC~\cite{luszczek_hpcc2006} &	RandomAccesses	&	&	svd	&	&	nfs	\\	\cline{1-1}	\cline{2-2}		\cline{3-4}		\cline{6-6}
\multicolumn{4}{|l||}{\multirow{7}{*}{}} 				&	&	redis	\\			\cline{6-6}
\multicolumn{4}{|l||}{}				                &	&	sql	\\						\cline{6-6}
\multicolumn{4}{|l||}{}				                &	&	tpcc	\\						\cline{6-6}
\multicolumn{4}{|l||}{}				                &	&	vacation	\\						\cline{6-6}
\multicolumn{4}{|l||}{}				                &	&	ycsb	\\					\cline{5-5}	\cline{6-6}
\multicolumn{4}{|l||}{}				                &	ZipML~\cite{kara2017fpga} & 	SGD	\\					\cline{5-5}	\cline{6-6}
\multicolumn{4}{|l||}{}				                &	Stream~\cite{mccalpin_stream1995} &	STREAM	\\	\cline{1-1}	\cline{2-2}	\cline{3-3}	\cline{4-4}	\cline{5-5}	\cline{6-6}
\end{tabular}%
}
\bigskip

\end{table*}

\pagebreak
